\documentclass[10pt,preprint]{emulateapj}

\usepackage{graphicx}
\usepackage{epstopdf}
\usepackage{xspace}

\newcommand{\spitzer}{\textit{Spitzer}\xspace}
\newcommand{\msun}{\hbox{$\hbox{M}_{\odot}$}}
\newcommand{\herschel}{\textit{Herschel}\xspace}
\newcommand{\herscheltit}{\textit{Herschel}}
\newcommand{\lsun}{L$_{\odot}$}

\newcommand{\mdotenv}{$\dot{M}_{\rm env}$}

\newcommand{\mdotdisk}{$\dot{M}_{\rm disk}$}

\newcommand{\lbol}{L$_{\rm bol}$}
\newcommand{\tbol}{T$_{\rm bol}$}
\newcommand{\lsl}{L$_{\rm smm}$/L$_{\rm bol}$}
\newcommand{\lsmm}{L$_{\rm smm}$}
\newcommand{\degree}{\mbox{$^{\circ}$}}

\begin{document}

\title{A \herscheltit$^\star$ and APEX Census of the Reddest Sources
  in Orion: Searching for the Youngest Protostars$^{\star\star}$}

\thanks{$^\star$\herschel is an ESA space observatory with
    science instruments provided by European-led Principal
    Investigator consortia and with important participation from
    NASA. }

\thanks{$^{\star\star}$Based (in part) on observations collected at
  the European Organisation for Astronomical Research in the Southern
  Hemisphere, Chile, proposals E-284.C-0515, E-086.C-0848,
  E-088.C-0994, and E-090.C-0894.}

\author{Amelia M.\ Stutz\altaffilmark{3},
John J.\ Tobin\altaffilmark{4},
Thomas Stanke\altaffilmark{5},
S.\ Thomas Megeath\altaffilmark{6},
William J.\ Fischer\altaffilmark{6},
Thomas Robitaille\altaffilmark{3},
Thomas Henning\altaffilmark{3},
Babar Ali\altaffilmark{7},
James di Francesco\altaffilmark{8},
Elise Furlan\altaffilmark{9,7},
Lee Hartmann\altaffilmark{10},
Mayra Osorio\altaffilmark{11},
Thomas L. Wilson\altaffilmark{12},
Lori Allen\altaffilmark{9},
Oliver Krause\altaffilmark{3},
P.\ Manoj\altaffilmark{13}
}

\altaffiltext{3}{Max Planck Institute for Astronomy, K\"onigstuhl 17,
  D-69117 Heidelberg, Germany; stutz@mpia.de}

\altaffiltext{4}{Hubble Fellow, National Radio Astronomy
  Observatory, Charlottesville, VA 22903, USA}

\altaffiltext{5}{ESO, Karl-Schwarzschild-Strasse 2, 85748 Garching bei
  M\"unchen, Germany}

\altaffiltext{6}{Department of Physics and Astronomy, University of
  Toledo, 2801 W.\ Bancroft St., Toledo, OH 43606, USA}

\altaffiltext{7}{NHSC/IPAC/Caltech, 770 S. Wilson Avenue, Pasadena, CA
  91125, USA}

\altaffiltext{8}{National Research Council of Canada, Herzberg
  Institute of Astrophysics, 5071 West Saanich Rd., Victoria, BC, V9E
  2E7, Canada; University of Victoria, Department of Physics and
  Astronomy, PO Box 3055, STN CSC, Victoria, BC, V8W 3P6, Canada}

\altaffiltext{9}{National Optical Astronomy Observatory, 950 N. Cherry
  Avenue, Tucson, AZ, 85719, USA}

\altaffiltext{10}{Department of Astronomy, University of Michigan, 830
  Dennison Building, 500 Church Street, Ann Arbor, MI 48109, USA}

\altaffiltext{11}{Instituto de Astrof\'{i}sica de Andaluc\'{i}a, CSIC,
Camino Bajo de Hu\'{e}tor 50, E-18008, Granada, Spain}

\altaffiltext{12}{Naval Research Laboratory, 4555 Overlook Ave.\ SW,
Washington, DC 20375, USA}

\altaffiltext{13}{Department of Physics and Astronomy, 500 Wilson
  Blvd., University of Rochester, Rochester, NY 14627, USA}

\begin{abstract}
We perform a census of the reddest, and potentially youngest,
protostars in the Orion molecular clouds using data obtained with the
PACS instrument onboard the \herschel\ Space Observatory and the
LABOCA and SABOCA instruments on APEX as part of the Herschel Orion
Protostar Survey (HOPS).  A total of 55 new protostar candidates are
detected at 70~\micron\ and 160~\micron\ that are either too faint
($m_{24} > 7$~mag) to be reliably classified as protostars or
undetected in the \spitzer/MIPS 24~\micron\ band.  We find that the 11
reddest protostar candidates with log~$\lambda F_{\lambda}70/ \lambda
F_{\lambda}24 > 1.65$ are free of contamination and can thus be
reliably explained as protostars.  The remaining 44 sources have less
extreme 70/24 colors, fainter 70~\micron\ fluxes, and higher levels of
contamination.  Taking the previously known sample of
\spitzer\ protostars and the new sample together, we find 18 sources
that have log~$\lambda F_{\lambda}70/ \lambda F_{\lambda}24 > 1.65$;
we name these sources "PACS Bright Red sources", or PBRs.  Our
analysis reveals that the PBRs sample is composed of Class~0 like
sources characterized by very red SEDs (\tbol\,$ < $\,45~K) and large
values of sub--millimeter fluxes (L$_{\rm smm}$/L$_{\rm bol} >
0.6\%$).  Modified black--body fits to the SEDs provide lower limits
to the envelope masses of 0.2~M$_{\odot}$ to 2~M$_{\odot}$ and
luminosities of 0.7~\lsun\ to 10~\lsun.  Based on these properties,
and a comparison of the SEDs with radiative transfer models of
protostars, we conclude that the PBRs are most likely extreme Class~0
objects distinguished by higher than typical envelope densities and
hence, high mass infall rates.
\end{abstract}


\section{Introduction}

The onset of the star formation process is broadly characterized by a
dense collapsing cloud envelope surrounding the nascent protostar. The
dense cloud or protostellar envelope is opaque to radiation shortward
of about $\sim\,$10~\micron\ and most of the radiation from these
sources is reprocessed and emitted in the far--infrared
(FIR). Furthermore, bipolar outflows from the protostar and disk carve
out envelope cavities that enable a fraction of the protostellar
luminosity to escape in the form of scattered light emission,
predominantly at wavelengths shortward of $\sim\,$10~\micron.

The earliest phase of protostellar evolution, the Class~0 phase
\citep{andre93}, is thought to be short compared to the Class~I phase
\citep{lada87}, with combined Class 0 and Class I lifetimes
of $\sim\,$0.5~Myr \citep[][]{evans09}; these estimates assume a
constant star formation rate and a typical age for Class~II objects
(pre--main sequence stars with disks) of 2 Myr.  At the onset of
collapse and immediately before the Class~0 phase, protostars may go
through a brief first hydrostatic core (FHSC) phase where the forming
protostellar object becomes opaque to its own radiation for the first
time \citep{larson69}. The FHSC are expected to be very
low--luminosity and deeply embedded.  A population of very low
luminosity protostars (VeLLOs) were also recently identified by
\spitzer\ \citep[e.g.,][]{dunham08,bourke06}, defined to have
model--estimated internal source luminosities of less than
0.1~$L_{\sun}$.  VeLLOs, however, appear more evolved than FHSCs with
features consistent with Class~0 and I protostars.  Furthermore, while
several FHSC candidates have been identified recently \citep[e.g., see
][]{enoch10,chen10,pineda11,pezzuto12}, it has proven observationally
difficult to distinguish such sources from the young Class~0
protostellar phase.  It is therefore currently difficult to identify
the very earliest phases of the formation of a protostar.

Before the launch of \textit{Spitzer} and the advent of extremely
sensitive mid--infrared surveys, conventional wisdom held that a
Class~0 protostar should not be detectable at wavelengths shortward of
10~\micron\ due to the envelope opacity \citep{williams11}.  Outflows,
however, can carve out cavities in the protostellar envelopes at a
very early age and are expected to widen with evolution
\citep{arce06}. Indeed, recent simulations have shown that even the
extremely young FHSC sources may be capable of driving outflows
\citep{commercon12,price12}.  Regardless of evolutionary state, the
outflow cavities enable near-- to mid--infrared light from the
protostar and disk to escape and scatter off dust grains in the cavity
or on the cavity walls.  This phenomenon has been well--known for
Class~I sources \citep{kenyon93a,padgett99}, but Class~0 protostars
were only well--detected in the mid--infrared with
\spitzer\ \citep[][]{noriega04,jorgensen07,stutz08}.  The scattered
light from Class~0 protostars is often brightest at wavelengths
$\sim\,$3.6~\micron\ or 4.5 \micron\ due to the dense envelope
obscuration at shorter wavelengths
\citep[e.g.,][]{whitney03a,tobin07}.

The combination of results from recent space--based and ground--based
surveys have resulted in well--sampled spectral energy distributions
from the near--infrared to the (sub)millimeter for large samples of
protostellar objects
\citep[e.g.,][]{hatchell07,enoch09,laun10,fischer10,laun12}.  These
SEDs are dominated by scattered light between $\sim\,$1~\micron\ and
10~\micron, optically thick thermal dust emission from $\sim\,$10 to
$\sim\,$160~\micron, and optically thin dust emission at wavelengths
longward of $\sim\,$160~\micron.  Radiative transfer models of
protostellar collapse have become increasingly important to interpret
these data since these can account for the varying temperature and
density profiles in the envelopes surrounding the protostar
\citep[e.g.,][]{whitney03a,whitney03b}.

The large number of free model parameters --- such as the combination
of outflow cavities, rotationally flattened envelopes
\citep{ulrich76,cassen81,terebey84}, and varying viewing angles ---
can however make the best--fit SED model parameters highly degenerate
\citep[e.g.,][]{whitney03a}.  For example, sources viewed at nearly
edge--on orientations can be substantially more obscured than sources
at the same evolutionary state viewed from a less extreme vantage
point.  Thus, standard diagnostics such as bolometric temperature or
mid--infrared spectral index can yield vastly different results
depending on the source inclination \citep[e.g.,][]{dunham10}.
Radiative transfer models can help break some of these degeneracies
but ambiguities can remain as to whether a source has a very dense
envelope or if it is simply viewed edge--on.

While much has been learned about the Class~0 phase from observations
and modeling, there are relatively few Class~0 objects present in
nearby star--forming clouds and globules \citep{evans09} compared to
the numbers of Class~I and Class~II sources. One of the principal
goals of recent star formation surveys has been to understand the
evolution of protostellar sources. \citet{young05} generated models
for the smooth luminosity evolution of protostellar objects that will
become 0.3~\msun\, 1~\msun\ and 3~\msun\ stars.  These models,
however, over predict the luminosities of most protostars located in
nearby star--forming regions, a fact that is taken as evidence for
episodic accretion \citep{kenyon90,evans09,dunham10}.  However,
\citet{offner11} show that the observed luminosity functions of
protostars can be explained through a dependence of the mass accretion
rate on the instantaneous and final mass of the protostar.  The low
resolutions and sensitivities of previous FIR instrumentation have
made the detection of protostars in more distant and richer
star--forming regions difficult and subject to substantial
confusion. Thus, studies of protostellar evolution have been limited
to combining all known Class~0 protostars from the nearby regions into
a single analysis \citep[e.g.,][]{myers98,evans09} to achieve a more
robust sample size.

The advent of the \textit{Herschel Space Observatory} \citep{pil10}
has tremendously improved resolution and sensitivity to FIR radiation,
where protostars emit the bulk of their energy. These improvements
enable the study of protostellar populations to be extended to more
distant, richer regions of star formation that have more statistically
significant samples of protostars in the Class~0 and I phases
\citep[e.g.,][]{ragan12}. The \herschel\ Orion Protostar Survey (HOPS)
is a \herschel\ Open Time Key Programme (OTKP)
\citep[e.g.,][]{stanke10,fischer10,ali10,manoj12} targeting
$\sim\,$300 of the \spitzer\ identified Orion protostars with PACS
\citep{pog10} 70~\micron\ and 160~\micron\ photometry and PACS
spectroscopy \citep[53~\micron\ to 200~\micron; ][]{manoj12} for a
subset of 30 protostars. Orion is the richest star--forming region
within 500~pc of the Sun \citep{megeath12}, at a distance of
$\sim\,$420 pc \citep[average
  value;][]{menten07,hirota07,sandstrom07}. The large sample of
protostars in Orion enables studies of protostellar evolution to be
carried out for a single star forming complex where all protostars lie
at nearly the same distance with a statistically significant sample,
comparable to or larger than all the nearby regions combined.  The
large sample may also enable short timescale phenomena
\citep[e.g.,][]{fischer12} to be detected such as brief periods of
high envelope infall rate in the earliest phases of star formation.
Even with the increased numbers of protostars in Orion, however, it is
unclear if we would expect to detect FHSCs given the faintness of
these sources and short lifetimes of less than 10~kyr
\citep{commercon12}.

The PACS imaging of the HOPS program has the potential to identify
protostars that were not detected by \spitzer\ due to a combination of
opacity of the envelope and/or confusion with nearby sources.  Indeed,
the PACS 70~\micron\ band of \herschel\ is ideal for detecting such
protostars, with the highest angular resolution, limiting the blending
of sources. Also, the lower opacity relative to MIPS
24~\micron\ allows the reprocessed warm inner envelope radiation to
escape.  Finally, and most importantly, a 70~\micron\ point source is
strong evidence for an embedded protostar because external heating
cannot raise temperatures high enough to emit at this
wavelength. Thus, some cores in sub--millimeter surveys that were
previously identified as starless may in reality be protostellar.

Using \herschel, we have serendipitously identified a sample of
70~\micron\ point sources that were not identified in the previous
\spitzer\ protostar sample \citep{megeath12}. Furthermore, we have
identified a subset of these that have the reddest 70~\micron\ to
24~\micron\ colors of all protostars in the combined Orion sample.
These sources may have the densest envelopes and are possibly the
youngest detected Orion protostars and we name them ``PACS Bright Red
sources'' or PBRs.

We will describe the methodology of identifying these sources and
classify them as either being protostellar, extragalactic
contamination, or spurious detections coincident with extended
emission.  We will discuss the observations and data reduction in
\S~2, the source finding and classification methods in \S~3, the
observed properties of the new sources in \S~4, the PBRs in \S~5, the
comparison of the cold PBRs to models in \S~6, some relevant model
degeneracies in \S~7, and, finally, our results in \S~8. Throughout
this work, all positions are given in the J2000 system.

\section{Observations, Data Reduction, and Photometry}

In this work, we present \herschel\ scan map observations of a sub--set
of the HOPS fields containing candidate protostars.  In addition, we
present a subset of our APEX LABOCA and SABOCA observations of these
fields.  A summary of the HOPS \herschel\ PACS survey observations is
presented in Tables~\ref{tab:all_aors_1} and
\ref{tab:all_aors_2}. Here we discuss the observations, data
processing, and photometry extraction.

\subsection{\herschel\ PACS}

The PACS data were acquired simultaneously at 70~\micron\ and
160~\micron\ over 5$\arcmin\times5\arcmin$ or 8$\arcmin\times8\arcmin$
field sizes. The field sizes and centers were chosen to maximize
observing efficiency by allowing each field to include as many of the
target \spitzer--selected protostars \citep{megeath12} as possible
while minimizing redundant coverage.  The observations were
acquired at medium scan speed (20$\arcsec$\,sec$^{-1}$), and are
composed of two orthogonal scans with homogeneous coverage.  

The PACS data were reduced using the \herschel\ Interactive Processing
Environment (HIPE) version 8.0 build 248 and 9.0 build 215.  We used a
custom built pipeline to process data from their raw form (the
so--called Level 0 data) to fully calibrated time lines (Level 1) just
prior to the map--making step.  Our pipeline uses the same processing
steps as described by \citet{pog10} but also include the following
additions and modifications. First, we used a spatial redundancy based
algorithm to identify and mask cosmic ray hits.  Second, we mitigated
instrument cross--talk artifacts by masking (flagging as unusable)
detector array columns affected by cross--talk noise.  This technique
is effective but at the expense of loss of signal from the affected
detector array columns.  Third, we used the "FM6" version of the
instrument responsivity, which has a direct bearing on the absolute
calibration of the final mosaics.

The Level 1 data were processed with ``Scanamorphos'' \citep{roussel12}
version 14.0.  The final maps were produced using the {\it galactic}
option and included the turn--around (non--zero acceleration) data.
The final map pixel scales used in this work are 1.0$\arcsec$/pix at
70~\micron\ and 2.0$\arcsec$/pix at 160~\micron. 

The photometry was performed in the following fashion. We first
derived customized aperture corrections to the 70~\micron\ and
160~\micron\ data using the \herschel\ Science Center (HSC) provided
observations of Vesta (to be discussed in more detail in Fischer et
al., in preparation).  At 70~\micron, we used radii of sizes
$9.6\arcsec$, $9.6\arcsec$, and $19.2\arcsec$, for the aperture and
sky annuli respectively. For these parameters we derived an aperture
correction of $0.7331$, where the measured flux in the aperture is
divided by this correction to obtain a total point--source flux. At
160~\micron, we used aperture radii of $12.8\arcsec$, $12.8\arcsec$,
and $25.6\arcsec$, for the aperture and sky annuli
respectively. Similarly, we derived an aperture correction of
$0.6602$.  The encircled energy fractions provided by the {\tt
  photApertureCorrectionPointSource} task in HIPE do not account for
the effect of applying an inner sky annulus that is close to the size
of the source aperture and small compare to the PSF. Our apertures
therefore account for 3--4\% of the source flux that is removed.
Furthermore, our adopted aperture sizes are smaller than the PACS
instrument team recommendation but were chosen to minimize
contribution from nebulosity (extended, non point--like emission)
often surrounding the protostars in Orion.  Given the complex
structure in the images and at times crowded fields, our aperture
photometry may suffer from blending and contamination.  The
photometric errors include a 10\% systematic error floor added in
quadrature to the standard photometric uncertainties.  These errors
represent systematic uncertainties in our photometry and aperture
correction, as well as the overall calibration uncertainty of PACS.
We note that the reported HSC point-source calibration uncertainties
for PACS are $\sim3$\% at 70~\micron\ and $\sim5$\% at 160~\micron,
and were derived from isolated photometric standards.  Therefore, our
final uncertainties are conservative.

We include the 100~\micron\ Gould Belt Survey \citep[GBS; e.g.,][ see
  also Schneider et al., in preparation, for Orion B, and Roy et al.,
  in preparation, and Polychroni et al., in preparation, for Orion
  A]{andre10,konyves10,mensch10} data of Orion in this work for the
PBRs analysis. Given the sparsely covered SEDs of our sources, these
data provide important information regarding the shape of the thermal
SED of cold envelope sources.  These data were acquired using the
medium scan speed (20$\arcsec$\,sec$^{-1}$) and cover an area much
larger than the HOPS fields.  These data were processed in a similar
way to the HOPS processing described above.  Following the above
70~\micron\ and 160~\micron\ analysis, we used aperture radii of sizes
$9.6\arcsec$, $9.6\arcsec$, and $19.2\arcsec$, for the aperture and
sky annuli respectively.  For these parameters, we derived an aperture
correction of $0.6944$.  As with the 70~\micron\ and
160~\micron\ data, we also assume a conservative 10\% systematic error
floor.

\subsection{APEX SABOCA and LABOCA}

We obtained sub--millimeter (smm) continuum maps using the LABOCA and
SABOCA bolometer arrays on the APEX telescope.  LABOCA
\citep{siringo09} is a $\sim\!$250 bolometer array operating at
870~\micron, with a spatial resolution of $\sim\!19\arcsec$ at FWHM.
We used a combination of spiral and straight on--the--fly scans to
recover extended emission. Data reduction was done with the BOA
software \citep{schuller12} following standard procedures, including
iterative source modeling.  SABOCA \citep{siringo10} is a 37 bolometer
array operating at 350~\micron, with a resolution of
$\sim\!7.3\arcsec$ FWHM.  The observing and data reduction procedures
were similar to those used for LABOCA. For both cameras, observations
were carried out between 2009 November and 2012 June, and are still
ongoing to complete our Submillimeter Orion Survey.  Conditions
were generally fair over the course of our observing campaign.  The
observations will be summarized in more detail by Stanke et
al.\ 2013, in preparation.

The beam sizes of the final reduced maps are $7.34\arcsec$ and
$19.0\arcsec$ FWHM for the SABOCA and LABOCA observations,
respectively.  The photometry was extracted in the same way for both
wavelengths. When possible, if there was a strong source detection, we
re-centered using the 70~\micron\ catalog source coordinates.
Given the contributions of flux due to surrounding cold material, such
as filaments and other extended envelope structure, it is likely that
a single photometric measure can suffer from large systematic effects.
We have measured source fluxes in three ways: \\
\noindent 1.\ We measured the source peak flux per beam.  \\
\noindent 2.\ We measured source flux over an aperture with radius
equal to the FWHM at the corresponding wavelength (r = $7.34\arcsec$
and $19.0\arcsec$ at 350 and 870~\micron, respectively), using a sky
annulus with inner and outer radii equal to $[1.5,2.0]\times$FWHM,
corresponding to $11.0\arcsec$ and $14.7\arcsec$ at 350~\micron, and
$28.5\arcsec$ and $38.0\arcsec$ at 870~\micron.  \\
\noindent 3.\ We measured the flux over the same aperture size as the
previous method without any sky subtraction.  In the case that a
source was not strongly detected and we were not able to re--center,
the 70~\micron\ catalog source coordinates was used, along with method
3, and the photometric point was flagged as an upper limit.  By
re--centering whenever possible, we accounted for possible pointing
offsets between data--sets, which can be significant.  The calibration
error dominated the error budget for well detected sources; we
therefore adopted a flux error equal to 20\% and 40\% of the measured
flux for LABOCA and SABOCA respectively.  The photometric fluxes are
presented in Table~\ref{tab:photapex}.

\subsection{\spitzer\ IRAC and MIPS}

The IRAC and MIPS imaging and photometry presented here are taken from
the 9 degree$^2$ survey of the Orion A and B cloud obtained during
the cryogenic \spitzer\ mission.  The data analysis, extraction of the
IRAC 3.6~\micron, 4.5~\micron, 5.8~\micron, and 8~\micron\ and MIPS
24~\micron\ photometry, and the compilation of a point source catalog
containing the combined 2MASS, IRAC and MIPS photometry are described
in \citet{megeath12}; also see \citet{kryukova12} for a detailed
description of the MIPS 24~\micron\ photometry.  In total, 298405 point
sources were detected in at least one of the \spitzer\ bands, and
8021 sources were detected at 24~\micron\ with uncertainties $\le
0.25$~mag. The \spitzer\ images used in this work are taken from the
mosaics generated from the Orion Survey data using Cluster Grinder for
the IRAC data \citep{gutermuth09} and the MIPS instrument team's Data
Analysis Tool for the 24~\micron\ data \citep{gordon05}. The MIPS data
are saturated toward the Orion Nebula and parts in the NGC2024
region; we exclude these saturated regions from our analysis.

The identification of protostars with the \spitzer\ data was based
primarily on the presence of a flat or rising spectral energy
distributions between 4.5~$\mu$m and 24~$\mu$m
\citep{kryukova12,megeath12}.  In addition, Megeath et al.\ identified
objects which have point source detections only at 24~$\mu$m but which
also showed other indicators of protostellar nature such as the
presence of jets in the IRAC bands. To minimize contamination from
galaxies, the \spitzer--identified protostars were required to have
24~$\mu$m magnitudes brighter than $7$th magnitude; fainter than
7~magnitudes, the number of background galaxies begins to dominate
over the number of embedded sources \citep{kryukova12}.  Given the
imposed 24~$\mu$m magnitude threshold, the faintest and reddest
protostars may not be included in the \spitzer\ sample.  In total, the
\citet{megeath12} catalog contains 488 protostars.  Of these, 428 are
classified as bona fide protostars, 50 are faint candidate protostars,
and 10 are red candidate protostars.  The faint candidate protostars
are sources with 24~\micron\ magnitudes higher than 7.0.  The red
protostars are sources that are only detected as a point source at
24~\micron\ and are thus not classifiable. Due to their location
within high extinction regions and/or association with jets or compact
scattered light nebulae in the IRAC bands, , they have been included
in the catalog.  (Indeed, this last category was added to the Megeath
et al.\ catalog after the \herschel\ data revealed that a relatively
large number of such sources would likely be confirmed as protostars.)
In addition, the Megeath et al.\ catalog identified 2992 objects as
pre--main sequence stars with disks.  In what follows,
we use the protostar catalog of 488 \spitzer\ sources to catalog
previously identified sources in the HOPS images.

In contrast to the full Megeath et al.\ catalog, the HOPS protostar
sample represents a sub--set of that catalog.  The HOPS sample is not
a complete representation of the Megeath et al.\ catalog because the
HOPS survey targeted a smaller area then the \spitzer\ survey and
required an extrapolated 70~\micron\ flux that would be detectable
with PACS.  However, the HOPS catalog represents our best
pre--\herschel\ knowledge of the protostellar content within the HOPS
fields.  The HOPS sample is therefore the best one against which to
bench--mark and compare the new \herschel\ protostellar candidates.
The HOPS sample constitutes $\sim\,300$ previously identified
\spitzer\ protostars all having \herschel\ PACS
70~\micron\ detections.  Of these, $\sim\,250$ HOPS sources have
PACS~160~\micron\ detections.  This sample will be analyzed in detail
in Fischer et al., in preparation.

In contrast to the full Megeath et al.\ catalog, the HOPS protostar
sample are those protostars specifically targeted by the HOPS program.
The majority of the HOPS sample consists of \spitzer--identified
protostars with 24~\micron\ detections; hence, protostars in regions
that are saturated in the 24~\micron\ images of Orion, namely the
brightest regions of the Orion Nebula and NGC2024, are not
included. These protostars were also required to have a predicted
70~\micron\ flux $>$~20~mJy as extrapolated from their 3~\micron\ to
24~\micron\ SEDs. In addition, protostar candidates with only
24~\micron\ detections were included if there was independent
information of their protostellar nature. The HOPS sample represents
the best pre--\herschel catalog of protostars that were expected to be
detected with \herschel/PACS and were not found in bright nebulous
regions.  There are $\sim\,$300 protostars in the HOPS catalog which
have been detected at 70~\micron\ and $\sim\,$250 protostars detected
at 160 um.  The uncertainty in there absolute number of protostars is
due to the ongoing process of eliminating contamination from the
sample.

\begin{figure}[t]
  \begin{center}
    \scalebox{0.49}{\includegraphics{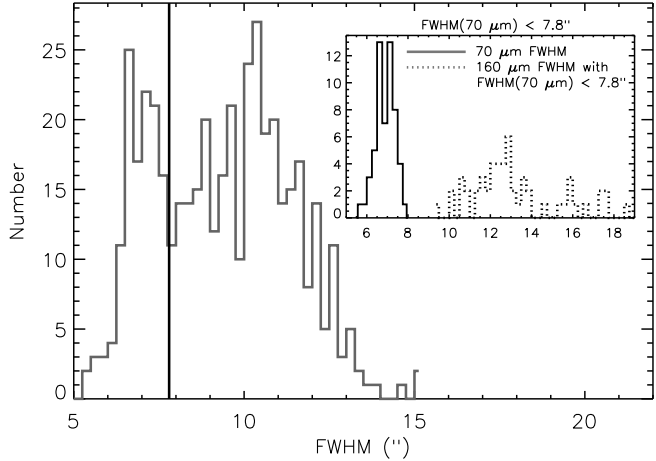}}
    \caption{Frequency of 70~\micron\ FWHM values for all sources
      detected in our HOPS images that have not been previously
      identified in the \spitzer\ catalog.  The black line shows our
      adopted 7.8$\arcsec$ FWHM threshold, above which we reject all
      sources. {\it Inset:} Same as above for the subset of sources
      with both FWHM(70~\micron) $<$ 7.8$\arcsec$ and
      160~\micron\ detections (see \S~3). The dotted histogram
      indicates the 160~\micron\ FWHM distribution.}
    \label{fig:fwhm}
  \end{center}
\end{figure}

\section{Identification of new candidate \herschel\ protostars}

To find protostars which were not reliably identified with \spitzer,
we must first isolate a sample of sources that are detected in the
PACS 70~\micron\ band but are either fainter than 7.0 magnitudes or
undetected in MIPS 24~\micron\ waveband.  To identify such sources in
each HOPS field we first generate a 70~\micron\ source catalog using
the PhotVis tool~\citep{gutermuth08}.  The PhotVis tool uses a sunken
Gaussian filtering to extract sources that are of order the size of
the Gaussian FWHM, an input parameter.  We choose this parameter to be
the size of the 70~\micron\ PSF FWHM, or 5$\arcsec$.  PhotVis also
requires a SNR threshold as input; we adopt a low value of 7 to
balance the recovery of as many candidate sources as possible while
still rejecting noise spikes.

Furthermore, we must reject unreliable sources near the edges of maps
where the lower coverage causes elevated noise levels.  The
Scanamorphos scan map image cubes include a weight map for each field.
Within Scanamorphos, the weight map is computed over the same
projection as the sky map, and is defined as one over the variance in
the white noise \citep{roussel12}.  Each weight map is then normalized
by the average map value \citep{roussel12}.  We find that for the HOPS
data--set, weight map values of $\sim\!$20 are confined to the outer
higher noise edges of our scan maps.  We therefore use the weight maps
to reject edge sources from the catalog at this phase of the analysis.
We accomplish this by requiring that the mean value of the weight map
in a $9\times9$ pixel area centered on the candidate source have a
value of at least 20. For reference, all HOPS 70~\micron\ scan maps
have weight map values greater than 60 over most of the map areas.

The resulting preliminary source catalog includes all sources in the
70~\micron\ images, i.e., previously identified \spitzer\ sources, new
candidate protostars, nebulosity, and other undesirable features and
artifacts in the images.  We then cross--correlate this PACS
70~\micron\ preliminary catalog with the existing \spitzer\ catalog to
eliminate all previously identified protostars in each field that are
brighter than the previously adopted 24~\micron\ cutoff of 7~mag
\citep{megeath12}.  Therefore our sample includes by definition only
sources that are faint or undetected in the previous
\spitzer\ catalog.

To identify previous source detections, we require that a source be
matched to within a positional offset of 8$\arcsec$ when
cross--correlated with the \spitzer\ catalog.  This threshold is
conservatively large compared to the \spitzer\ astrometry and is meant
to encompass two main sources of astrometric error.  First, it is
possible that the absolute coordinates of a source may shift as
function of wavelength (although this effect is expected to be
relatively small) since different wavelengths may trace different
material near the protostars.  Second, the \herschel\ pointing
accuracy, which is of order $\sim\,2\arcsec$ ($1\sigma)$, dominates
the positional accuracy for most sources when comparing the
\spitzer\ catalog source coordinates to the
\herschel\ 70~\micron\ coordinates.  To match coordinates robustly, we
therefore adopt a conservatively large 8$\arcsec$ threshold.  We
find that inspection of the matched sources by eye shows that this
threshold works well and provides a low rate of mismatched or
duplicate sources.

Our final goal is to obtain a sample of previously unidentified and
uncharacterized \herschel\ protostar candidates.  These sources should
be characterized by a point--like appearance at 70~\micron. Therefore,
after rejecting all \spitzer\ protostars as described above, we next
apply a simple FWHM (or apparent size) filter to the remaining
70~\micron\ sources.  The distribution of 70~\micron\ azimuthally
averaged FWHM values is shown in Figure~\ref{fig:fwhm} as the solid
black histogram.  We find a clear peak in the distribution at low FWHM
values, indicating a population of point--like sources.  Based on this
distribution, we adopt a FWHM threshold of 7.8$\arcsec$, meant to
select 70~\micron\ point--sources.  We find 127 sources that fulfill
the criteria listed here: 85 of these have 24~\micron\ detections
while 42 do not.

In a further step, we then require that all 70~\micron\ sources also
have a 160~\micron\ detection and not upper limits.  The
70~\micron\ FWHM distribution of this sub--set of sources is shown in
the inset of Figure~\ref{fig:fwhm} along with the corresponding
160~\micron\ FWHM distribution.  Our final sample consists of 55
candidate \herschel\ protostars with both 70~\micron\ and
160~\micron\ detections.  Of these, 34 have
\spitzer\ 24~\micron\ detections fainter than 7.0 magnitudes and 21 do
not have any 24~\micron\ detection.

\begin{figure}[t]
  \begin{center}
    \scalebox{0.49}{\includegraphics{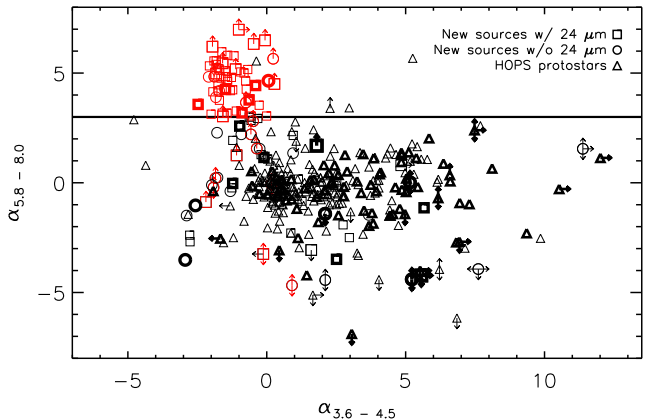}}
    \caption{IRAC color indices $\alpha_{3.6-4.5}$ and
      $\alpha_{5.8-8.0}$ for sources detected in at least one IRAC
      band.  The \herschel\ protostar candidate sample is shown as
      squares (sources with 24~\micron\ detections) and circles
      (sources without 24~\micron\ detections), while the HOPS
      protostar sample is shown as triangles. Thick bold points are
      sources with robust 870~\micron\ detections.  Candidate
      protostars with $\alpha_{5.8-8.0} > 3.0$ or upper limits
      consistent with this threshold are flagged as possible
      extragalactic contamination and highlighted in red. Note that
      some candidate protostars have both $\alpha_{5.8-8.0} > 3.0$ and
      a robust 870~\micron\ detection; these sources are not
      considered extragalactic contamination.}
    \label{fig:alpha}
  \end{center}
\end{figure}  

\begin{figure}[t]
  \begin{center}
    \scalebox{0.49}{\includegraphics{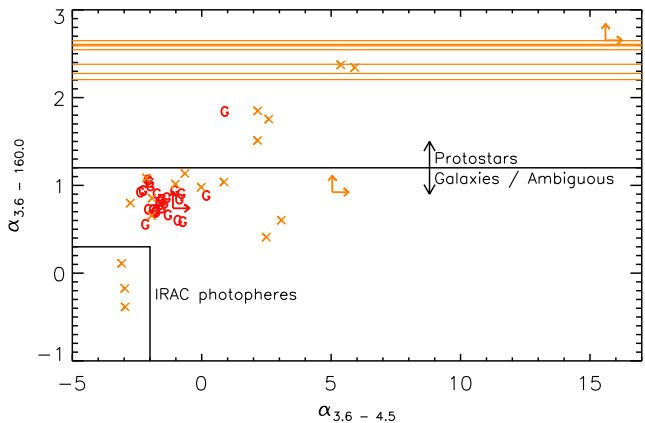}}
    \caption{IRAC and PACS 160~\micron\ color indices
      $\alpha_{3.6-4.5}$ and $\alpha_{3.6-160}$ for the candidate
      protostar sample.  Sources previously characterized as
      extragalactic based on their $\alpha_{5.8-8.0} > 3$ index (or
      limit) are indicated in red with a ``G''.  Sources with
      $\alpha_{5.8-8.0} < 3$ are indicated in orange. The orange lines
      indicate the lower limits of $\alpha_{3.6-160}$ for sources with
      no IRAC detections.  Sources with $\alpha_{3.6-160} > 1.2$ are
      considered highly probable protostars, while other sources are
      flagged as described in the text.}
    \label{fig:alpha160}
  \end{center}
\end{figure}  

\begin{figure*}[t]
  \scalebox{0.375}{\includegraphics{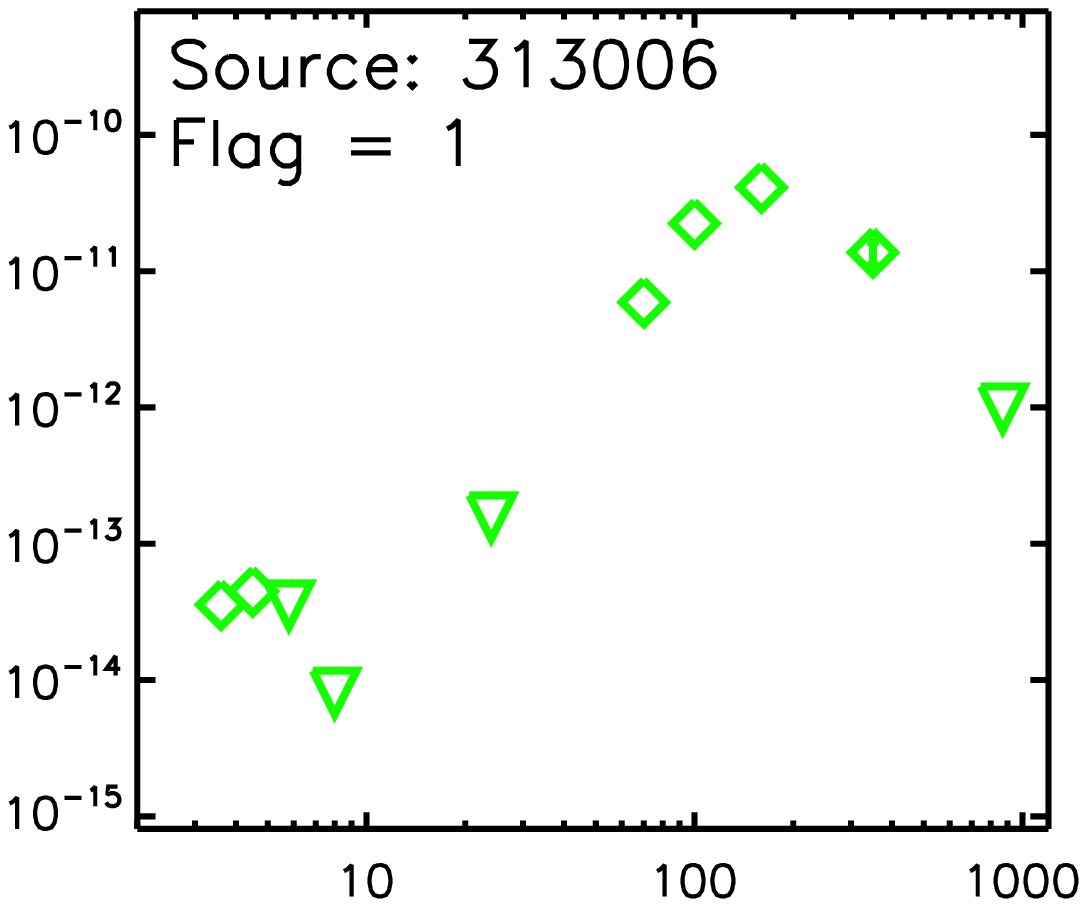}\includegraphics{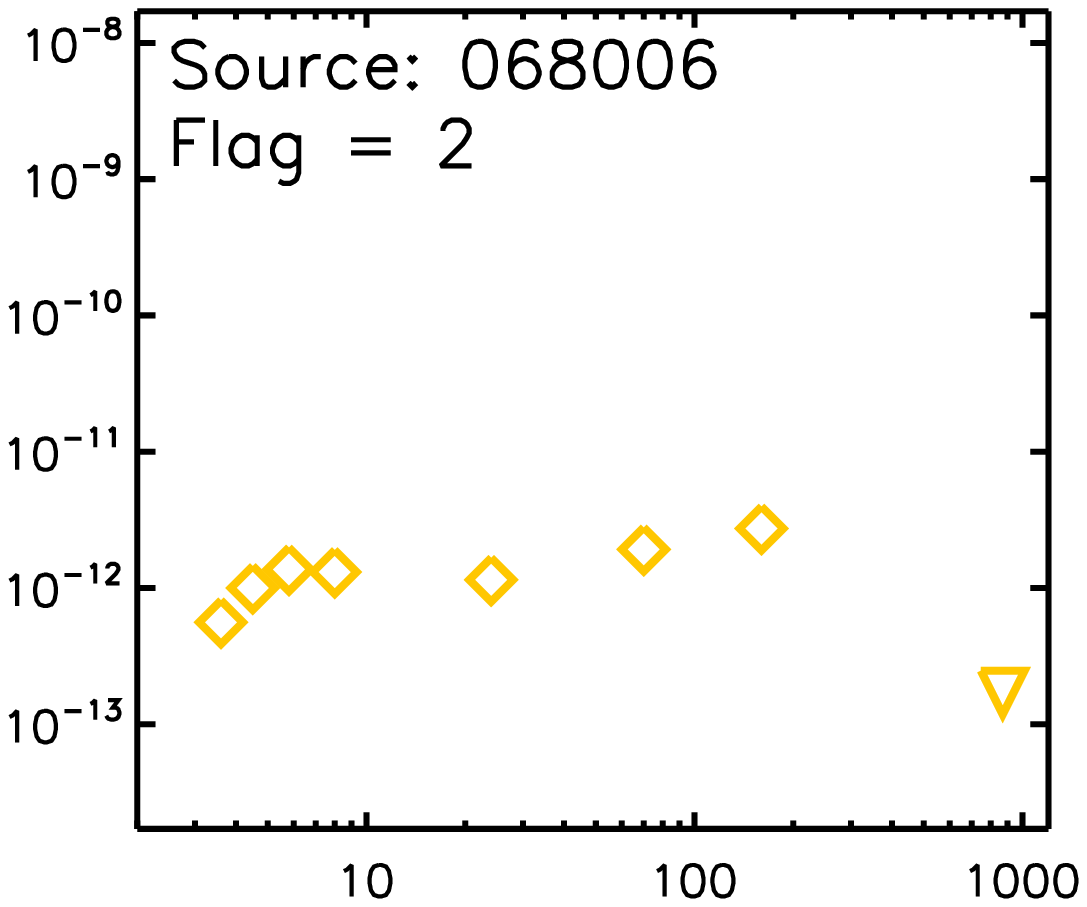}\includegraphics{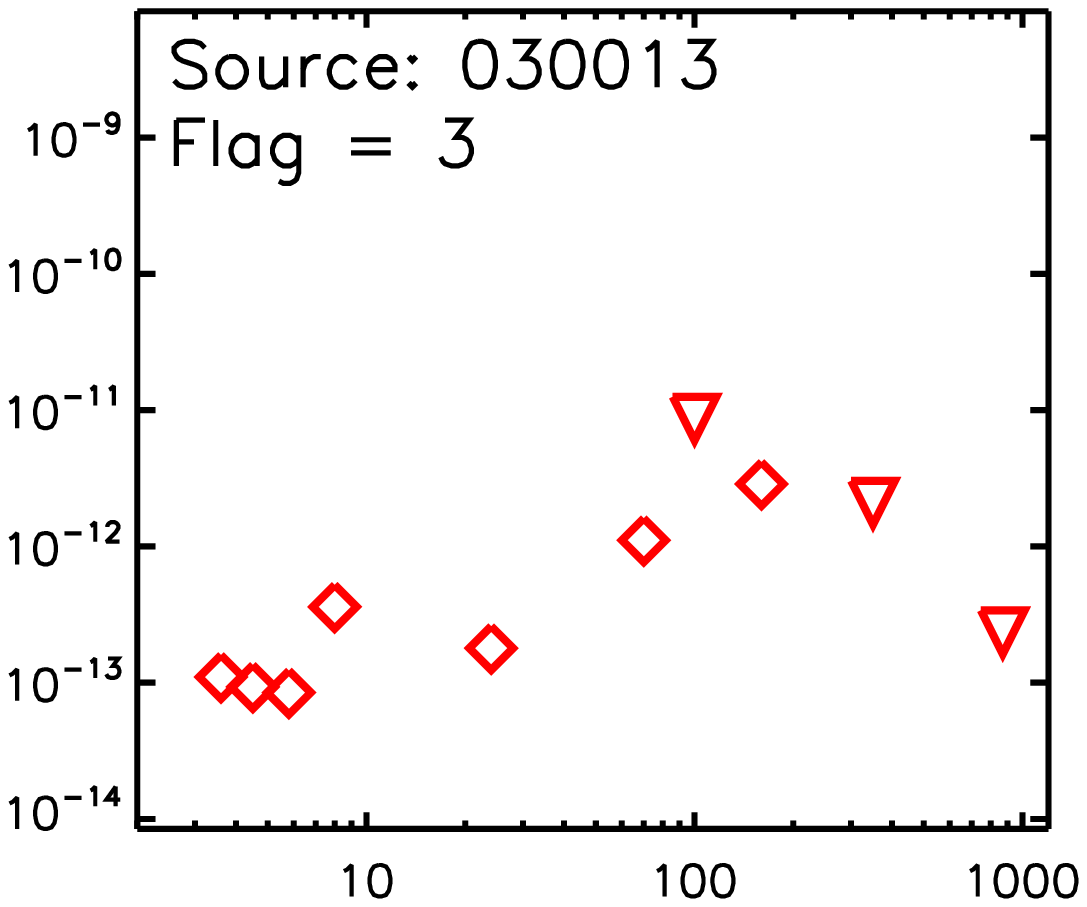}\includegraphics{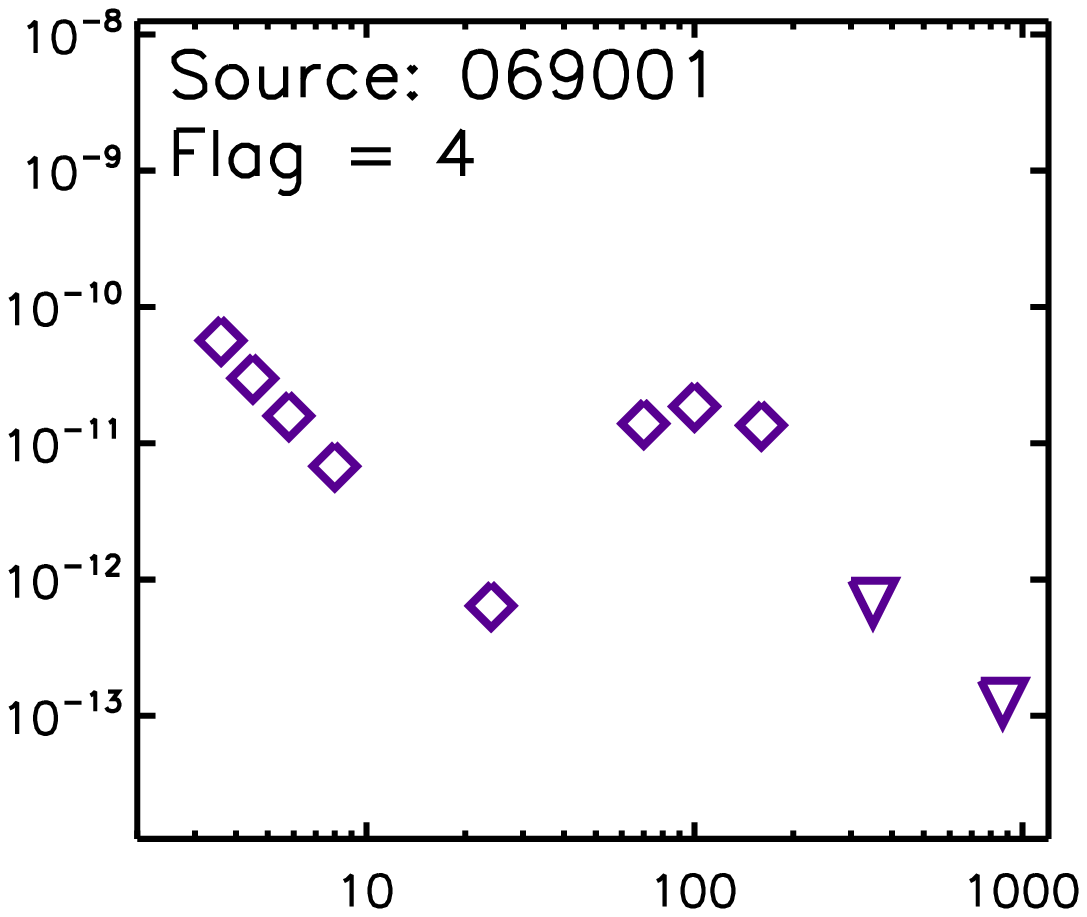}}

  \scalebox{0.375}{\includegraphics{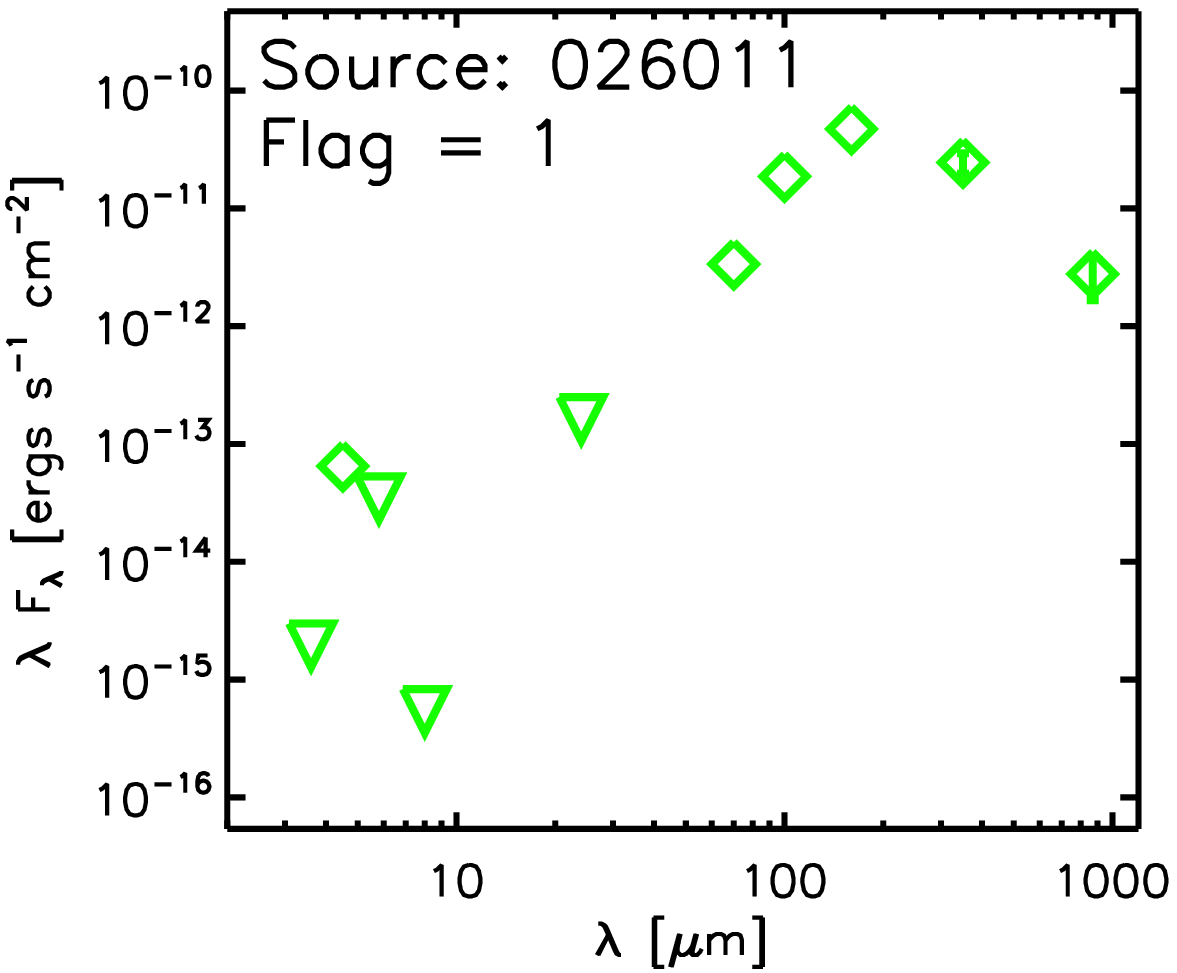}\includegraphics{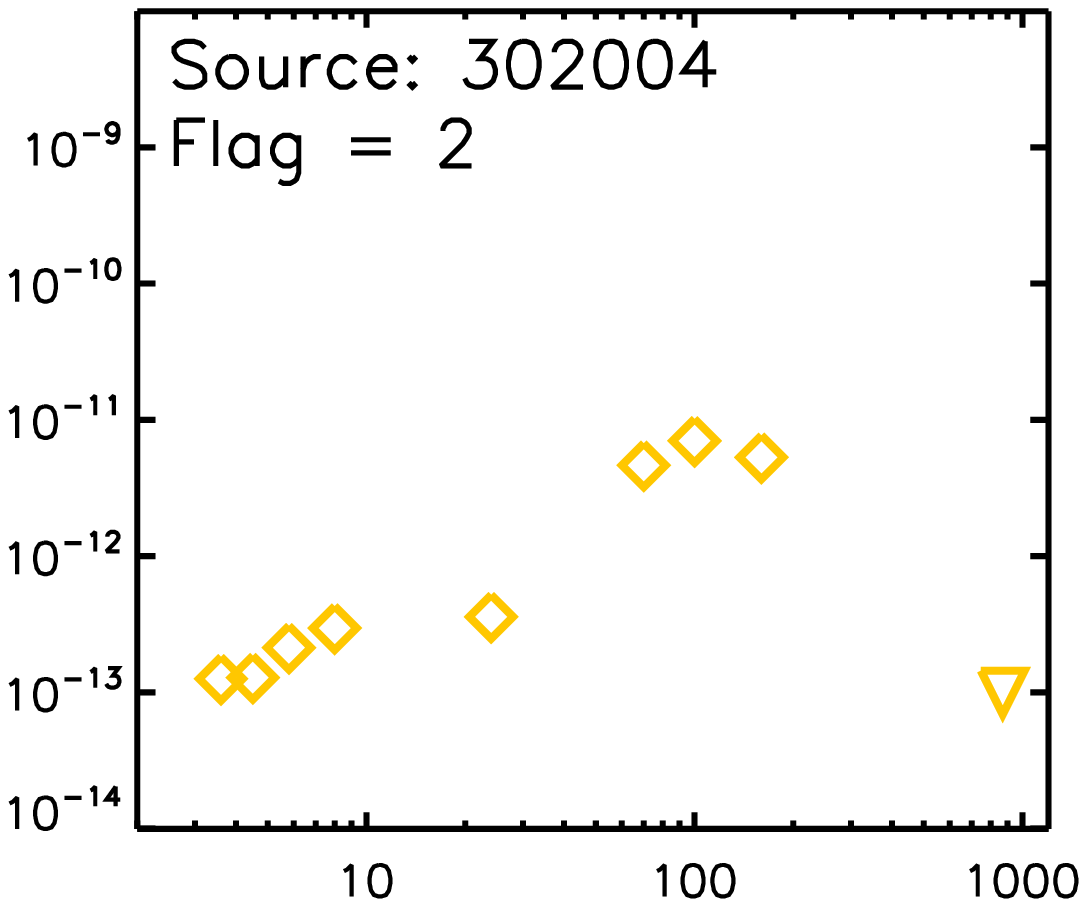}\includegraphics{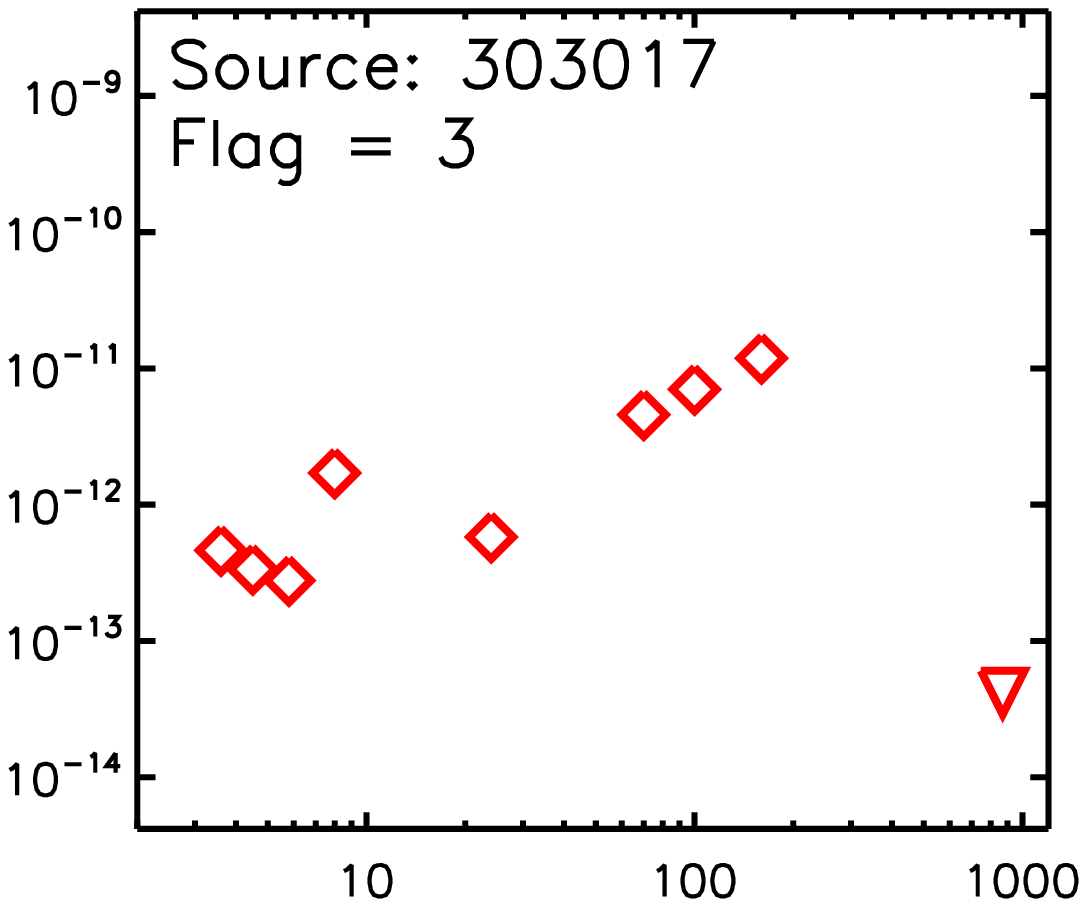}\includegraphics{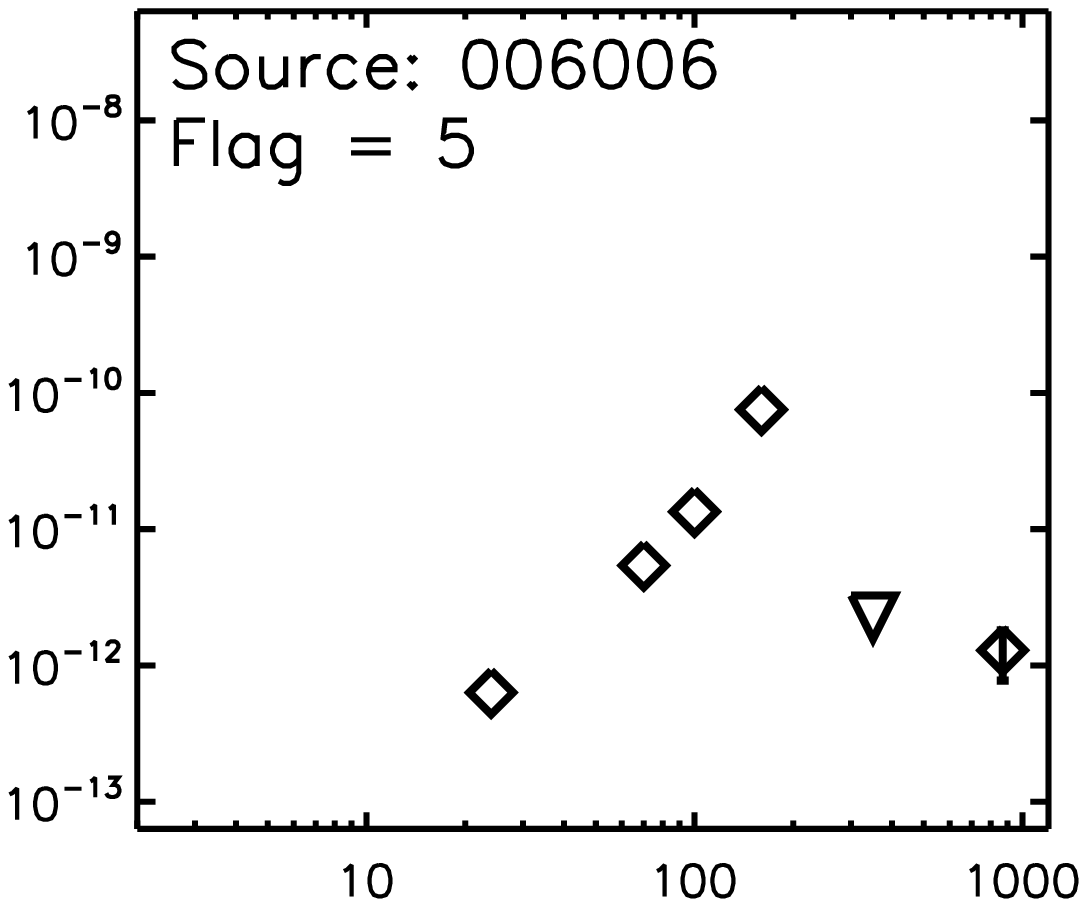}}
  \begin{center}
    \caption{Example SEDs for each of the 5 categories described in
      \S~3.2: green SEDs (flag $=1$) are considered to be reliable
      protostar candidates, yellow SEDs (flag $=2$) are less likely to
      be protostellar, red SEDs (flag $=3$) are considered
      extragalactic contamination based on the 8~\micron\ PAH feature,
      the purple SED (flag $=4$) is an example of one of 3 sources
      that may be explained as transition disks, and, finally, the
      black SED (flag $=5$) is unclassifiable due to the lack of IRAC
      coverage.  Errors are smaller than the size of the points,
      except for the 350~\micron\ and 870~\micron\ points; upper
      limits are indicated with triangles.}
    \label{fig:sedclass}
  \end{center}
\end{figure*}    

\subsection{\spitzer\ non--detections}

A search for newly detected protostars using \herschel\ requires us to
determine upper limits at 24~\micron\ for those sources that are not
detected by \spitzer.  To determine these limits, we adapt the method
developed by Megeath et al.\ to assess the spatially varying
completeness of the \spitzer\ Orion Survey data.  The completeness of
the 24~\micron\ data depends strongly on the presence of nebulosity
and point source crowding.  To account for these factors, we measure
the fluctuations of the 24~\micron\ signal in an annulus centered on
the position of the \herschel\ point--source using the the root median
square deviation, or RMEDSQ \citep[see Equation 1 of][]{megeath12}.
We then use the results from the artificial star tests \citep[see
  Appendix of][]{megeath12} to determine the magnitude at which 90\%
of the point sources would be detected for the observed level of
fluctuations.  We convert this magnitude into a flux density to
obtain 24~\micron\ upper limits.

Several of the identified protostars show IRAC emission but are not
included in the Megeath et al.\ point source catalog because they are
spatially extended.  To obtain homogeneously extracted IRAC fluxes for
the entire sample of sources, we measure fluxes using an aperture of
2 pixels, with a sky annulus of 2 to 6 pixels, corresponding to
aperture radii of 2.44\arcsec, 2.44\arcsec, and 7.33\arcsec,
respectively, with a pixel scale of 1.22\arcsec~pixel$^{-1}$.  We use 
the PACS 70~\micron\ source coordinates as starting guesses, and
attempt to re--center at each IRAC wavelength.  If the re-centering
fails, as for sources with no IRAC detections, we take the integrated
flux in that aperture at the original PACS 70~\micron\ source
coordinate to be the upper limit.  The aperture corrections and
photometric zero--points are those given by Kryukova et al.

\subsection{Contamination in the sample} 

Galaxies often exhibit infrared colors similar to those of young
stellar objects (YSOs) due to the presence of dust and hydrocarbons in
the galaxies \citep{stern05}.  Extensive work has been done towards
characterizing the extragalactic ``contamination'' in
\spitzer\ surveys of star--forming regions and mitigating it through
photometric criteria designed to separate galaxies from bona fide YSOs
\citep{gutermuth09,gutermuth08,harvey07}. These authors show that
star--forming galaxies can be distinguished from YSOs by the galaxies'
stellar--like emission in the IRAC 3.6~\micron\ and 4.5~\micron\ bands
and their bright polycyclic aromatic hydrocarbon (PAH) emission in the
IRAC 5.8~\micron\ and 8.0~\micron\ bands
\citep{gutermuth09,gutermuth08,winston07,stern05}.  However, we note
that some Active Galactic Nuclei (AGN) dominated galaxies may not
exhibit PAH emission; therefore an analysis based only on the IRAC
colors may not capture all possible sources of contamination
\citep{robitaille08}.

To analyze the IRAC colors of our sample, we define $\alpha = d
log(\lambda F_\lambda) / d log(\lambda)$.  In Figure~\ref{fig:alpha},
we plot $\alpha_{5.8-8}$ vs.\ $\alpha_{3.6-4.5}$ for the sample of new
\herschel\ sources with coverage in all four of the \spitzer/IRAC
bands and detections in at least one band, compared to the HOPS
protostar sample.  This color index is relatively insensitive to
reddening since the extinction in the 5.8~\micron\ and 8~$\mu$m bands
of IRAC are very similar \citep[e.g.,][]{flaherty07,gutermuth08}.
Figure~\ref{fig:alpha} shows a cluster of sources with high values of
$\alpha_{5.8-8}$ (i.e., $\alpha_{5.8-8} \ge 3$; solid horizontal line)
yet $\alpha_{3.6-4.5}$ values of an SED that is declining or flat with
increasing wavelength.  These sources show the characteristics of
star--forming galaxies with bright PAH emission (resulting in high
values of $\alpha_{5.8-8}$) but values of $\alpha_{3.6-4.5}$ that are
dominated by starlight.  In our adopted scheme, $\alpha_{5.8-8} \geq
3$ corresponds to a color of $[5.8]-[8] \geq 2.17$; this threshold is
higher than the $[5.8]-[8] \geq 1$ threshold used by
\citet{gutermuth08} to isolate galaxies and thus ensures that most
protostellar candidates will be less likely to be miss--identified
extragalactic sources \citep{allen04,megeath04}.  We identify the
cluster of sources with $\alpha_{5.8-8} \geq 3$ and $\alpha_{3.6-4.5}
\leq 0.5$ as likely extragalactic contamination.  We note that nebular
contamination of the photometry can cause PAH--like $\alpha_{5.8-8}$
values, and thus may cause us to over--estimate the extragalactic
contamination.  Of the 55 sources identified here as protostellar
candidates we flag 23 as possible extragalactic contamination based on
this criteria.  However, other sources of contamination, such as AGN
lacking PAH emission, may remain in our sample.

Inspection of the SEDs of the remaining 32 sources show a range of SED
slopes and shapes.  It is possible that such sources may also be
extragalactic contamination by AGN, for example.  To address this, we
further refine the analysis presented above by analyzing the
3.6~\micron\ to 160~\micron\ SED shapes with the spectral index
$\alpha_{3.6-160}$.  As illustrated in Figure~\ref{fig:alpha160}, the
sources flagged as extragalactic based on the $\alpha_{5.8-8}$ index
(red points) generally have $\alpha_{3.6-160} \lesssim 1.2$.  We
therefore calibrate the $\alpha_{3.6-160}$ relative the reliable
extragalactic candidates with robust IRAC detections and expect that
extragalactic sources will have $\alpha_{3.6-160} \lesssim 1.2$.
Using this criterion, we refine our source classification as follows.
All sources with $\alpha_{3.6-160} > 1.2$ (and $\alpha_{5.8-8} < 3$
when IRAC detections exist) are flagged as high probability protostars
(flag $= 1$ in Table~\ref{tab:pbrscat}).  Sources having values of
$0.5 \lesssim \alpha_{3.6-160} \lesssim 1.2$ but that originally
classified as candidate protostars based on a low value of
$\alpha_{5.8-8}$ are flagged as less likely to be of a protostellar
nature (flag $= 2$ in Table~\ref{tab:pbrscat}).  Furthermore, by
definition, sources originally classified as extragalactic based on
their PAH signature at 8~\micron\ remain classified as such (flag $=3$
in Table~\ref{tab:pbrscat}).  Sources with $\alpha_{3.6-160} < 0.3$
and $\alpha_{3.6-4.5} \sim -3$ are flagged as ``other'' (flag = 4 in
Table 3) since their SEDs are consistent with a stellar photosphere at
shorter wavelengths. Finally, one source has no IRAC coverage and
therefore is flagged with a value of 5.  In Figure~\ref{fig:sedclass}
we show example SEDs of each category.

Only one source (313006) originally flagged as extragalactic based on
its $\alpha_{5.8-8}$ limit (non--detection at 5.8~\micron\ and
8~\micron) was revised to a highly probable protostar (see
Figure~\ref{fig:alpha160} and top left panel of
Figure~\ref{fig:sedclass}).  In addition, as we note above, we find
three sources with SEDs that we label ``other'' (flag value $=4$)
which are inconsistent with the categories described above.  Source
069001 (see Figure~\ref{fig:sedclass}, top right panel) was previously
characterized by \citet{fang09} as a K7 star with a debris disk, with
a very poorly constrained age of $\sim0.06^{+4.66}_{-0.03}$~Myr.
These authors only include data up to 24~\micron.  The SED we observe
with \herschel\ may be consistent with a transition disk but not a
debris disk.  The remaining two sources in this category have similar
SEDs as that of 069001; while a transition disk explanation for all
three sources may appear likely depending on the age of the sources,
we cannot currently rule out other possibilities.  Nevertheless, all
of these sources have SEDs consistent with a stellar photosphere in
the IRAC bands, and hence these are likely to be fully formed stars
surrounded by circumstellar dust.

Interestingly, we find that the most reliable SED classification
criterion by far is that of sources that have neither IRAC nor
MIPS~24~\micron\ detections.  Of these, we find that all 6 sources 
have strong sub--millimeter detections and reside in dense and
filamentary environments.  This finding points to the critical
importance of obtaining high resolution sub--millimeter data to
constrain the properties of such sources.  In the following text, we
include all 55 \herschel--detected sources in our analysis and figures.

\begin{figure}[t]
  \begin{center}
    \scalebox{0.49}{\includegraphics{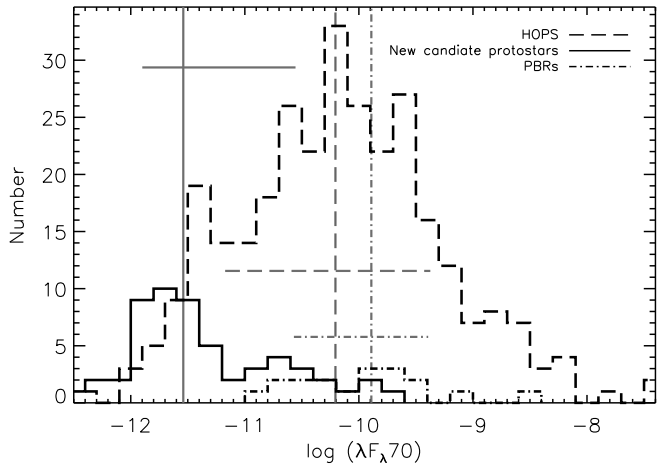}}
    \caption{Distributions of 70~\micron\ fluxes for the HOPS
      protostars (dashed histogram), new candidate protostars (solid
      histogram), and the 18 reddest sources drawn from both combined
      samples (dot--dashed histogram).  The corresponding grey lines
      indicate the median flux values (vertical lines) and the 68\%
      interval (horizontal lines).}
    \label{fig:70dist}
  \end{center}
\end{figure}    

\begin{figure*}
  \begin{center}
    \scalebox{0.81}{\includegraphics{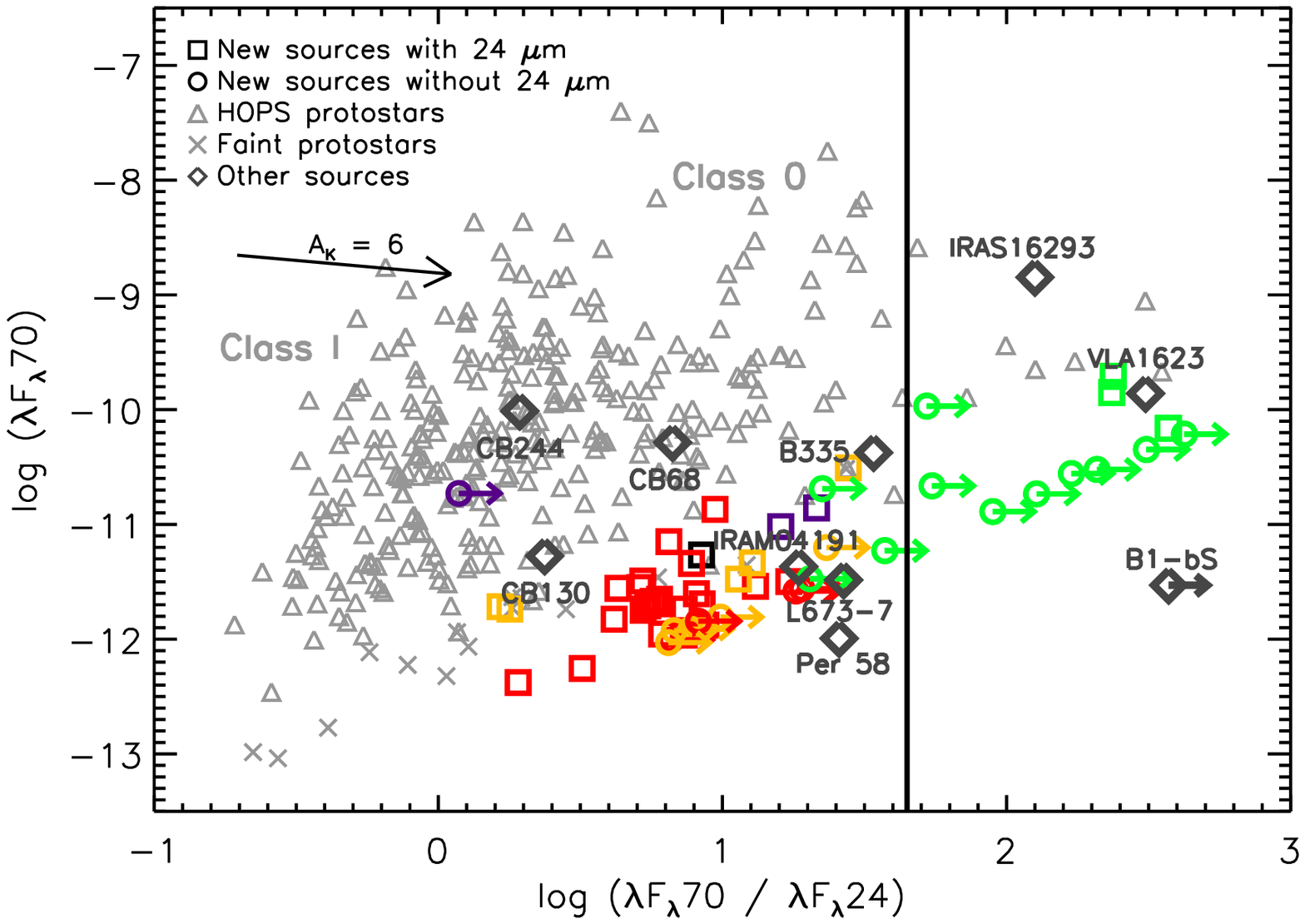}}
    \scalebox{0.81}{\includegraphics{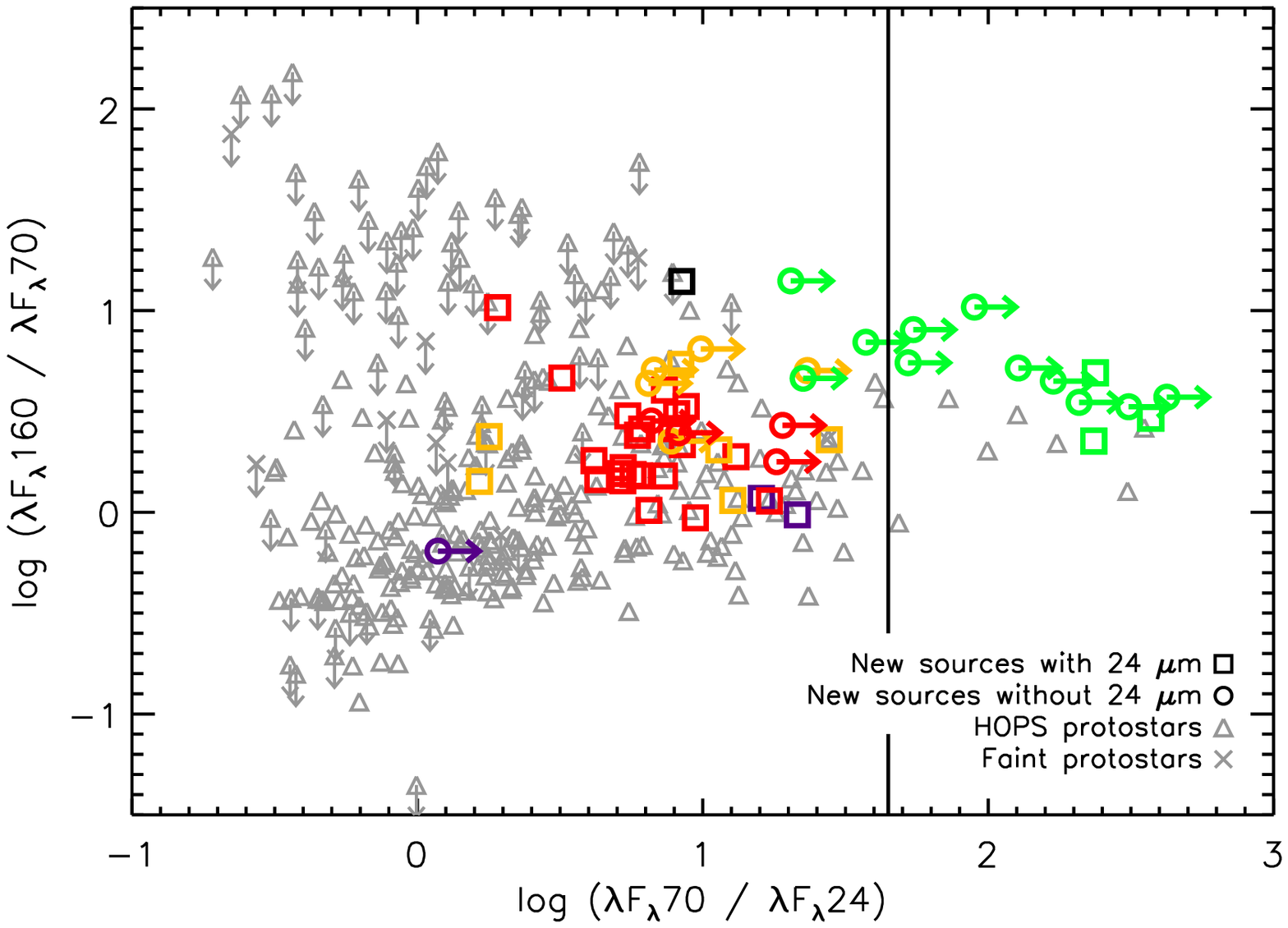}}
    \caption{{\it Top:} 70~\micron\ flux vs. 70~\micron\ to
      24~\micron\ flux ratio for HOPS--detected Orion protostars. Open
      squares and open circles indicate the new candidate protostars
      with and without MIPS 24~\micron\ detections, respectively.  The
      color of the symbols indicates the flag values shown in
      Table~\ref{tab:pbrscat} and discussed in \S~3.2: here, green
      indicates sources that are reliable protostar candidates (flag
      $=1$), orange indicates less reliable sources (flag $=2$), red
      indicates extragalactic contamination (flag $=3$), purple
      indicates other sources (flag $=4$), and black indicates the
      single source without IRAC coverage.  The solid vertical line
      indicates a 70~\micron\ to 24~\micron\ flux ratio of 1.65, our
      PBRs selection criterion (see \S~5).  Triangles and $\times$
      symbols indicate previously detected and characterized
      \spitzer\ protostars and faint protostellar candidates from
      \citet{megeath12}, respectively.  The arrow indicates the
      extinction vector for a value of $A_K = 6$. For comparison, we
      also indicate as large diamonds the scaled observed measurements
      for some well--known Class~0 sources, VeLLOs, and two FHSC
      candidates (see text). {\it Bottom:} 160~\micron\ to
      70~\micron\ flux ratio vs. 70~\micron\ to 24~\micron\ flux ratio
      for HOPS--detected Orion protostars.  The symbol and color type
      is the same as in the top panel.  The reddest sources are
      distinguished from the bulk of \spitzer--identified HOPS sources
      (triangles) both by brighter 160\micron\ fluxes and redder
      70~\micron\ to 24~\micron\ flux ratios.}
    \label{fig:cm}
  \end{center}
\end{figure*}    

\section{\herschel\ protostar candidates} 

We present the \herschel\ protostar candidate catalog in
Table~\ref{tab:pbrscat}. Here we include the PACS
70~\micron\ coordinates and flux measurements at 24~\micron,
70~\micron, and 160~\micron.  We indicate which sources are flagged as
reliable protostellar candidates and which are likely contamination,
based on the results from the previous section.  We also indicate if
the sources have a robust 870~\micron\ detection.  Furthermore, we
present the values of \lbol\ and \tbol\ and their corresponding
estimated statistical errors (see discussion in \S~5.1).  In
Figure~\ref{fig:70dist}, we show the 70~\micron\ flux distributions
for the sample compared to the distribution of HOPS protostars.  The
majority of the new candidate protostars have 70~\micron\ fluxes that
are lower than the previously identified \spitzer\ HOPS sample; this
is not surprising since the new candidate protostar sample is selected
to be faint or undetected at 24~\micron.  Furthermore, the peak at low
70~\micron\ flux values is dominated by extragalactic contamination,
as discussed above.

In Figure~\ref{fig:cm}, we show the MIPS 24~\micron, PACS 70~\micron,
and PACS 160~\micron\ colors of the new \herschel\ sources compared to
the HOPS sample of 70~\micron\ detected protostars.  The top panel
shows the 70~\micron\ flux vs.\ the log $(\lambda F_\lambda70 / \lambda
F_\lambda24)$ color (henceforth $70/24$ color), while the bottom panel
shows the log $(\lambda F_\lambda160 / \lambda F_\lambda70)$ color
(henceforth $160/70$ color) vs.\ the $70/24$ color, for our sample of
new protostar candidates compared to the colors of the
\spitzer--identified HOPS sample.  The \spitzer\ 24~\micron\ 7
magnitude limit, imposed on the HOPS sample for a reliable
protostellar identification, is apparent in the top panel as the
diagonal line approximately separating the new protostellar candidates
at fainter 70~\micron\ fluxes and redder $70/24$ colors from the
population of \spitzer--identified HOPS sources.  

For comparison, in the top panel of Figure~\ref{fig:cm}, we also show
the fluxes and colors of presumably typical and well--studied Class~0
sources: VLA1632--243 (J. Green and DIGIT team, private communication,
2012, and Green et al., in prep.), IRAS16293 \citep{evans09}, B335
\citep{stutz08, laun12}, CB68 \citep{laun12}, and CB244
\citep{stutz10,laun12}.  Furthermore, we also show the colors of
various VeLLOs: L673--7 \citep{dunham08}, IRAM04191 \citep{dunham06},
and CB130 \citep{laun12}.  We find that the observed colors of our
sample of candidate protostars appear consistent with the colors of
more near--by Class~0 and VeLLO sources but not with FHSC candidate
colors proposed in the literature \citep[e.g.,][]{commercon12}.  We
find that the majority of these previously known Class~0 and VeLLO
sources do not appear as red in their $70/24$ colors as the reddest
sources in our sample. The only exceptions to this trend are IRAS16293
and VLA1632--243, perhaps representing an extrema in the $70/24$ color
distribution that may be driven by their comparatively large envelope
densities.

We also show in Figure~\ref{fig:cm} the colors of two FHSC candidates
in Perseus: Per--Bolo~58 \citep{enoch10} and B1-bS \citep{pezzuto12}.
In this diagram, the $70/24$ color of Per--Bolo~58 appears generally
consistent with that of a VeLLO, as \citet{enoch10} point out.  As
such, this source may be an extremely low--mass protostar.  On the
other hand, the $70/24$ color of B1-bS is comparable to the very
reddest sources we find in Orion while the 70~\micron\ flux is
consistent with VeLLOs and fainter than the reddest sources in Orion
by more than one order of magnitude.  The faint but robust detection
of a 70~\micron\ point--source by \citet{pezzuto12} may indeed point
to the possible Class~0 or VeLLO nature of B1-bS.  We do however note
that Pezzuto et al.\ also detect a source with no
70~\micron\ counterpart, B1--bN, which may therefore represent a
more robust FHSC candidate.  Regardless of the elusive nature of FHSC
candidates, when comparing our new candidate protostar colors to FHSC
models by \citet{commercon12}, we find that our sources do not appear
to be consistent with predicted or expected FHSC colors, with the
caveat that distinguishing FHSCs from VeLLOs with continuum
observations alone is likely difficult. 

\section{PACS Bright Red Sources}

Up to this point, we have discussed two distinct and well--defined
samples of sources in Orion: i) the sample of candidate protostars
identified with \herschel\ that have PACS 70~\micron\ and
160~\micron\ detections but MIPS 24~\micron\ magnitudes greater than
7.0~mag, ii) the sample of protostars that were reliably identified
with \spitzer\ \citep[24~\micron\ magnitudes $\leq
  7.0$~mag][]{megeath12,kryukova12}.  The protostar catalog target
list used for the HOPS program consists mostly of the
\spitzer\ identified protostars, but also contains some of the
previously known protostars with $m(24) > 7.0$~mag (Fischer et al., in
preparation).  

\begin{figure*}[t]
  \begin{center}
    \scalebox{0.73}{\includegraphics{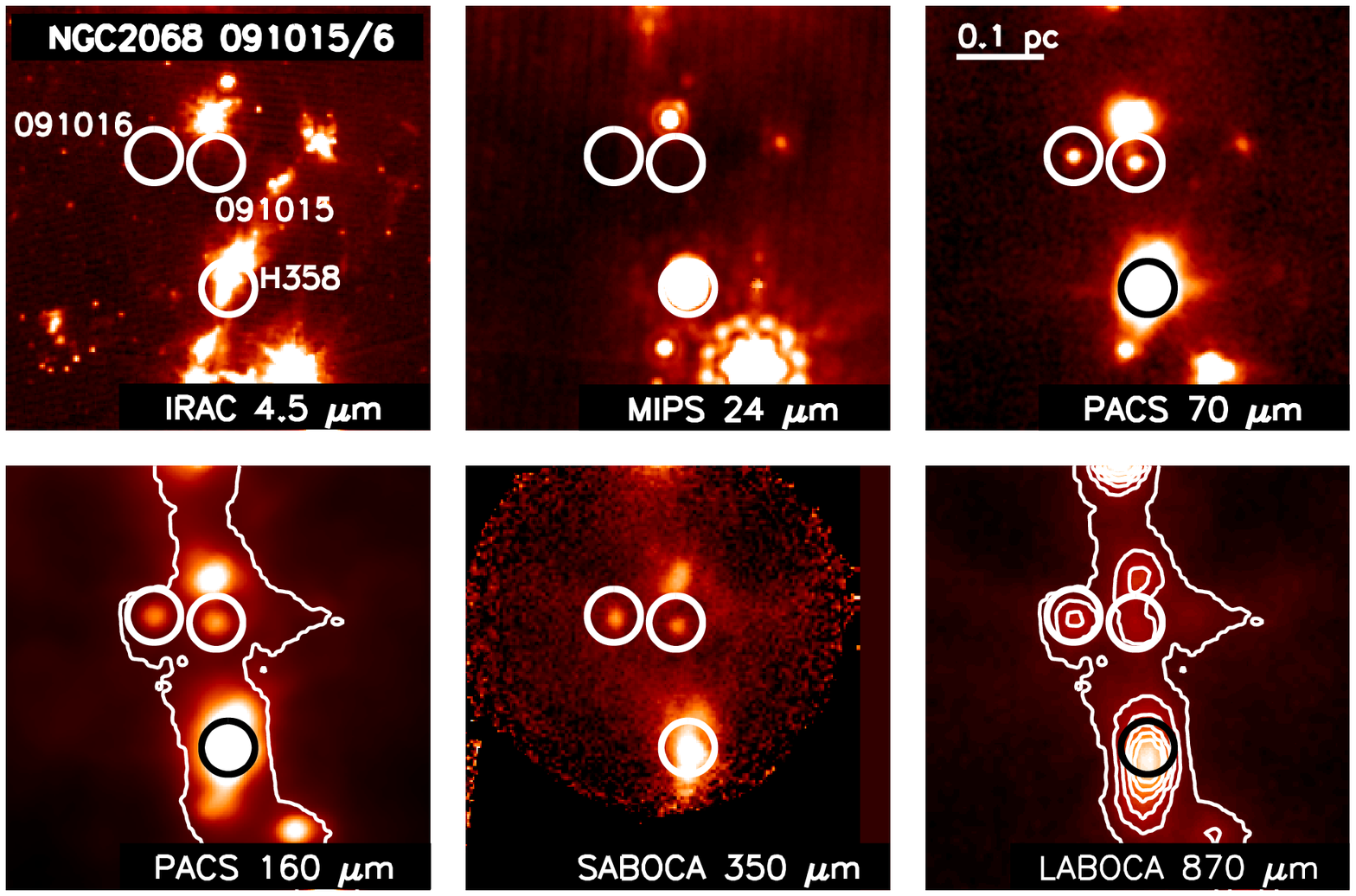}}
    \caption{$4\arcmin \times 4\arcmin$ images of three PBR sources,
      091015 and 091016 (top), and HOPS358 (bottom), at the indicated
      wavelengths, shown on a log scale.  North is up, east is to the
      left.  The circles indicate the location of the PACS
      70~\micron\ point--sources.  Contours indicate the
      870~\micron\ emission levels at \{0.25, 0.5, 0.75, 1.0, 1.25\}
      Jy\,beam$^{-1}$. The lowest 870~\micron\ contour is
      over--plotted on the 160~\micron\ image.  No IRAC emission
      coincident with 091015 or 091016 is detected; however, these
      sources clearly reside in dense filamentary material traced by
      the sub-mm emission. The HOPS358 photometry is likely blended.}
    \label{fig:img7}
  \end{center}
\end{figure*}   

\begin{figure*}[t]
  \begin{center}
    \scalebox{0.73}{\includegraphics{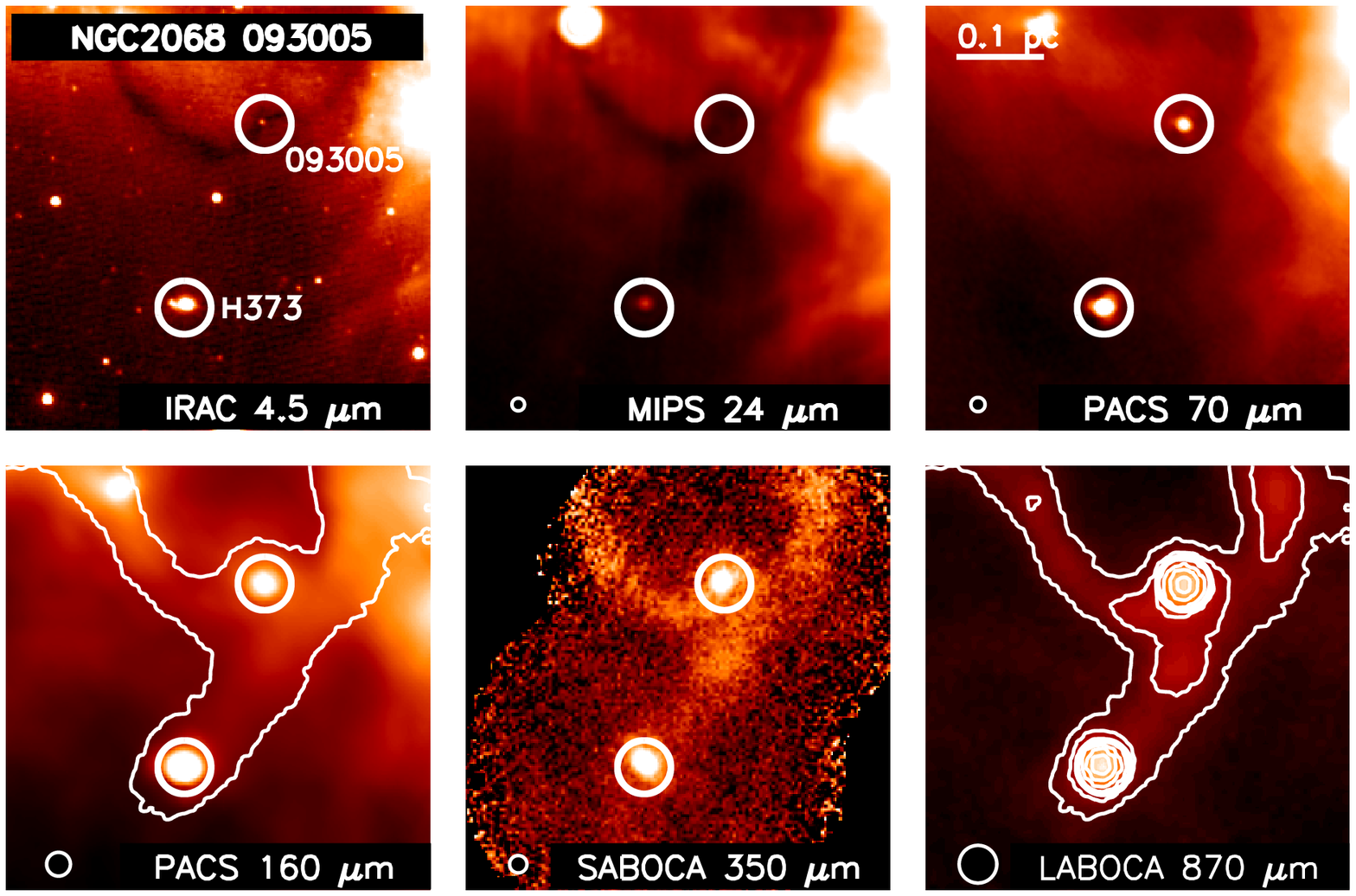}}
    \caption{Same as Figure~\ref{fig:img7}, showing $4\arcmin \times
      4\arcmin$ images two PBRs: 093005 (top) and HOPS373 (bottom).
      Contours indicate the 870~\micron\ emission levels at \{0.25,
      0.5, 0.75, 1.0, 1.25, 1.5\} Jy\,beam$^{-1}$. Source 093005 is
      the reddest PBRs shown in Figure~\ref{fig:cm} and lies at the
      intersection of three filaments traced by an 8 and
      24~\micron\ absorption feature and the 870~\micron\ emission.}
    \label{fig:img8}
  \end{center}
\end{figure*} 

In what follows, we focus our analysis on the Orion protostars that
have $70/24 > 1.65$.  Of the 18 known protostars that satisfy this
limit, 11 are identified with \herschel; hence this color regime is
dominated by our newly identified protostars.  Accordingly,
\herschel\ has provided us for the first time with a far more complete
sample of these red sources within the field of the HOPS survey.
Given their red colors and their brightness in the PACS wavelength
bands, we refer to this sample of protostars as PACS Bright Red
sources, or PBRs.  The coordinates, \spitzer photometry , and
\herschel\ photometry of the sample are listed in
Table~\ref{tab:phot}, while the APEX 350~\micron\ and
870~\micron\ photometry are presented in Table~\ref{tab:photapex}.
Since the APEX photometry are non--trivial to extract due in large
part to contamination by cold surrounding material and, at
870~\micron\ specifically, by the large beam size, we present three
measures of the source flux, as described above.

\subsection{Observed properties PBRs}

\begin{figure*}
  \scalebox{0.355}{\includegraphics{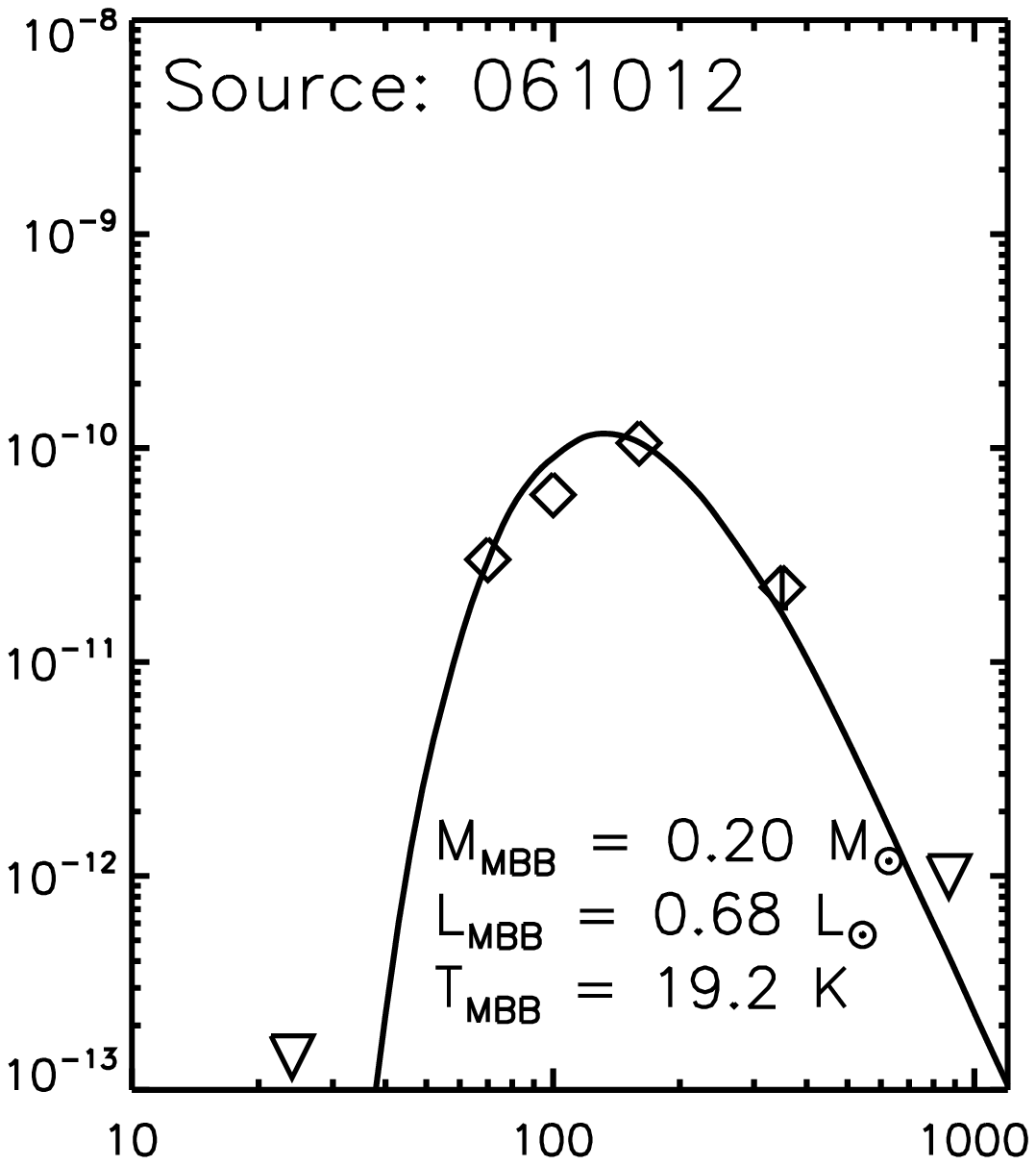}\includegraphics{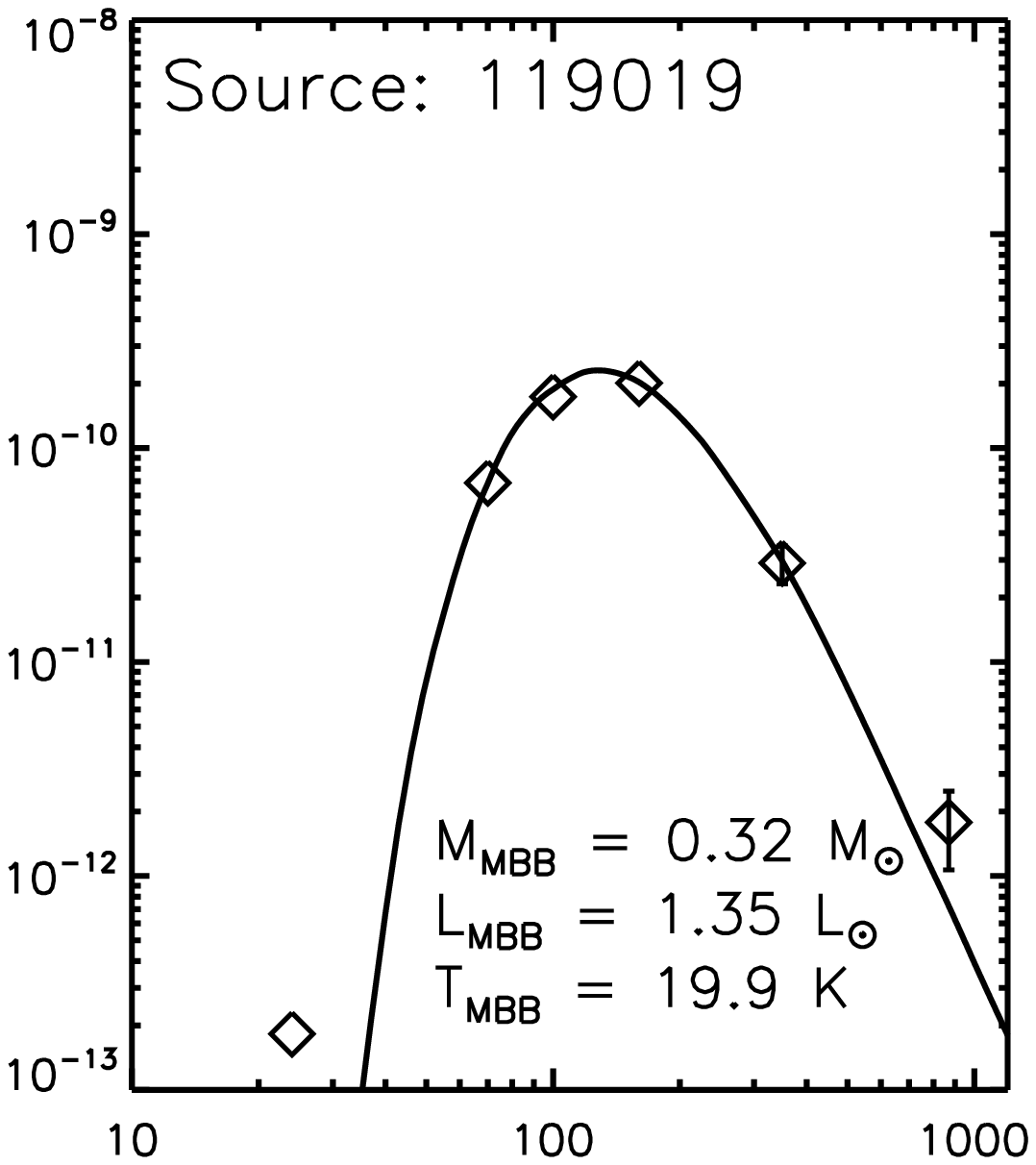}\includegraphics{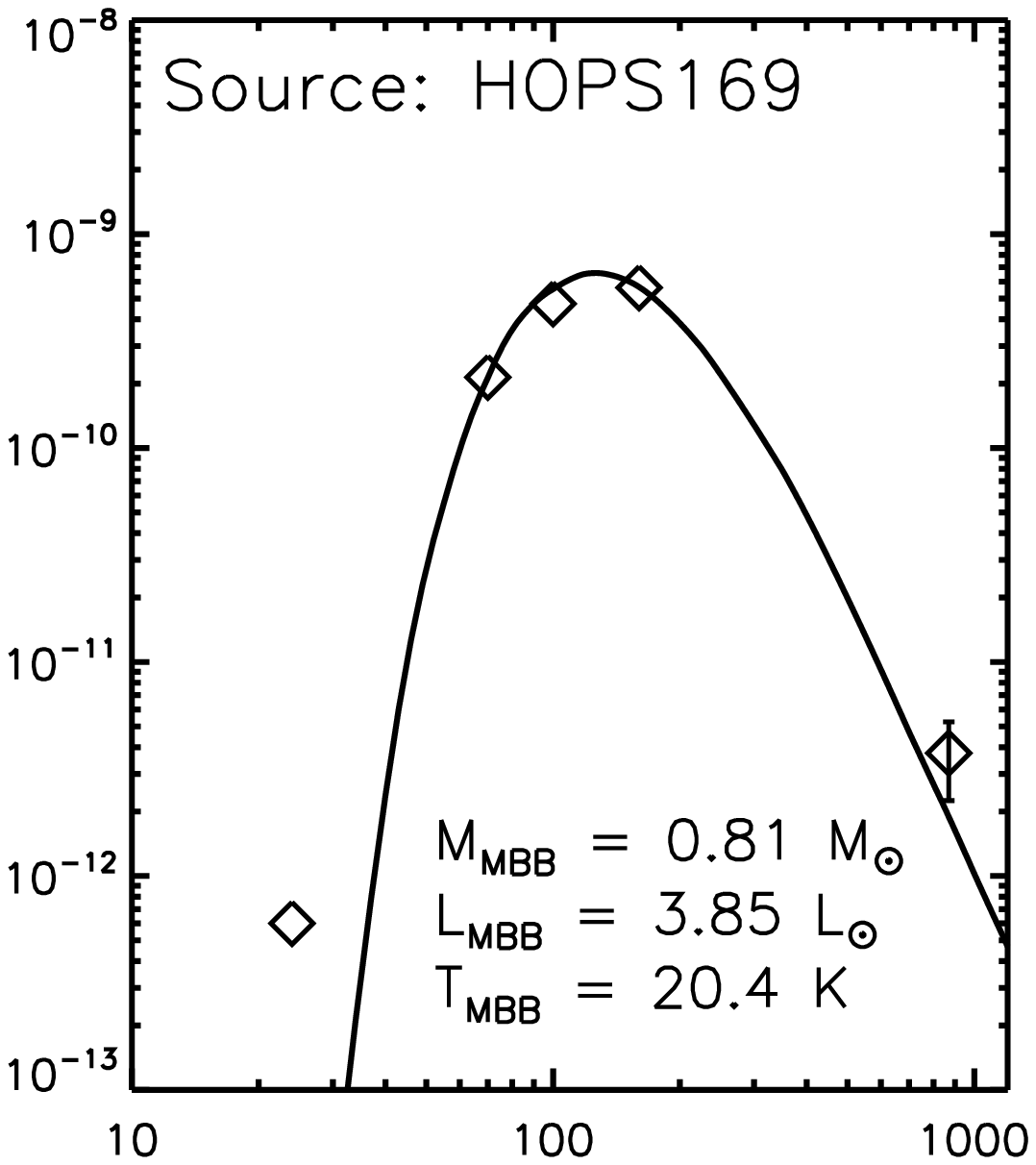}\includegraphics{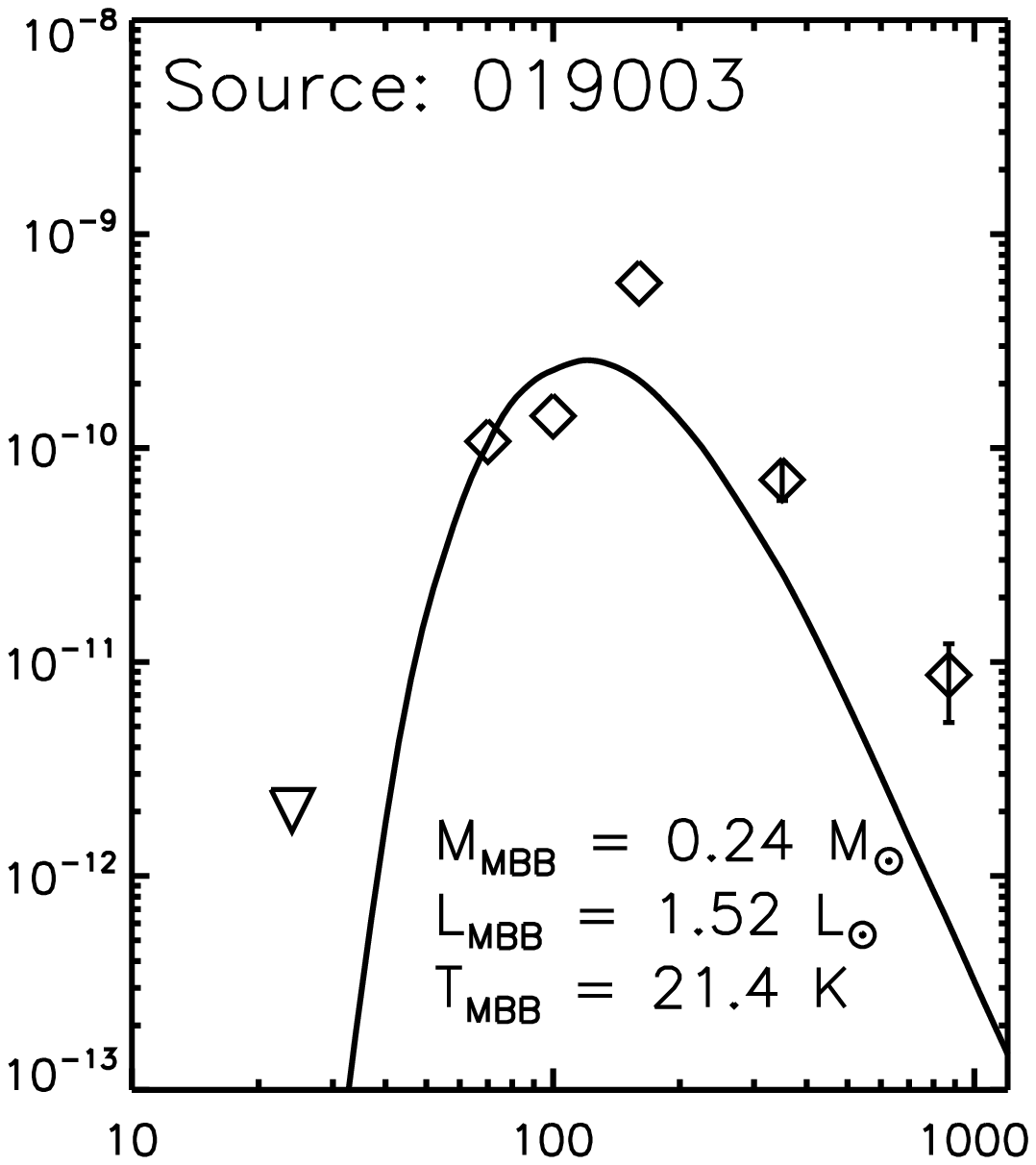}}
  \scalebox{0.355}{\includegraphics{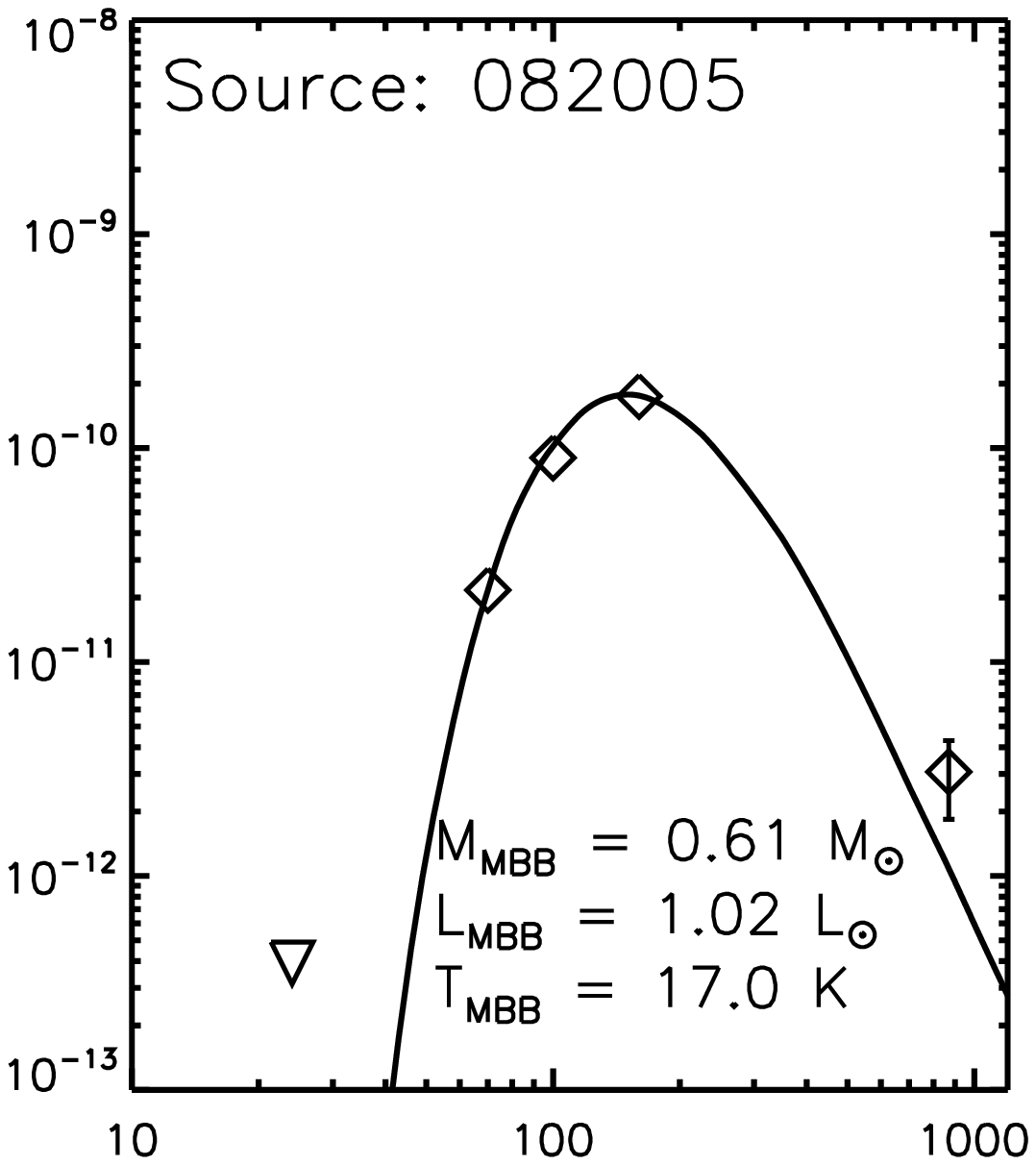}\includegraphics{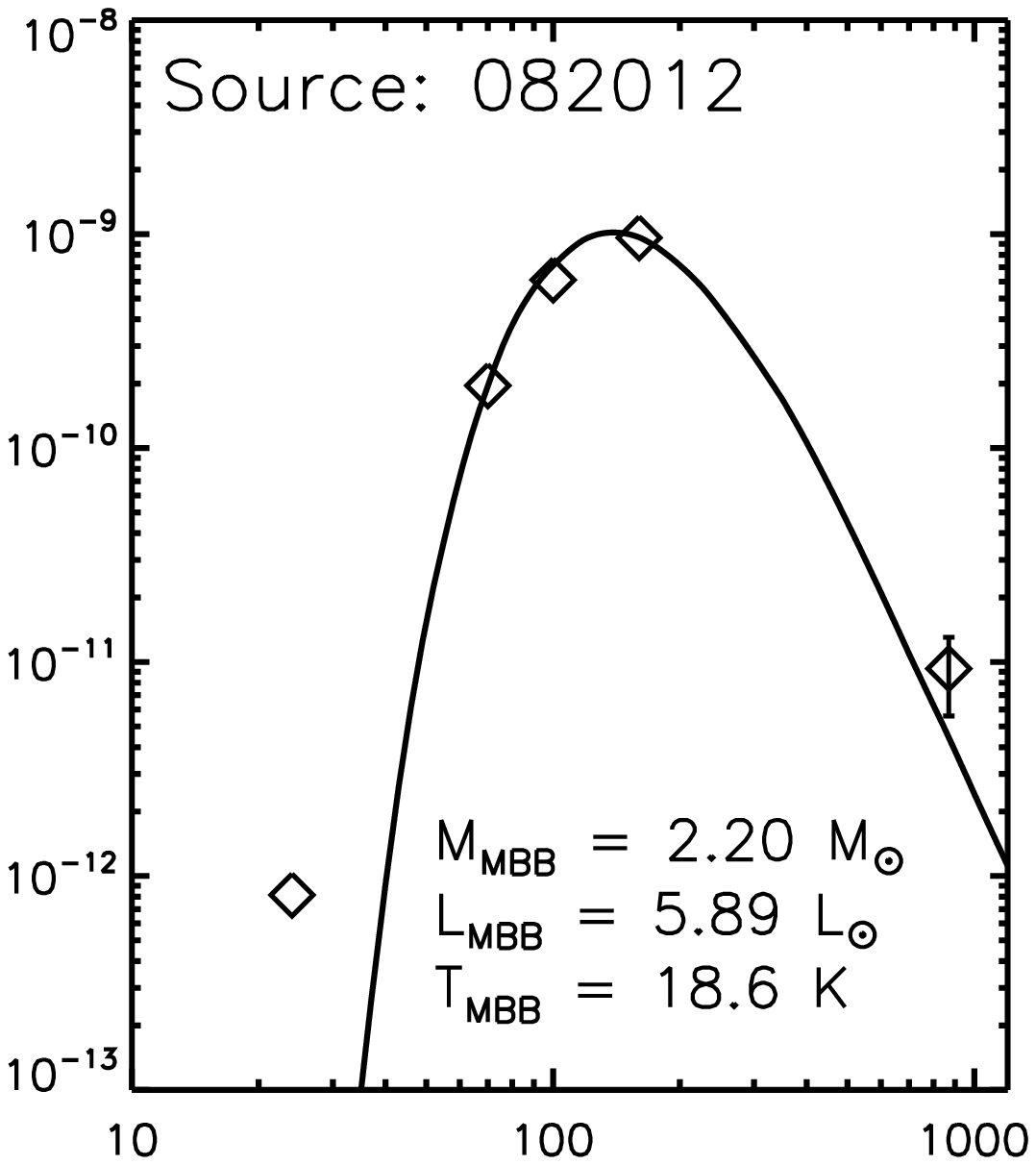}\includegraphics{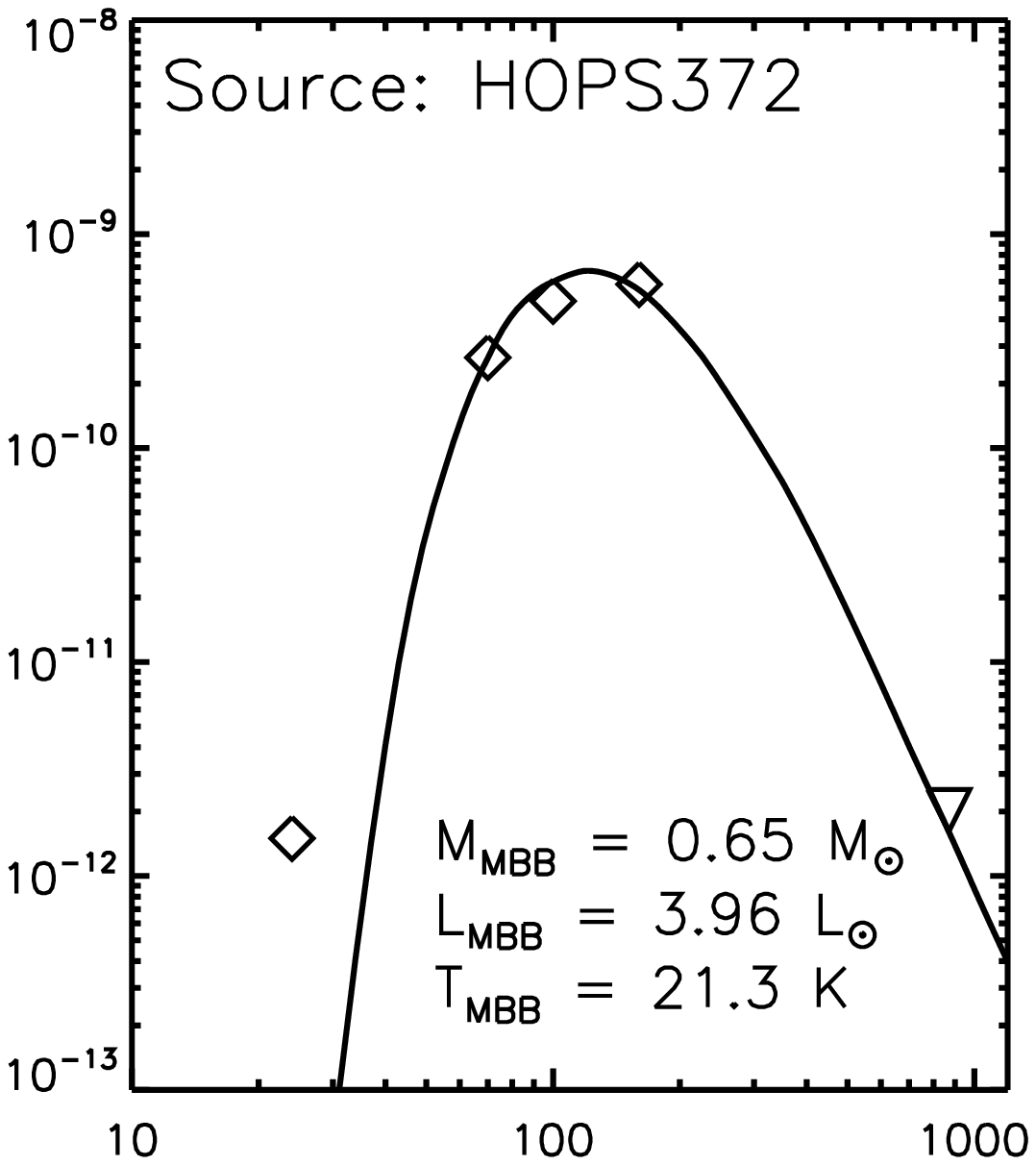}\includegraphics{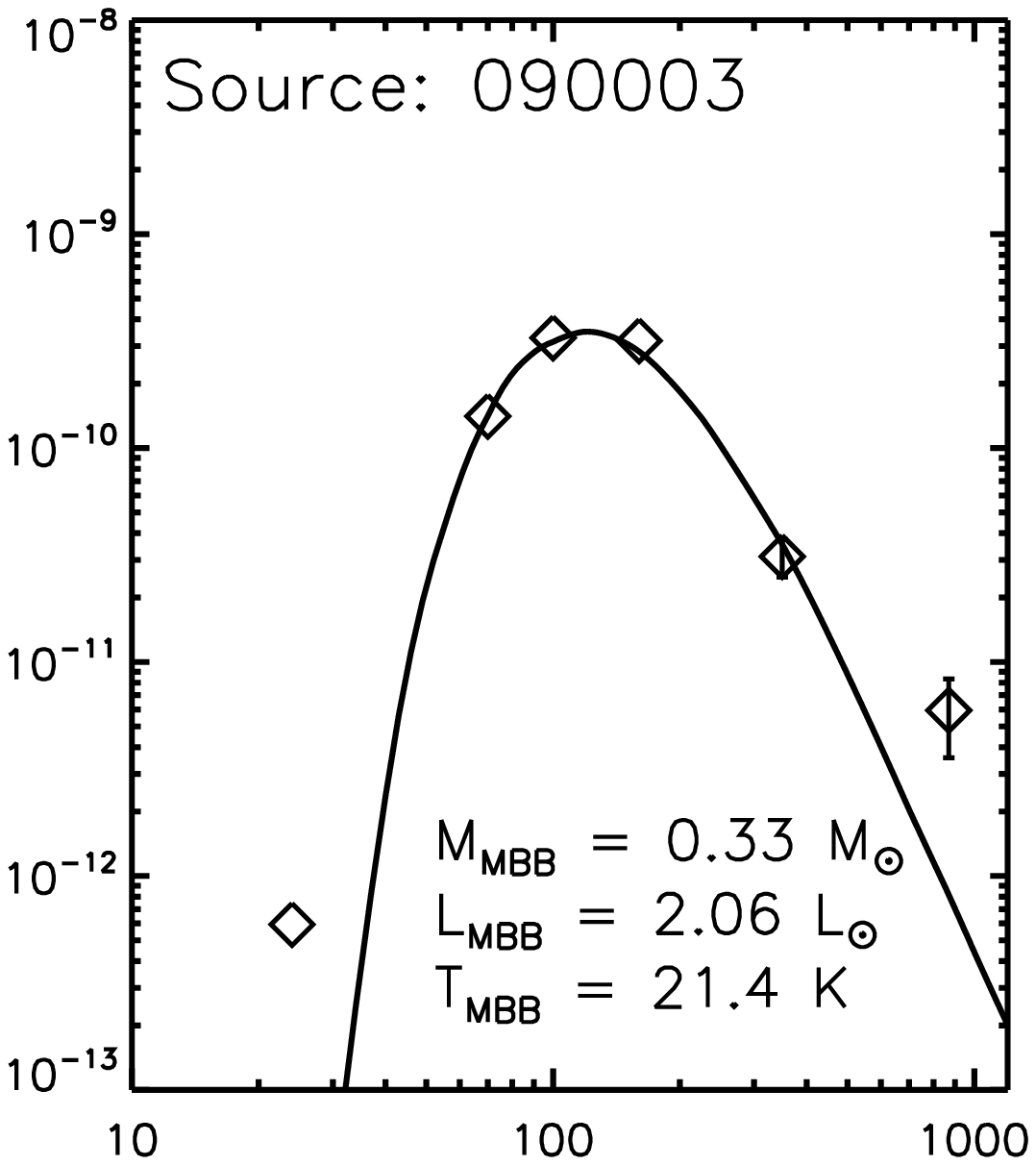}}
  \scalebox{0.355}{\includegraphics{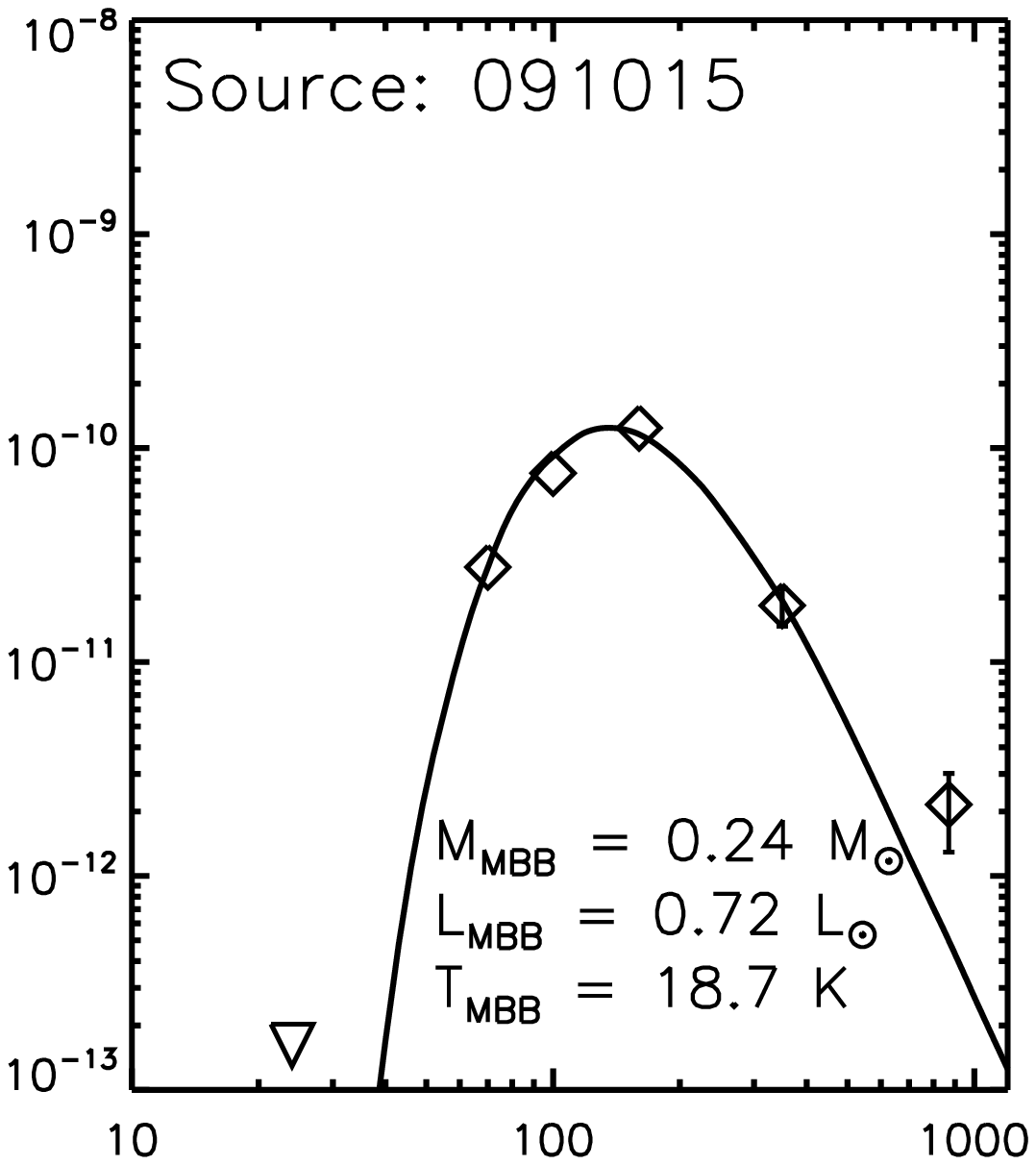}\includegraphics{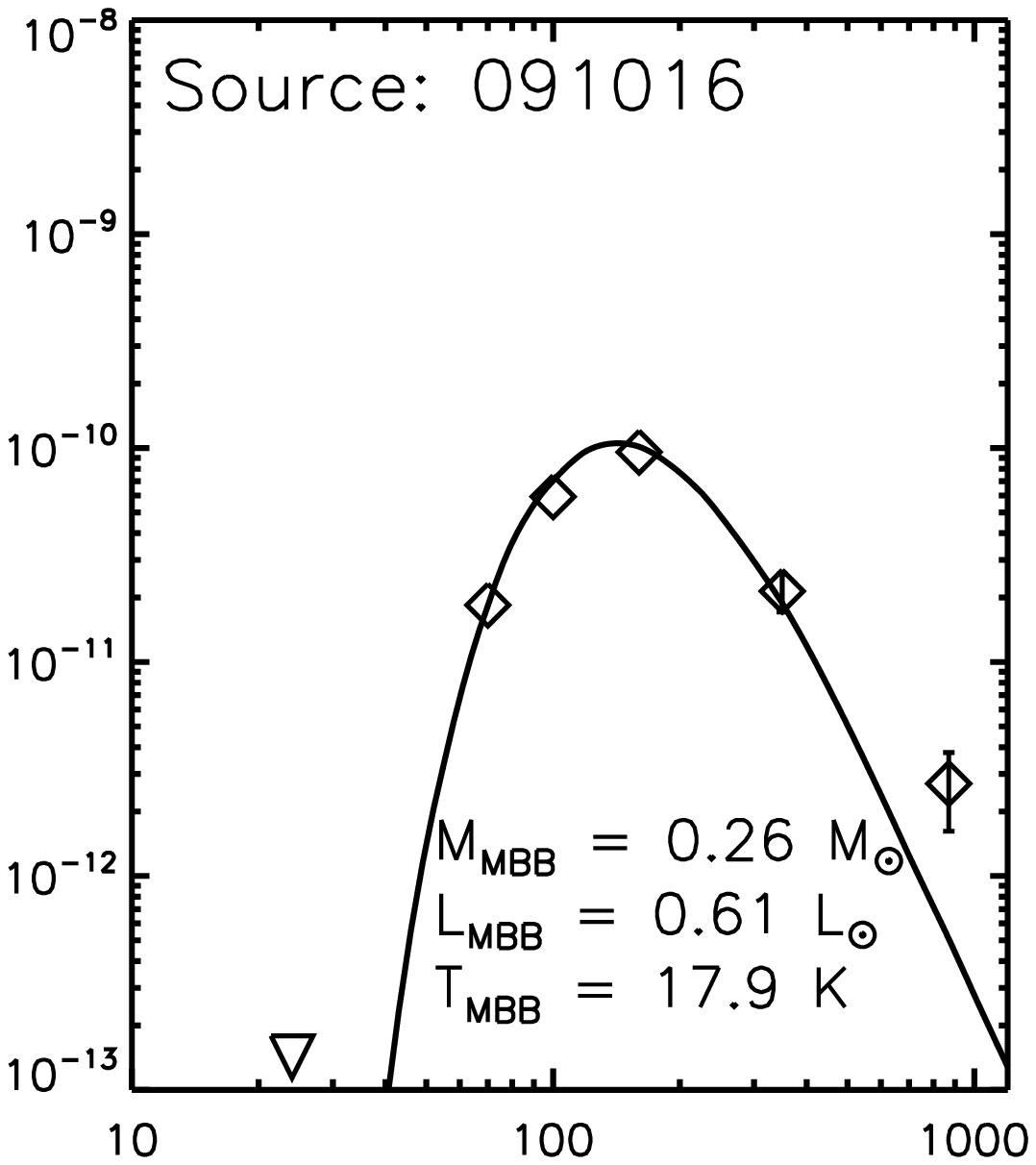}\includegraphics{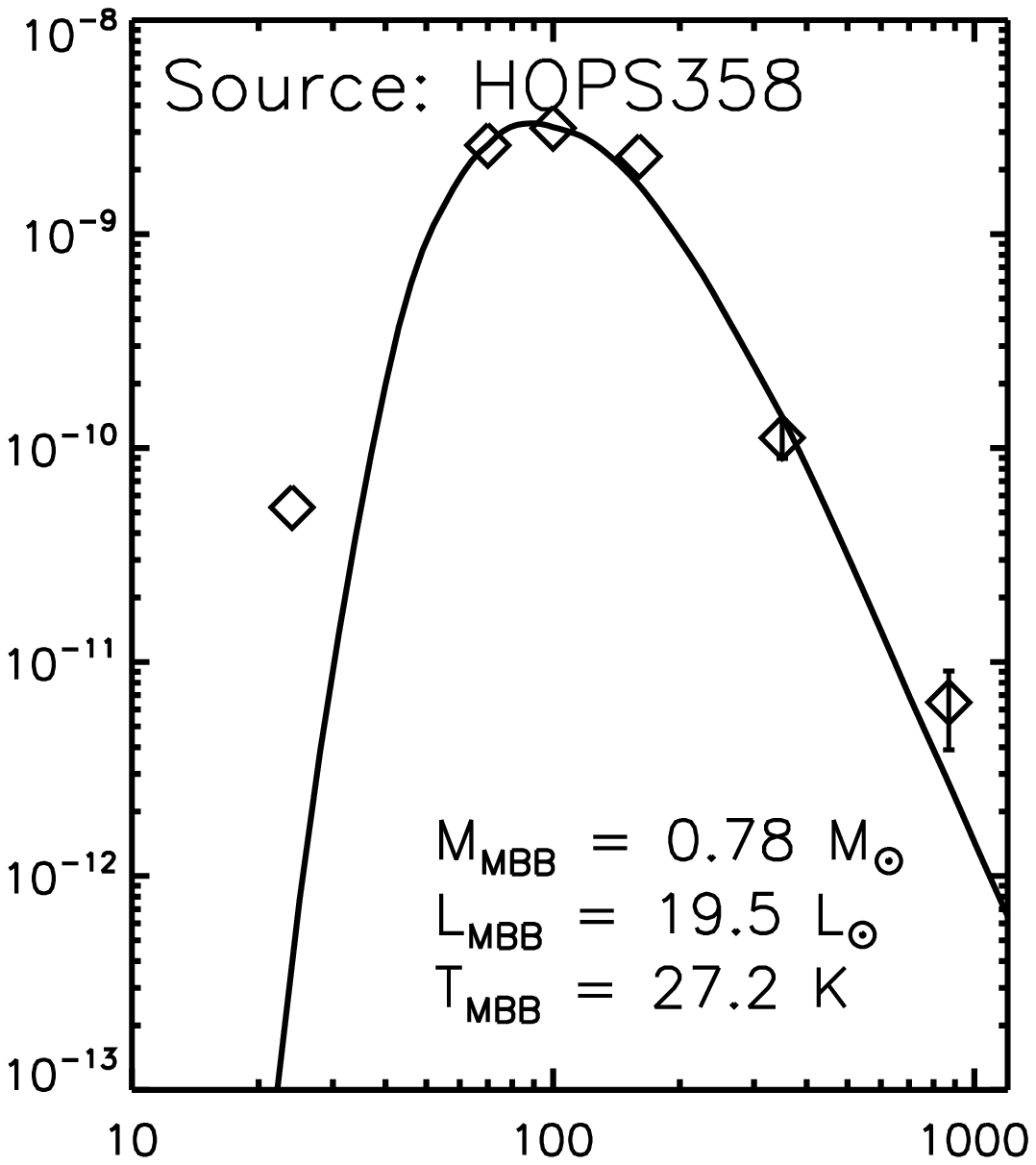}\includegraphics{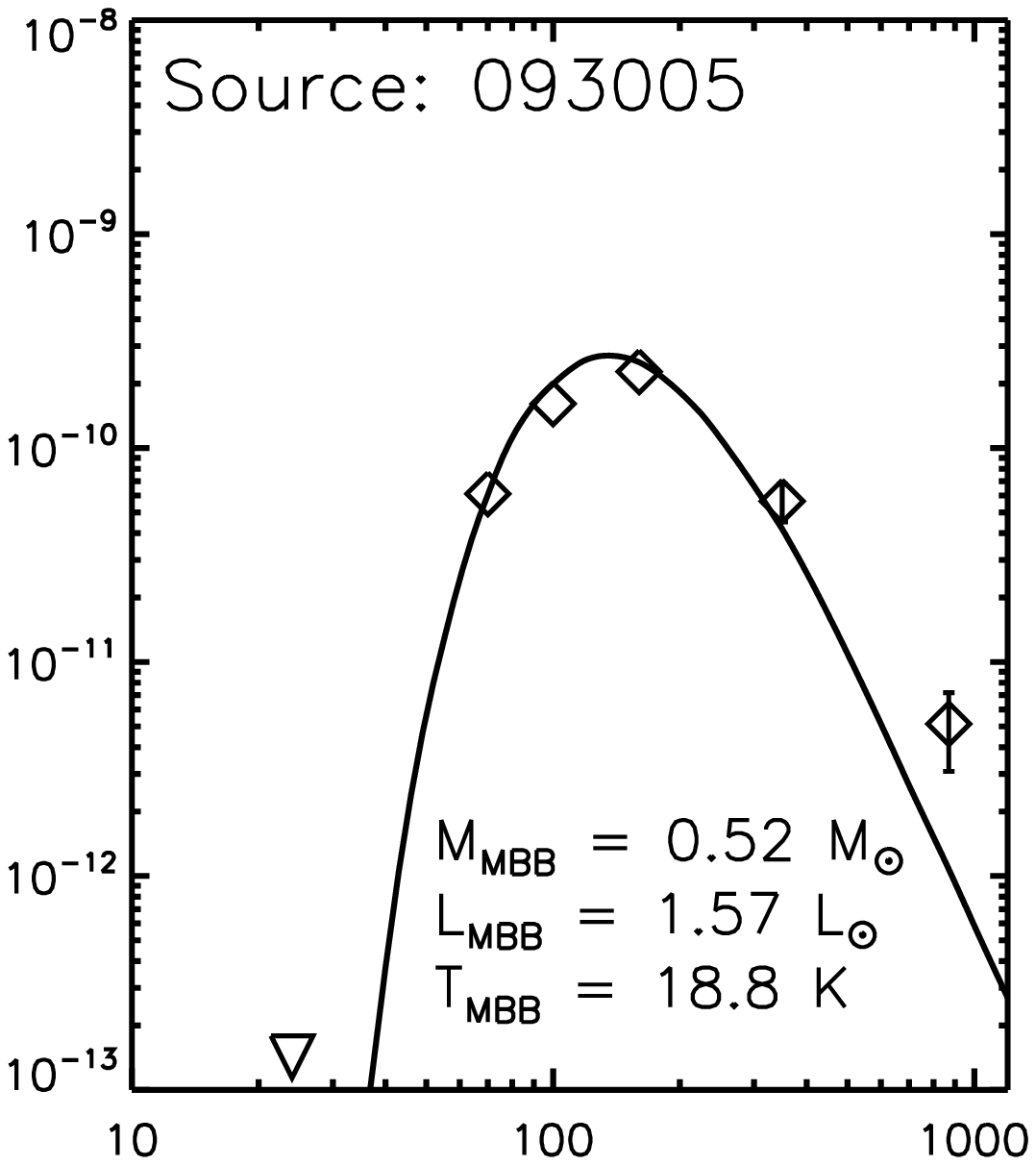}}
  \scalebox{0.355}{\includegraphics{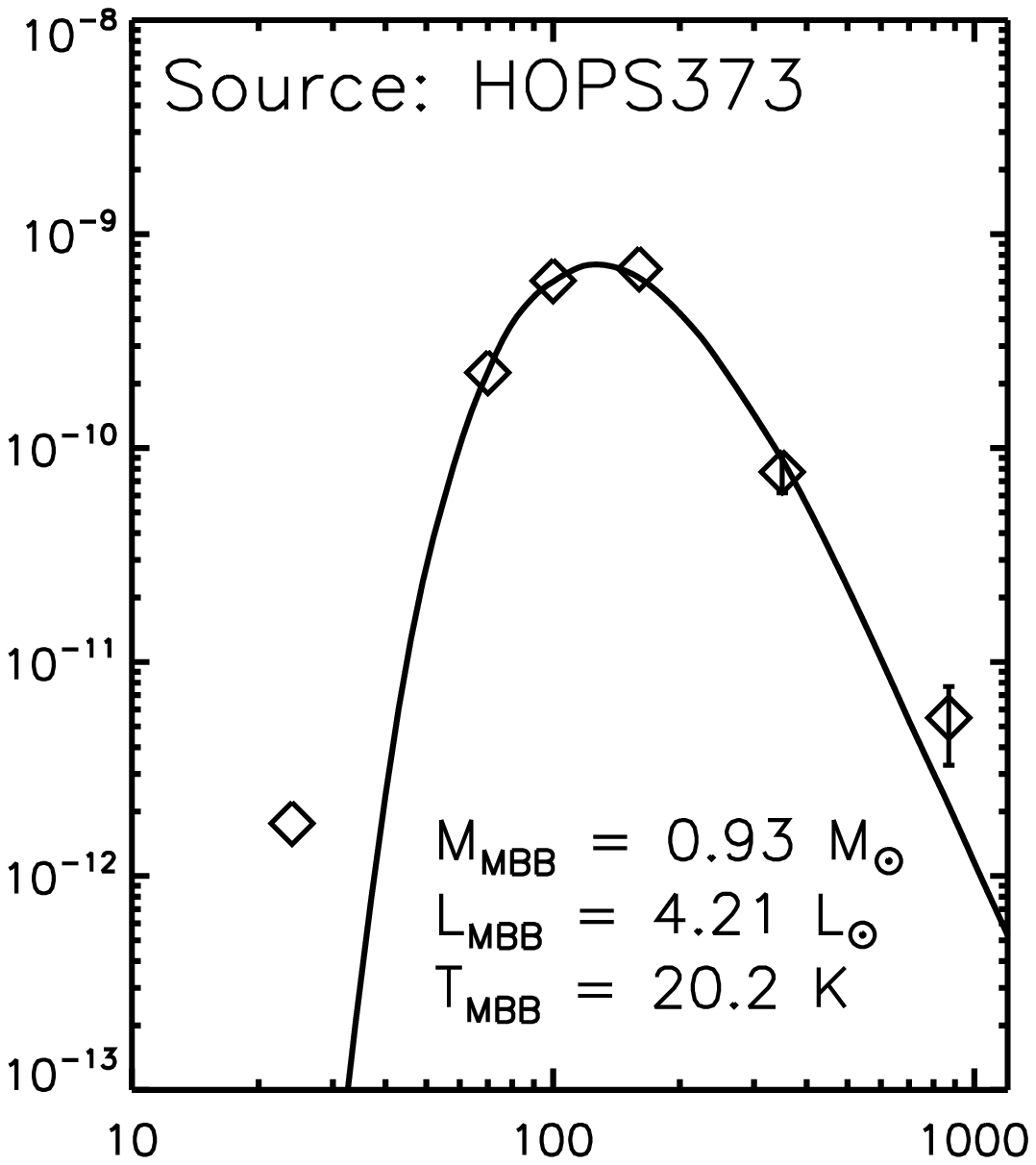}\includegraphics{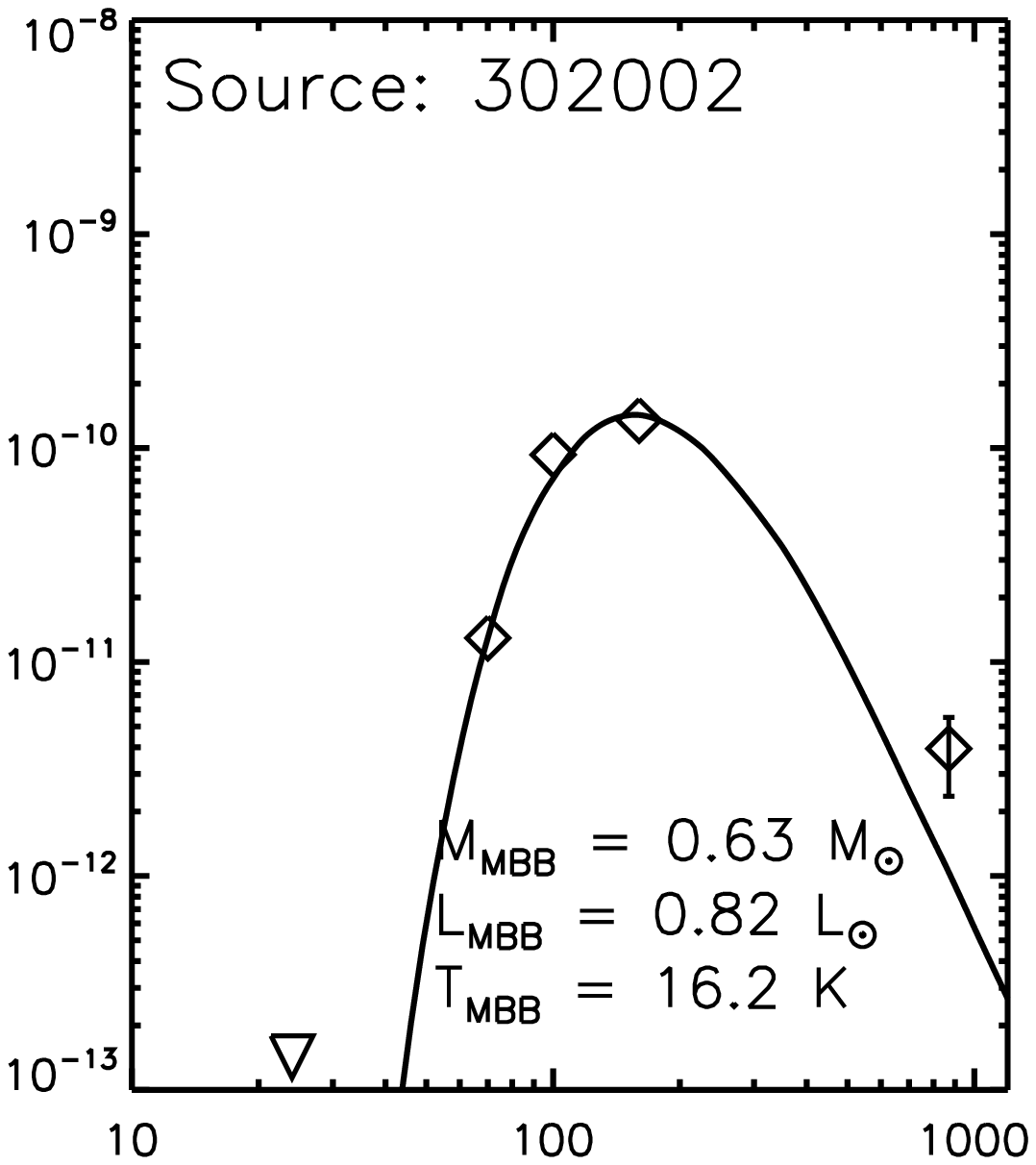}\includegraphics{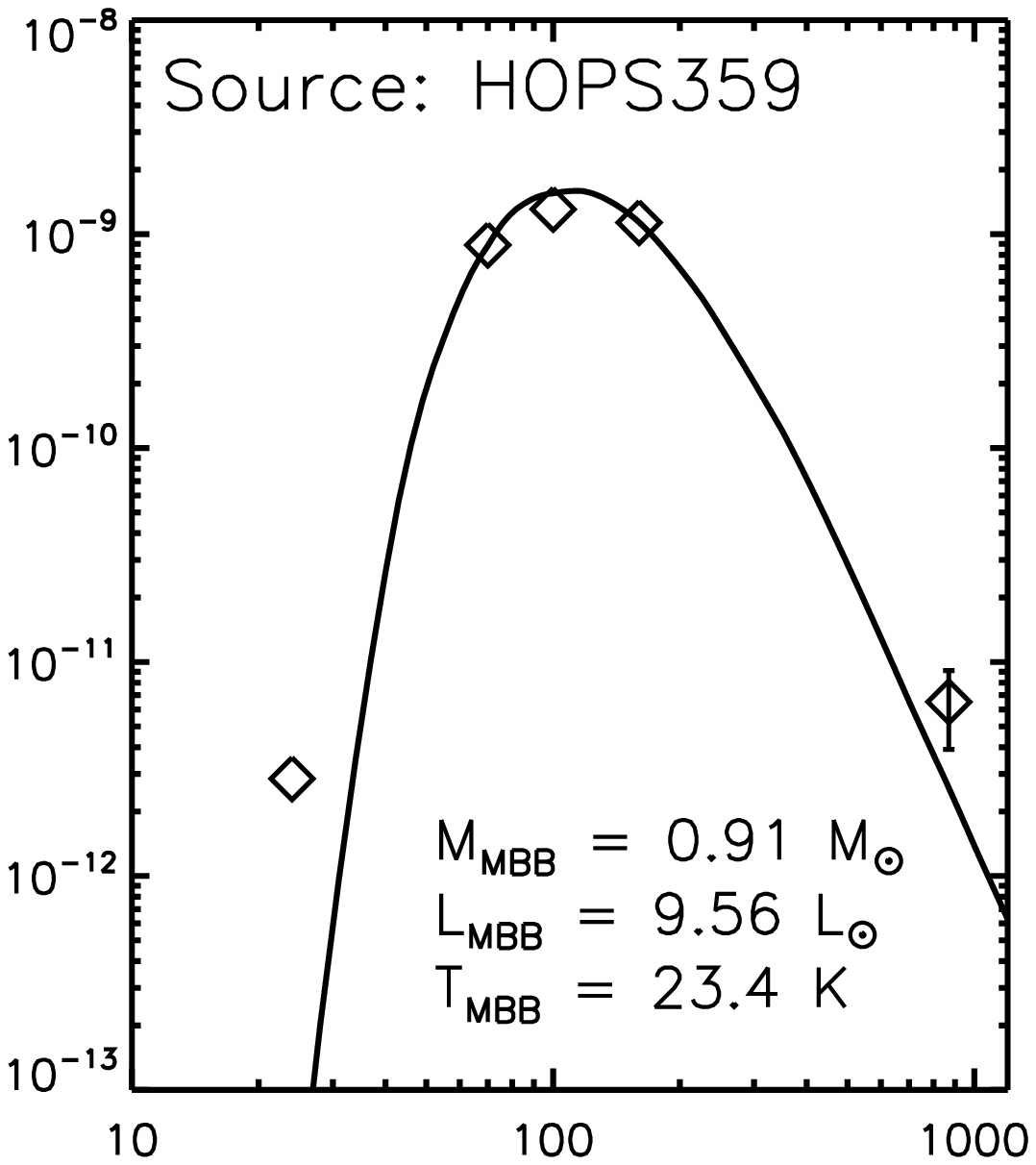}\includegraphics{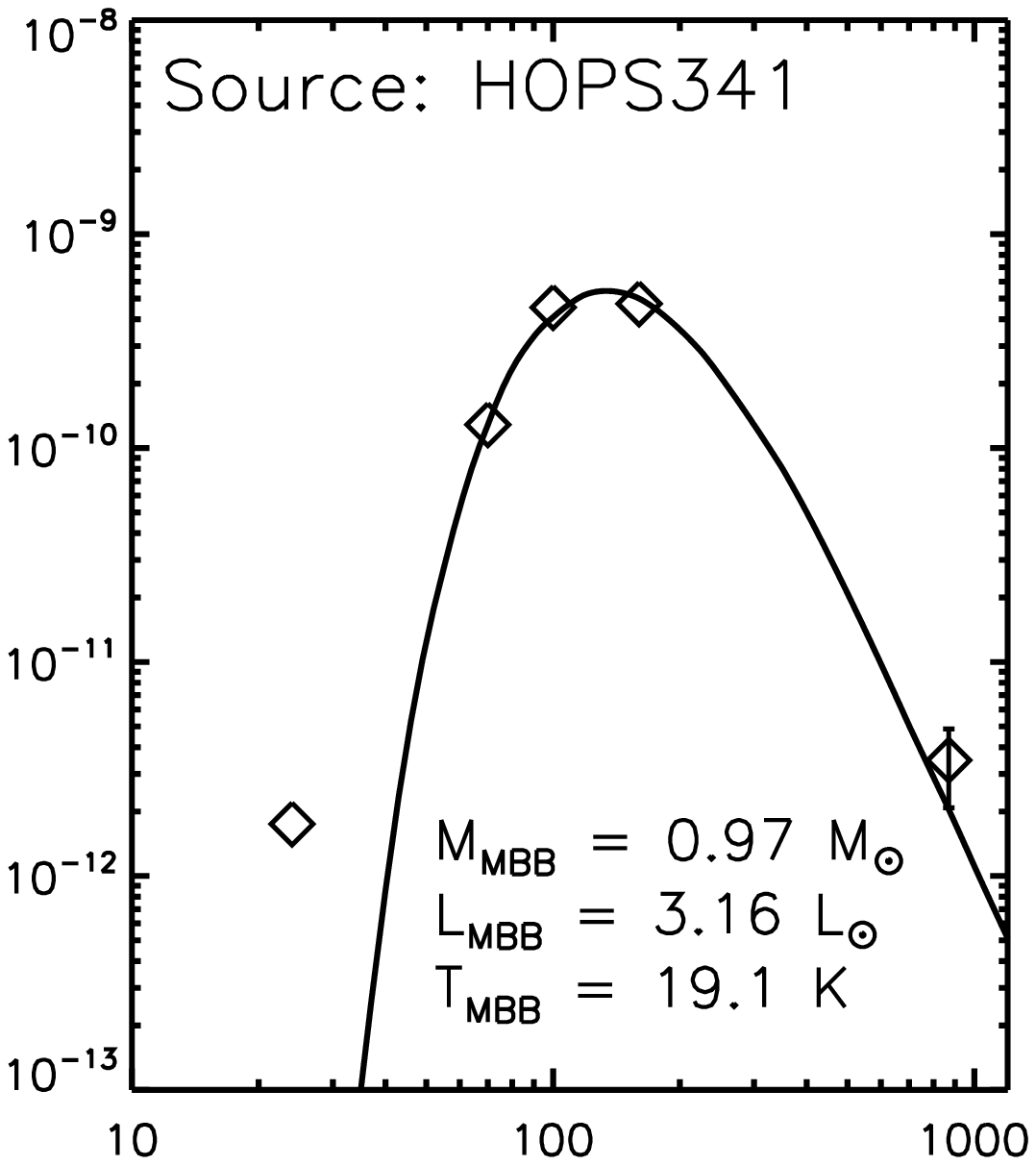}}
  \scalebox{0.355}{\includegraphics{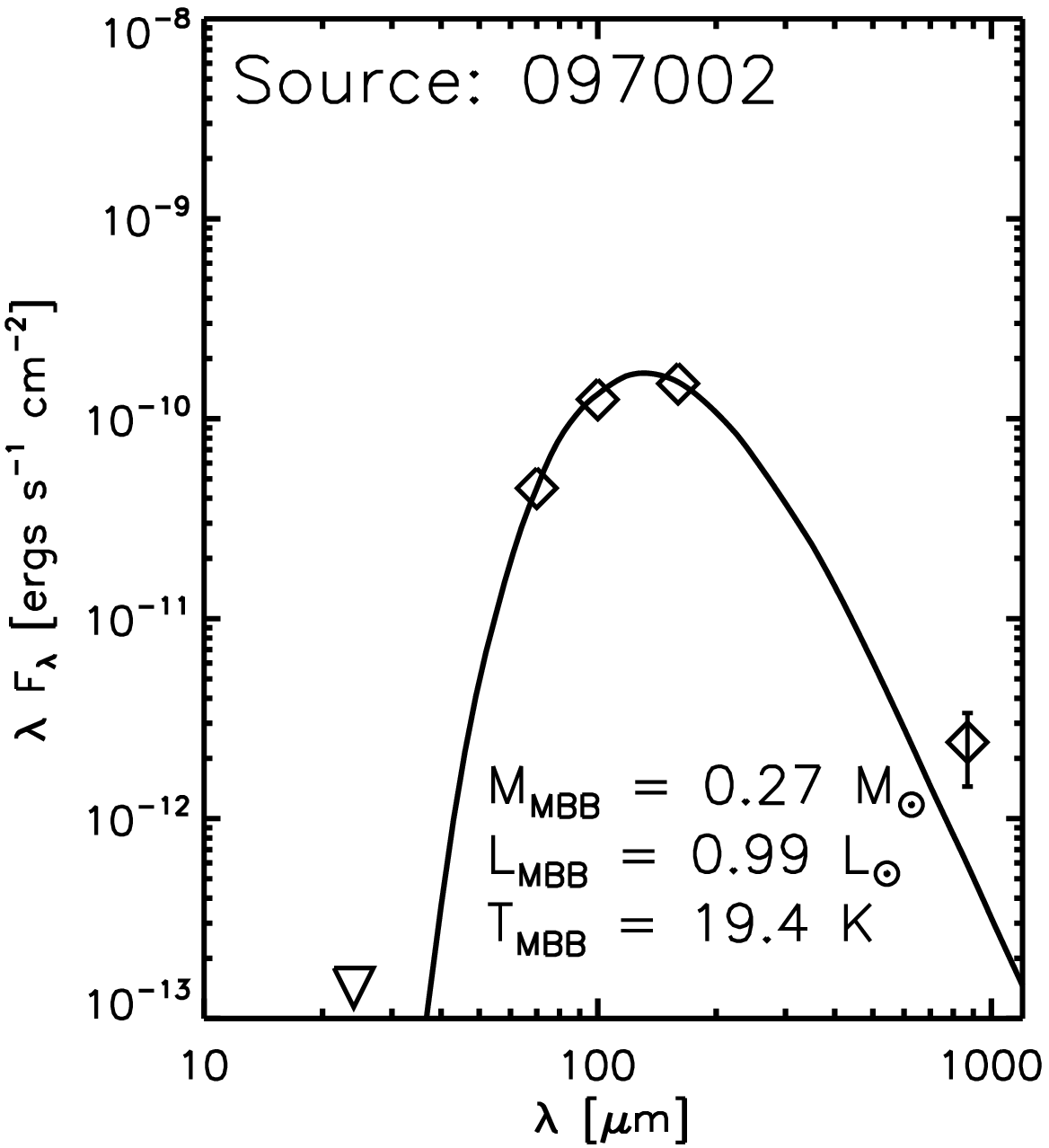}\includegraphics{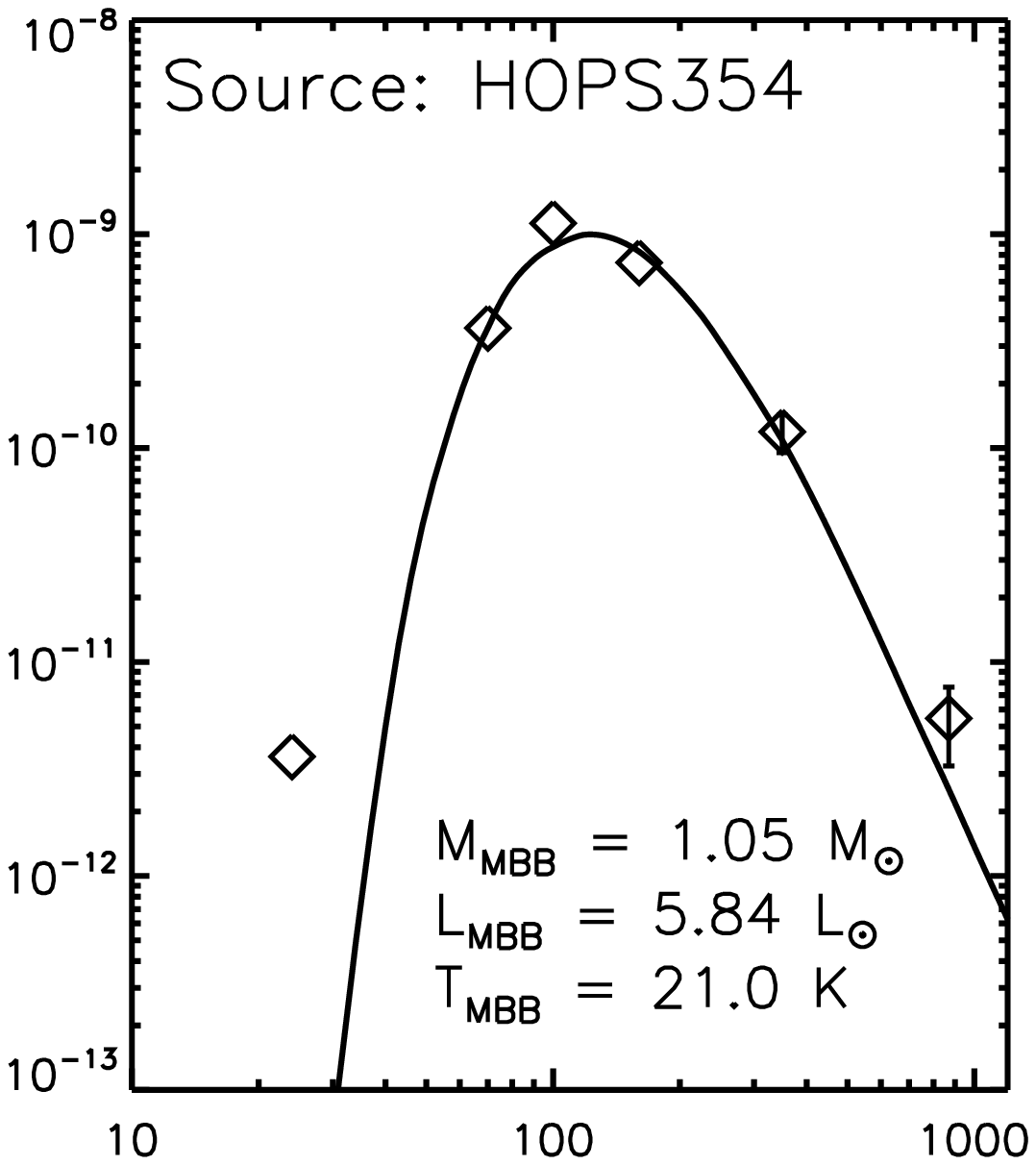}}
  \begin{center}
    \caption{SEDs of the 18 PBRs are shown. The errors are smaller
      than the symbol size except for the 350 and 870~\micron\ data
      points. The black curve shows the modified black--body fit to
      the observed SED, with the indicated best fit parameters (see
      \S~6.2). Note that PBRs 19003 (shown in the top right) is
      located in a complex field where the photometry may be strongly
      contaminated.}
    \label{fig:allseds}
  \end{center}
\end{figure*}

As discussed above, we select 18 PBRs in Orion with observed $70/24$
colors greater than $1.65$.  We show 4.5~\micron\ to
870~\micron\ images of 5 example PBRs in Figures~\ref{fig:img7}
to~\ref{fig:img8} (see Appendix for the full sample images).
Furthermore, in Figure~\ref{fig:allseds} we show the full set of 18
PBRs observed SEDs from 24~\micron\ to 870~\micron.  Inspection of the
observed SEDs confirms that the PBRs sample is composed of cold,
envelope dominated sources with peak emission always located at
$\lambda > 70$~\micron.  In addition, the peak of the SEDs, and thus
the temperatures, are well--constrained for all PBRs because we have
obtained APEX sub--millimeter coverage for all sources.

In Table~\ref{tab:obs}, we present some basic properties of the PBRs.
In particular, we find that 12/18 sources exhibit
\spitzer\ 4.5~\micron\ emission indicative of outflow activity.  We
also include some references to previous detections (see Appendix).
Furthermore, 4/18 sources have significant levels of
4.5~\micron\ emission that are indeed consistent with a high
inclination.  The majority of sources, however, do not give clear
indications of their inclinations at any observed wavelength, and
therefore we cannot make any statements about their possible
orientations based on their appearance in the images.  We find
indications from the 4.5~\micron\ image morphology that two sources
(HOPS341 and HOPS354) are binaries, while seven sources have a nearby
source within $30\arcsec$.  Two reside in more crowded regions, and
seven sources appear truly isolated.  We find that a significant
fraction (13/18) of sources appear to reside in filamentary regions,
i.e., the extended 870~\micron\ emission appears significantly
elongated.

The four sources with significant indications of a high inclination
orientation are HOPS169, 302002, HOPS341, and HOPS354 (see Appendix
for Figures~\ref{fig:img3}, \ref{fig:img9}, \ref{fig:img11}, and
\ref{fig:img13}).  Inspection of their 4.5~\micron\ images reveals
that their outflows appear well collimated and relatively narrow.
Indeed, we might expect that sources that that have denser envelopes,
and are therefore presumably younger, may have more narrow cavity
opening angles \citep[e.g., ][]{arce06}.  As an additional check on
our density analysis (see above), we use this inclination information
for an independent check of the envelope densities of these four
sources.  Despite the relatively sparsely sampled SEDs, we fix the
inclination to $87^{\rm o}$ and fit the source SEDs. We find that even
when we fix the model inclination to $\theta = 87\degree$, we still
obtain envelope densities significantly above the $\rho_1$ value
found in the previous section.

\subsection{Observational evolutionary diagnostics}

We calculate L$_{\rm bol}$, T$_{\rm bol}$ \citep{myers93}, and L$_{\rm
  smm}$/L$_{\rm bol}$ \citep{andre93,andre00}.  The errors in L$_{\rm
  bol}$ and T$_{\rm bol}$ are derived with the same Monte Carlo method
as described in \S~6.2 for the modified black--body parameters.  We
exclude the IRAC upper limits from this analysis; including these
limits has an effect on our L$_{\rm bol}$ and T$_{\rm bol}$ estimates
that is smaller than our estimated errors.  We do, however, include
the 24~\micron\ upper limits; therefore the \lbol\ and \tbol\ values
should be considered upper limits for sources not detected at this
wavelength.  Furthermore, we investigate the effect of applying an
average foreground reddening correction to all the new
\herschel\ candidate protostars.  We find that dereddening the
observed fluxes with extinction levels of $A_V = 40$ magnitudes has no
effect on the derived parameters because the observed SEDs are
extremely red and cold.

\subsection{Spatial distribution of the three samples}

\begin{figure*}
  \begin{center}
    \scalebox{0.75}{\includegraphics{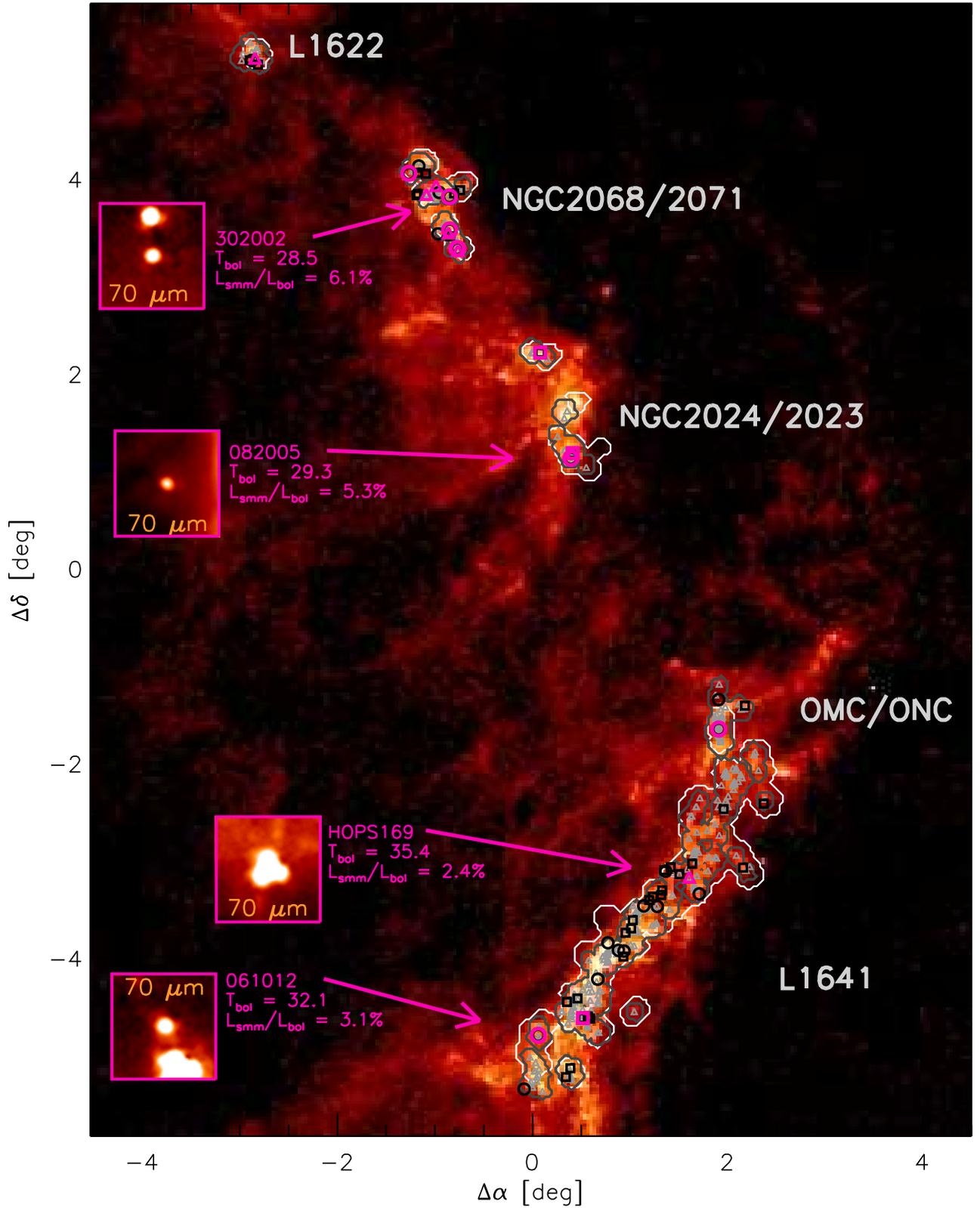}}
    \caption{The distribution of red protostars overlaid on the
      extinction map of Orion.  The dark grey contour shows the HOPS
      PACS coverage and the white contour shows the LABOCA
      870~\micron\ coverage.  The triangles (light grey) indicate
      the positions of the HOPS protostars while the black points
      (squares and circles) indicate the positions of the new
      candidate protostars.  The PBRs are highlighted in fuchsia and
      the properties of four selected PBRs are shown along with their
      corresponding $1\arcmin\times1\arcmin$ HOPS 70~\micron\ images.}
    \label{fig:orionimg}
  \end{center}
\end{figure*}

We show the locations of the Herschel protostar candidates compared to
the locations of the HOPS sample in Figure~\ref{fig:orionimg}; these
positions are overlaid on the extinction map of Orion.  It is
immediately apparent that the spatial distribution of the new
candidate protostars and PBRs is non--uniform.  To investigate this
distribution further, we show the relative fraction of new sources as
a function of individual region in Table~\ref{tab:frac}.  The
over--all number of new candidate protostars and PBRs is dominated by
the Orion A cloud, and in particular L1641.  This is not surprising
since the L1641 region is quite large and contains more protostars
compared to other Orion regions.  The fractions of new candidate
protostars and PBRs compared to the total number of HOPS and new
candidate protostars is, however, 2 times larger in Orion B.  This
result is even more pronounced when we consider only the fractions of
PBRs, with fractions that are more than 10 times larger in Orion B.
The NGC2068 (also containing the NGC2071 nebula) and NGC2024 (also
containing the Horsehead or NGC2023 nebula) fields in Orion B have not
only the largest fraction of new candidate protostars, but also of
PBRs.  While these numbers and fractions are subject to counting
statistics and other possibly large sources of errors, the differences
between Orion A and Orion B appear to be significant.

About 5\% of the combined protostars and candidate protostars in Orion
are PBRs. If we consider the PBRs as representing a distinct phase in
the evolution of a protostar, and we assume a constant rate of star
formation, the fraction the variation suggests that the protostars
spend 5\% of their lifetime in the PBRs phase \citep[approximately
  25,000 years with the 0.5 Myr protostellar lifetime of ][]{evans09},
averaging over all Orion regions.  However, the assumed the duration
of the PBRs phase would vary greatly with location, from $5,000$~years
in the Orion A cloud to $80,000$~years in the Orion B cloud.  There
are two alternative explanations.  First, there might be environmental
reasons which would favor the formations of PBRs, or perhaps extend
the duration of the PBRs phase, in the Orion B cloud. Second, the ages
of all the protostars in the Orion B cloud may be systematically
younger than those in the Orion A cloud.  In this case, the regions
containing the PBRs in the Orion B could be undergoing very recent
bursts of star formation.

Studies of pre--main sequence stars in the Orion molecular clouds show
little evidence for significant age differences between the Orion A
and B clouds. \citet{flaherty08} determined an age of $\sim\,$2~Myr
for NGC2068 and NGC2071, while \citet{reggiani11}, \citet{hsu12} and
\citet{dario12} determine ages for the ONC and L1641 of
$\sim\,$2~---~3~Myr.  However, most of the protostars associated the
NGC2068 and NGC2071 regions are outside the clusters of pre--main
sequence stars and in dense filaments gas neighboring these clusters
\citep{motte01}.  A number of PBRs are in the LBS~23 clump (directly
south of NGC2068) and in the NGC2023 clump (in the NGC2024 field);
these are two of the 5 most massive, dense clumps found in Orion
\citep{lada91}.  Compared to the other massive clumps, both of these
regions have $1/10$ the numbers of young stars per unit gas mass and
hence may be quite young \citep{lada91b,lada92}.  Alternatively, the
gas in the LBS~23 and NGC2023 clumps may have dense gas filling
factors that are much higher than the other massive clumps
\citep{lada97}; hence, they sources in these regions may be forming in
a very different birth environment.  The regions bordering the
southern rim of the NGC2068 nebula and the northern rim of the NGC2071
nebula are also rich in protostars \citep[see ][]{megeath12}.  Thus,
the PBRs are found concentrated in sub--regions which may indeed be
quite young.  We will investigate these possibilities in future work.

\section{Determining the Physical Properties of PBRs through models}

In this section we describe our analysis of the physical properties of
the PBRs as inferred from their colors and SEDs. We first compare the
$70/24$ colors of the PBRs sample with those derived from a grid of
models which adopt the solution for a rotating envelope undergoing
collapse \citep{ulrich76,terebey84} with outflow cavities along the
rotation axis of the envelope \citep{whitney03a}. This analysis puts a
constraint on the minimum inner density of the protostellar envelope.
Next, we compare the observed SEDs to model SEDs generated using the
Hyperion \citep{robitaille11} radiative transfer code.  This set of
models assumes radial power--law gradients consistent with either a
collapsing core with a constant infall rate and a static isothermal
core.  The models also encompass various combinations of internal and
external heating.  Given the prohibitively large computational time
needed to explore the full range of parameters space using radiative
transfer models, and given our inability to distinguish between models
purely from five to six photometry points, we do not provide
individual model fits to each protostars.  Instead we fit a single
temperature modified black--body function to the observed SEDs at
70~\micron\ and longer wavelengths. The modified black--body fits
provide luminosities and an initial characterization of the envelope
masses of the PBRs sample.

\subsection{Axisymmetric models: interpreting the $70/24$ color}

We begin our analysis by using a simplified version of the
\citet{ali10} protostellar envelope model grid to predict observed
fluxes and colors for comparison with our PBRs.  The density
distribution of these models is that of the collapse of a spherical
cloud in uniform rotation \citep{ulrich76}, which is the inner region
of the \citet{terebey84} model of the collapse of the slowly--rotating
isothermal sphere.  This model is then modified by the inclusion of
outflow cavities of various shapes \citep{whitney03a,whitney03b}.
This schematic model envelope captures the dependence of the short
wavelength (24~\micron\ and 70~\micron) fluxes on inclination due to
rotation and bipolar cavities.

The model fluxes depend upon the mass infall rate, the angular
momentum of the mass currently falling in, the outflow cavity
properties, the inclination of the rotation axis to the line of sight,
and the luminosity of the central source, as well as the assumed dust
properties.  The thermal emission of the dusty envelope does not
depend directly on the mass infall rate but instead on the density of
the envelope.  The model assumes free--fall at a constant rate, which
results in a density profile with shape $\rho \propto r^{-3/2}$
\citep{terebey84} .  The overall scaling of the density is
characterized by $\rho_1$, the density at 1~AU in the limit of no
rotation:
\begin{equation}
  \rho_1 =  7.4 \times 10^{-15} \Bigg( \frac{\dot{M}_{\rm env}}{10^{-6} {M}_{\odot}\,{\rm yr}^{-1}} \Bigg) \Bigg(\frac{M_*}{0.5\,{M}_{\odot}}  \Bigg)^{-1/2} \rm{g\,cm}^{-3},
\end{equation}
The envelope mass infall $\dot{M}_{\rm env}$ rate is related to
$\rho_1$ via the free--fall velocity, which in turn depends upon the
unknown central mass ${M}_{*}$.  The actual model density structure
departs from $r^{-3/2}$ on small scales because of the angular
momentum of the infalling material.  This enters into the model
through the parameter $R_{\rm disk}$, the outer disk radius at which
infalling material currently lands \citep[see ][]{ulrich76}.

The rotation leads to a significant dependence of the SED on the
inclination of the rotation axis relative to the line of sight
\citep{kenyon93a}. This dependence is significantly enhanced by the
inclusion of outflow cavities \citep{whitney03a} which are assumed to
be aligned along the rotation axis.  Finally, the overall shape of the
SED is only weakly affected by the luminosity of the central source
 \citep[$L_*$;][]{kenyon93a} and so this is easily scaled.

\begin{figure*}
  \begin{center}
    \scalebox{0.511}{{\includegraphics{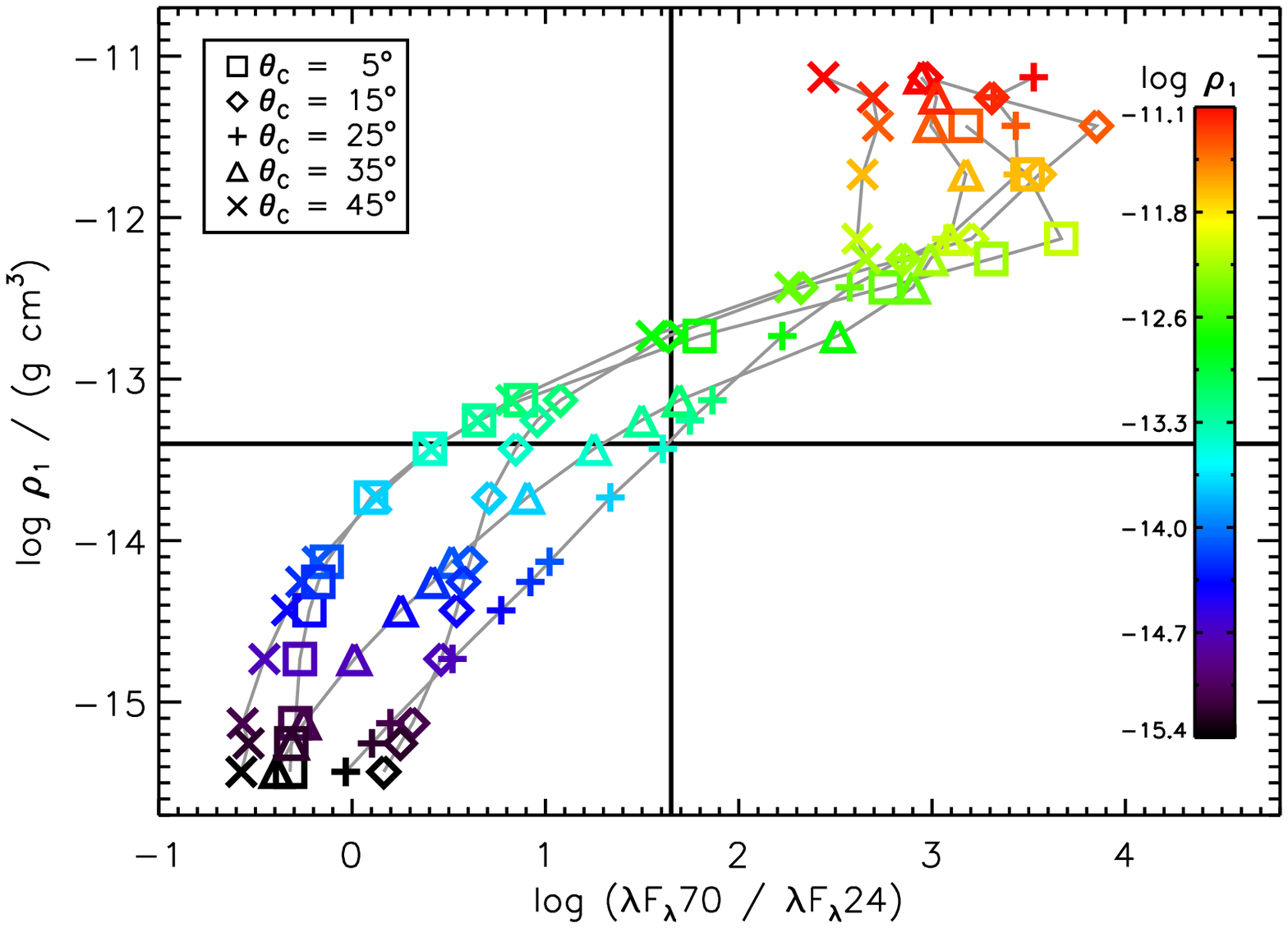}}{\includegraphics{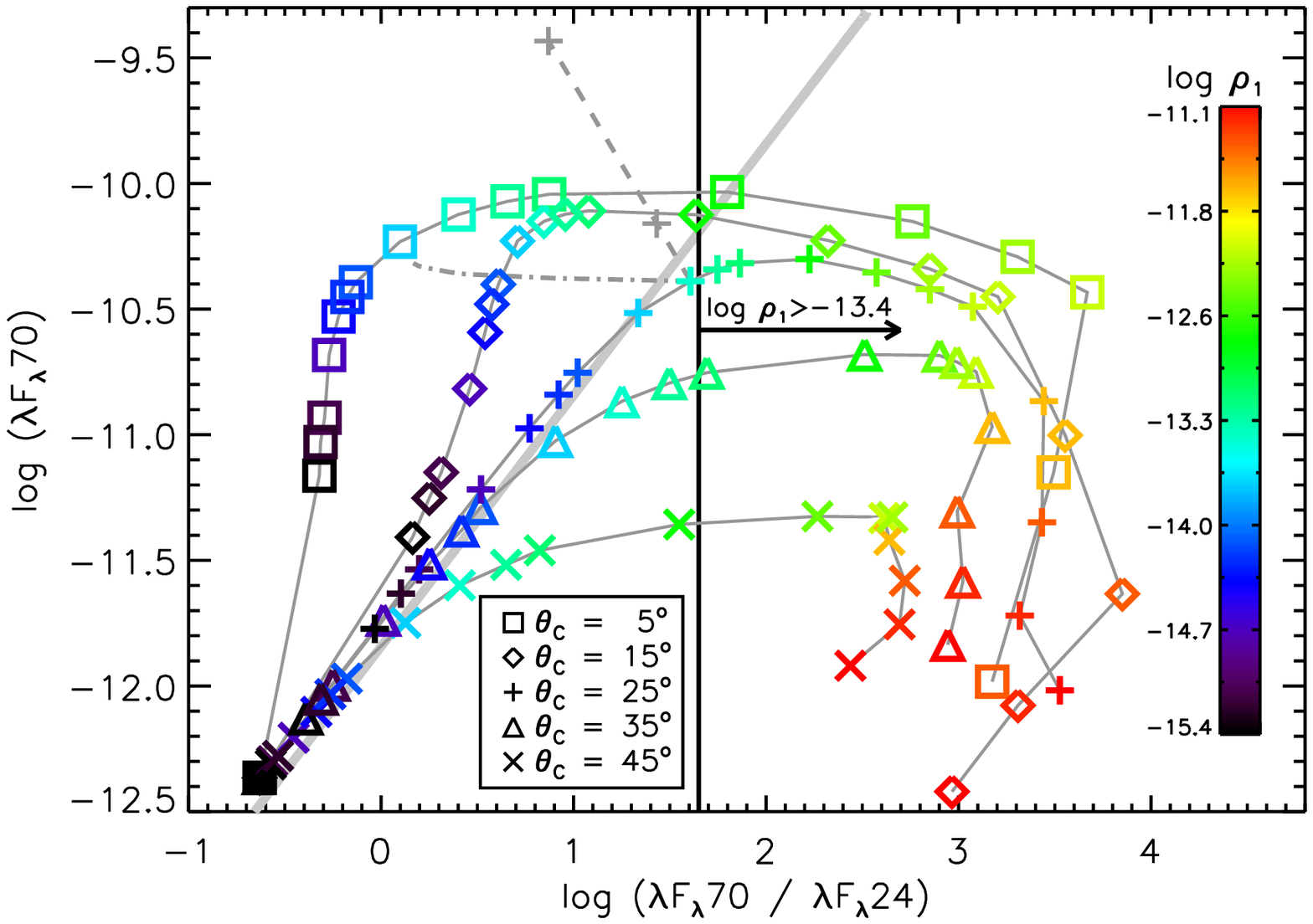}}}
    \caption{Left: Dependence of the model 70~\micron\ to
      24~\micron\ flux ratio for {\it high inclination} orientations
      ($87^{\rm o}$) on envelope density ($\rho_1$) for the 5 cavity
      opening angels in our grid. These model tracks assume an
      internal luminosity of 1.05~\lsun.  The $70/24$ color selection
      is indicated with the vertical black line, while the
      corresponding log $\rho_1 / (g\,cm^{-3}) \sim -13.4$
      (corresponding to an envelope infall rate of \mdotenv\ $= 7.5
      \times 10^{-6}$) threshold is indicated with a horizontal black
      line.  Right: Model 70~\micron\ flux vs. 70~\micron\ to
      24~\micron\ flux ratio for {\it high inclination} orientations
      ($87^{\rm o}$). For the model with a cavity opening angle of
      $\theta_{\rm C} = 25^o$ ($+$--symbols) and log $\rho_1 /
      (g\,cm^{-3}) \sim -13.4$, we show the effect of of increasing
      the internal luminosity by two orders of magnitude with the
      dashed grey curve while the effect of decreasing the inclination
      to $18^o$ is illustrated with the dot--dashed curve.  All models
      with higher assumed values of \mdotdisk\ and lower inclination
      will have envelopes that are denser than log $\rho_1 /
      (g\,cm^{-3}) \sim 13.3$, the median value of $\rho_1$ in our
      model grid; sources with redder 70 to 24~\micron\ colors cannot
      be explained by envelopes that are less dense than this
      threshold.  The light grey thick line corresponds to a
      24~\micron\ limit of 7 magnitudes, the imposed
      \spitzer\ protostar magnitude limit for identification of
      protostars.}
    \label{fig:modcolors}
  \end{center}
\end{figure*}

To roughly compare observed PBRs colors with those predicted by our
model grid in Figure~\ref{fig:modcolors} we show the effects of
varying the model inclination, envelope density, and cavity opening
angle on the $70/24$ color and 70~\micron\ flux.  As stated above,
these model tracks are based on a simplified version of the
\citet{ali10} model grid; we refer the reader to that publication for
details.  In brief, the model tracks that we consider here have the
same fixed parameters as those listed in Table~1 of \citet{ali10},
including a fixed central mass of $0.5$~\msun; in addition we have
fixed the cavity shape exponent to a value of $b = 1.5$, and the
envelope outer radius to $R_{\rm env,max} = 1 \times 10^4$~AU. We
have, however, expanded the envelope infall rate grid relative to the
\citet{ali10} grid to larger values (up to $\dot{M}_{\rm env} =
10^{-3}$~\msun~yr$^{-1}$ on a pseudo--logarithmic grid) and included a
model with no envelope.  As described above, for our model grid we
assume that the envelope density falls off as $\rho(r) \propto
r^{-3/2}$.  In addition, while our model grid is parametrized in terms
of $\dot{M}_{\rm env}$, with a fixed central mass of $M_* =
0.5$~\msun, we will refer to $\rho_1$ throughout this section.  

We note that the assumed central masses of protostellar sources remain
largely unconstrained observationally \citep[however see ][]{tobin12},
but may be lower than our assumed value.  The effect of a lower
central mass would be to lower $\dot{M}_{\rm env}$ for a given value
of $\rho_1$.  For example, if the central mass is $0.2$~\msun, the
infall rates reported for our models would be reduced by a factor of
0.6.  The assumed central mass, however, does not change the value of
$\rho_1$ corresponding to a given model SED.  Furthermore, our small
assumed disk radius does not strongly affect the trends shown here.

With this model grid we isolate the effects, albeit in a simplified
fashion, of varying the model inclination (viewing angle to the
protostar), envelope density ($\rho_1$), and cavity opening angle
($\theta_{\rm C}$) on the 70~\micron\ fluxes and $70/24$ colors.  In
Figure~\ref{fig:modcolors}, we show model tracks through
70~\micron\ flux vs.\ $70/24$ color space for high inclination
($87^{\rm o}$) viewing angles as a function of both $\theta_{\rm C}$
and envelope density ($\rho_1$).  By analyzing only the high
inclination models we obtain a lower limit on the envelope density
required to reproduce the red $70/24$ colors. 

We assume a fixed disk accretion rate of \mdotdisk\ $= 1.0\times
10^{-8}$~\msun~yr$^{-1}$, a disk outer radius of $R_{disk,max} =
5$~AU, and a central source luminosity of $1$~\lsun.  Increasing
\mdotdisk\ and $R_{disk,max}$ drives the models to both bluer $70/24$
colors and brighter 70~\micron\ fluxes, while decreasing the
inclination drives the models to bluer $70/24$ colors.  Therefore, we
find that no model in our grid can explain $70/24$ colors with
envelope densities less than log $\rho_1 / ({\rm g\,cm}^{-3}) \sim
-13.4$ (or equivalently an envelope infall rate of \mdotenv\ $= 5.4
\times 10^{-6}$~\msun~yr$^{-1}$).  While models with larger values of
$\rho_1$ and other combinations of parameters can be found for bluer
$70/24$ colors, this analysis sets an approximate lower limit on the
expected envelope densities of sources with $70/24 > 1.65$ of log
$\rho_1 / ({\rm g\,cm}^{-3}) \gtrsim -13.4$.  That is, high source
inclinations alone cannot explain the red $70/24$ colors of the PBRs,
which also require dense envelopes.  

We expect that this value of $\rho_1$ should lie well within the
Class~0 range; e.g., \citet{furlan08} found for a sample of 22 Class~I
sources in Taurus log $\rho_1 /({\rm g\,cm}^{-3}) \leq -13.2$, while
over half had $\rho_1$ values that are lower than log $\rho_1 /({\rm
  g\,cm}^{-3}) = -13.4$.  Note that the \citet{furlan08} data--set
included far better sampling of the source SEDs and, most critically,
\spitzer\ IRS spectroscopy, allowing for a more robust estimate of the
parameters of sources in their sample.  In contrast, our SEDs
generally are envelope dominated with few sources having robust
detections shortward of 24~\micron.  Therefore, any estimate of the
value of the envelope density will be necessarily imprecise and have a
back of the envelope character.  That said, our comparison with the
model grid and previous derived values of the envelope density leads
us to conclude that the $70/24 > 1.65$ color cut, while not uniformly
selecting a unique envelope density threshold, will preferentially
select Class 0 sources with log $\rho_1 / ({\rm g\,cm}^{-3}) \gtrsim
-13.4$, irrespective of source inclination.

Having demonstrated that the very red $70/24$ colors require large
values of $\rho_1$, as expected for SEDs peaking at long wavelengths,
we proceed by analyzing the PBRs in the context of more simplified
models.  In what follows, we carry out two independent analyses of the
source SEDs: a qualitative model image comparison using 1D models
generated with the Hyperion~\citep{robitaille11} radiative transfer
code, and modified black--body fitting to the observed SEDs at
70~\micron\ and longer wavelengths.  We attempt to maintain
consistency by using similar envelope dust models throughout the
analysis presented here.  For the Hyperion model image analysis we use
the \citet{ormel11} opacities.  These opacities are similar to the
commonly assumed \citet{ossen94} (``OH5''; see below) opacities but
include both the scattering and absorption components at short
wavelengths, needed for radiative transfer calculations.
Specifically, we use the ``icsgra2'' \citet{ormel11} model opacities,
which include icy silicates and bare graphites, with a coagulation
time of 0.1~Myr.  We assume the dust model from \citet{ossen94} for
the long--wavelength modified black--body fits, specifically, their
``OH5'' opacities, corresponding to column 5 of their Table~1.  These
opacities reflect grains having thin ice mantles with $10^5$ years of
coagulation time at an assumed gas density of $10^6$~cm$^{-3}$.

\subsection{Spherically symmetric models: Comparison to photometry 
derived from model images}

We use the Hyperion \citep{robitaille11} Monte Carlo radiative
transfer code to investigate some basic properties of the PBRs.  We
run a series of spherically symmetric models under a range of very
simple assumptions, and produce simulated images.  As stated above,
the dust model we assume is that of \citet{ormel11} ("icsgra2'').  To
explore which model assumptions might be reasonable for analyzing the
properties of the reddest sources in Orion, we investigate a few
limiting model scenarios.  We therefore disregard the details of the
individual source SEDs of the entire red sample and qualitatively
compare the SEDs of the two sources with the smallest and largest
values of L$_{\rm bol}$ and T$_{\rm bol}$ (sources 091016 and HOPS358,
respectively; see Table~\ref{tab:mbb}) with photometry derived from
model images.

We generate a series of model images that fall into four classes: i) a
starless core at a constant temperature of 10~K (referred to here as
the ``core'' model); ii) a starless core with an isotropic external
radiation field (``core$+$external''); iii) a core with an internal
source (``star'' model); and finally, iv), a core with both an
internal source and an isotropic external radiation field
(``star$+$external''). For each of these classes of models, we test a
range of density profile shapes and density normalizations.  For the
density profile shape we assume two values, $\alpha = 1.5, 2.0$, where
$\alpha$ is the radial density profile power law index: $\rho(r)
\propto r^{-\alpha}$. For the absolute value of the (gas) density
normalization at 1~AU we assume 5 values: $\log{(\rho_{1{\rm
      AU}}/(g\,cm^{-3})}) = -10$ to $-14$, in steps of $\delta
\log{(\rho_{1{\rm AU}}/(g\,cm^{-3})}) = 1.0$. For models with external
heating, the bolometric strength of the interstellar radiation field
(ISRF) is set to the value from \citet{mathis83} at the solar
neighborhood ($4\pi\,J_\nu=0.0217$\,ergs/cm$^2$/s).  The spectrum of
the radiation field is assumed to be that at the solar neighborhood
from \citet{porter05}, but reddened by $A_{\rm V} = 10$ using the
\citet{kim94} extinction law. The ISRF model includes contributions
from the stellar, PAH, and FIR thermal emission.  The inner radius of
the core is set to the radius at which the dust sublimates, assuming a
sublimation temperature of 1,600\,K, while the outer radius is set to
1\,pc. The central source is taken to have 10~\lsun\ and a spectrum
given by a Planck function at the effective temperature of the Sun
(5778\,K). The choice of the stellar temperature is arbitrary, and is
unimportant for the modeling presented here, since all sources are
deeply embedded and all stellar radiation is reprocessed --- only the
total bolometric luminosity is important \citep[see e.g., ][for a
  discussion of the R$_{\rm star}$ and T$_{\rm star}$
  degeneracy]{johnston12}.

The high levels of spatial filtering caused by our adopted
aperture photometry scheme require us to assume such a central source
luminosity to roughly match the flux levels in the observed SEDs.
Furthermore, while a 1\,pc sized envelope is larger than usually
assumed, the high levels of spatial filtering inherent in the
aperture photometry cause us to be insensitive to structure on scales
larger than the assumed aperture sizes.

The model images have a resolution of 1$\arcsec pix^{-1}$, or 420~AU
at our assumed distance.  We convolve these images with the
azimuthally averaged PSFs provided by \citet{aniano11}, except in the
case of the SABOCA 350~\micron\ and LABOCA 870~\micron\ images.  These
wavelengths are convolved with Gaussian PSFs with FWHMs equal to
7.4$\arcsec$ and 19$\arcsec$ respectively, i.e., the nominal beam
sizes for our observations.  All model image photometry is then
performed on the convolved model images using the same aperture and
sky annulus parameters as those applied to the data.  The use of such
photometric aperture parameters can cause large amounts of spatial
filtering due the small sizes of the apertures relative to the beam
sizes and the extent of the core emission (see below).

We show our extracted model SEDs in Figure~\ref{fig:hypsr} and a
subset of the corresponding model images in Figure~\ref{fig:modimg}.
In general, we find that models without internal sources are very
unlikely to match the observed PBRs properties, on the basis that
their SEDs are less luminous and peak at longer wavelengths than the
observed SEDs for all density profile shapes that we assume; see
Figure~\ref{fig:hypsr}, top panels.  While the "core+external'' models
(top right panel) suffer from severe spatial filtering, it is unlikely
that such models will well represent the data as these SEDs tend to
also peak at longer wavelengths.  For the two classes of models with
internal sources we find better agreement with the data; see lower
panels of Figure~\ref{fig:hypsr}.  While steeper envelope profiles may
imply somewhat higher envelope densities, the range in plausible
densities for $\alpha = 1.5$ is $\log \rho_1\,\sim\,-12$ to $-13$,
while for $\alpha = 2.0$ values of $log \rho_1\,\sim\,-12$ roughly
agree with the shapes the observed SEDs.  We note that we find very
little difference between the ``star+external''and the "star'' model
in the $\lambda \lesssim 160$~\micron\ regime, possibly indicating
that our assumed ISRF strength relative to the assumed internal source
luminosity may be underestimated compared to what we might expect to
find regions like Orion.

\begin{figure*}
  \begin{center}
    \scalebox{0.45}{\includegraphics{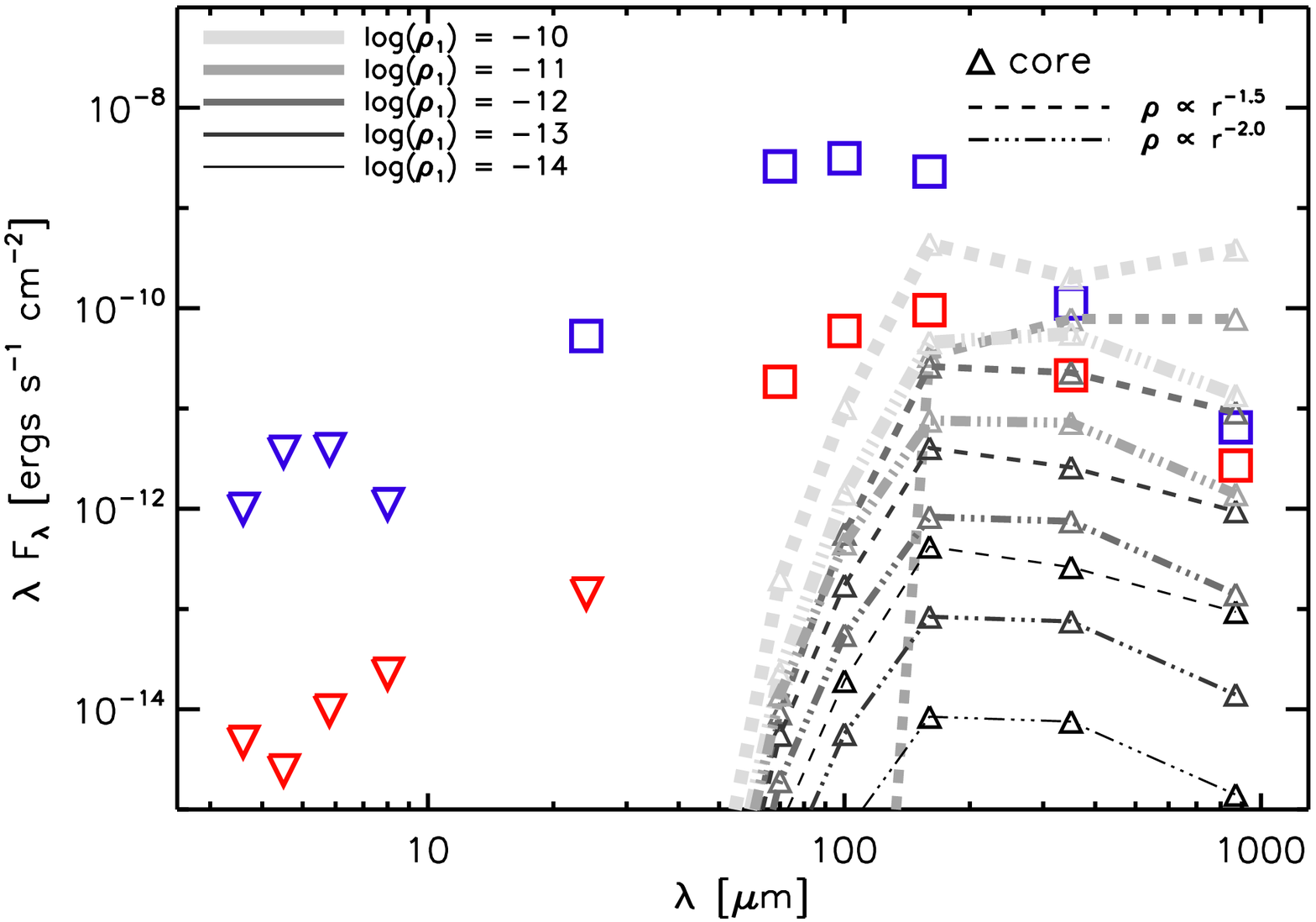}\includegraphics{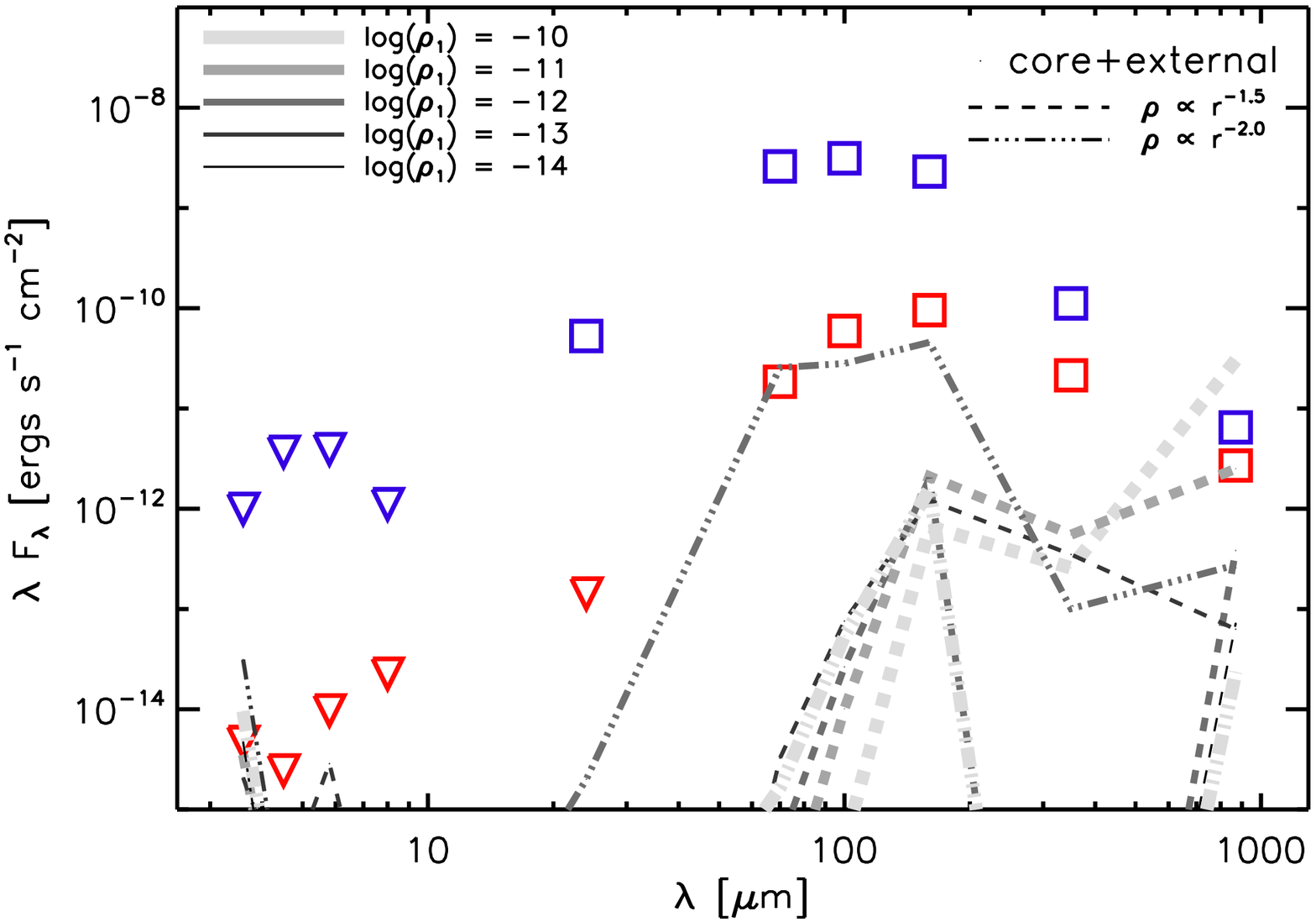}}
    \scalebox{0.45}{\includegraphics{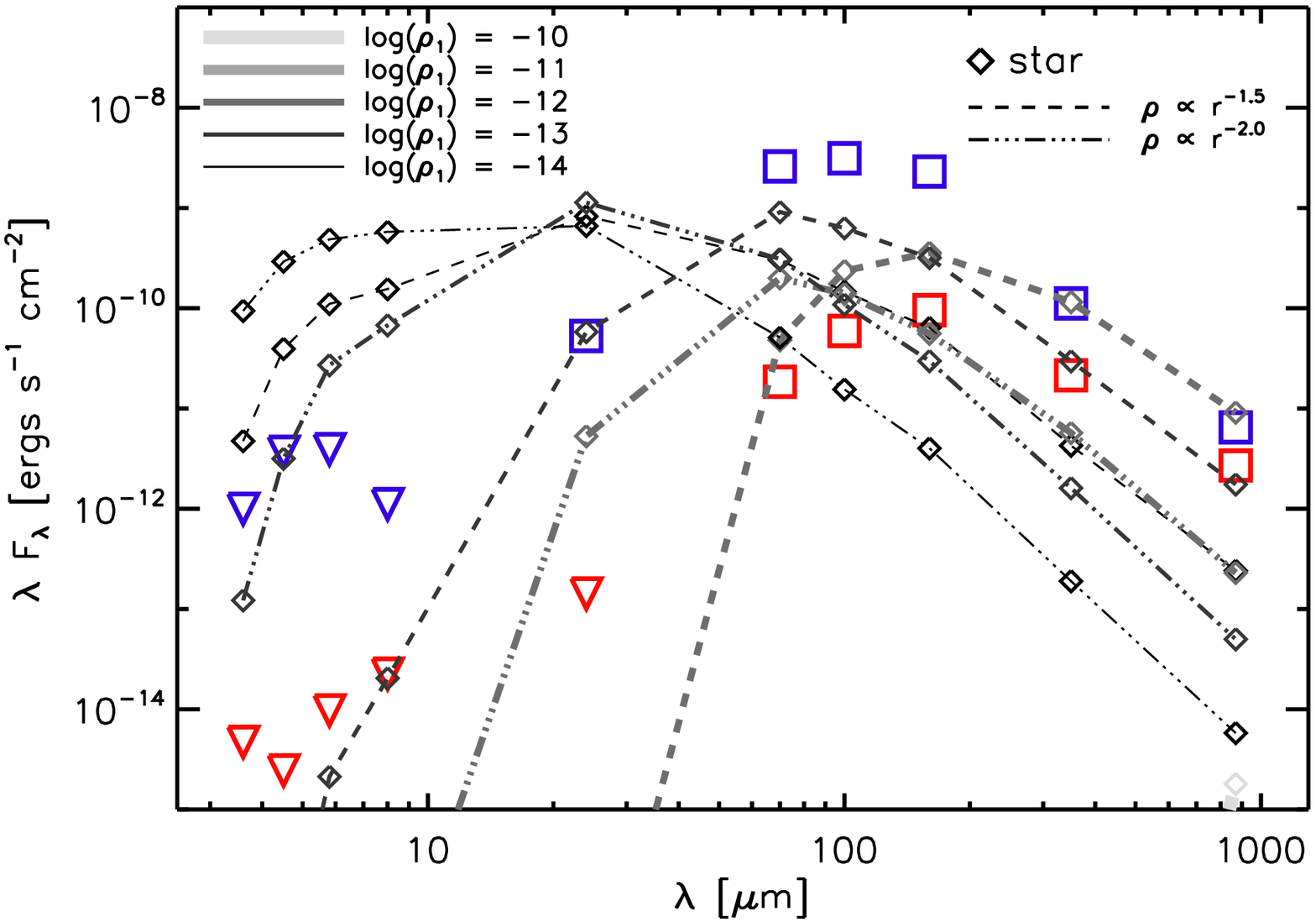}\includegraphics{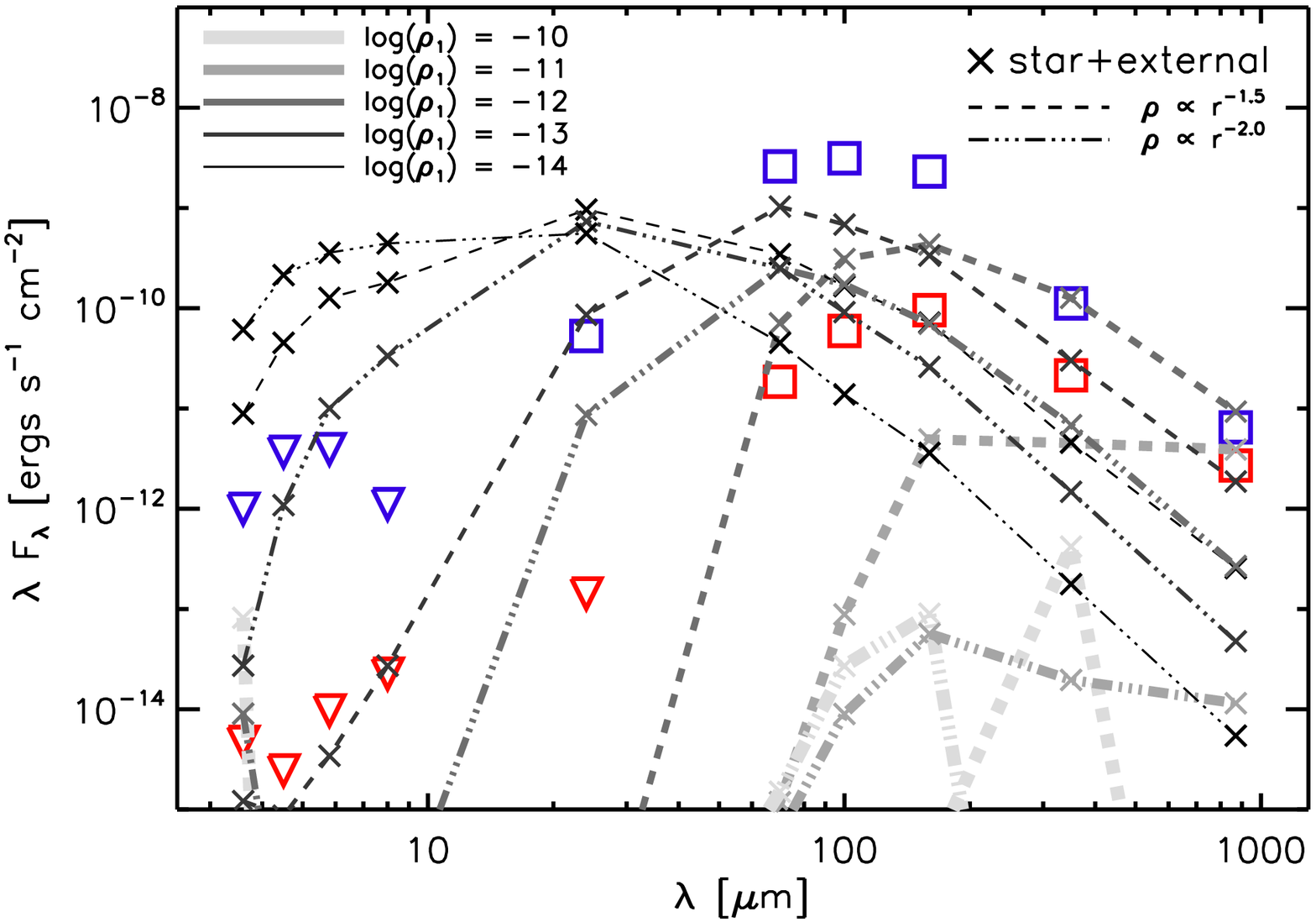}}
    \caption{All panels show the SEDs of HOPS358 (blue; L$_{\rm bol}$ =
      30.6~\lsun\ and T$_{\rm bol}$ = 44.2~K) and 091016 (red;
      L$_{\rm bol}$ = 0.65 and T$_{\rm bol}$ = 29~K), the extrema of
      the L$_{\rm bol}$ and T$_{\rm bol}$ distributions for the
      reddest sources, compared to spherical models.  The two line
      styles indicate the different assumptions for the density
      profile shape while the shade of grey indicates the different
      model density $\rho_1$ values.  The four panels correspond to
      the four different models that we test (see text): {\it Top
        left:} the ``core'' model, a starless core; {\it Top right:}
      the ``core+external'' model, a starless core with external
      heating; {\it Bottom left:} the "star'' model, a core with an
      internal source; and finally, {\it Bottom right:} the
      "star+external'' model, consisting of a core with an internal
      source which is irradiated by an external radiation field.
      Severe spatial filtering due to our aperture photometry on the
      model images can be seen most notably in the two externally
      heated models (left column).}
    \label{fig:hypsr}
  \end{center}
\end{figure*}    

\begin{figure*}[t]
  \begin{center}
    \scalebox{0.91}{\includegraphics{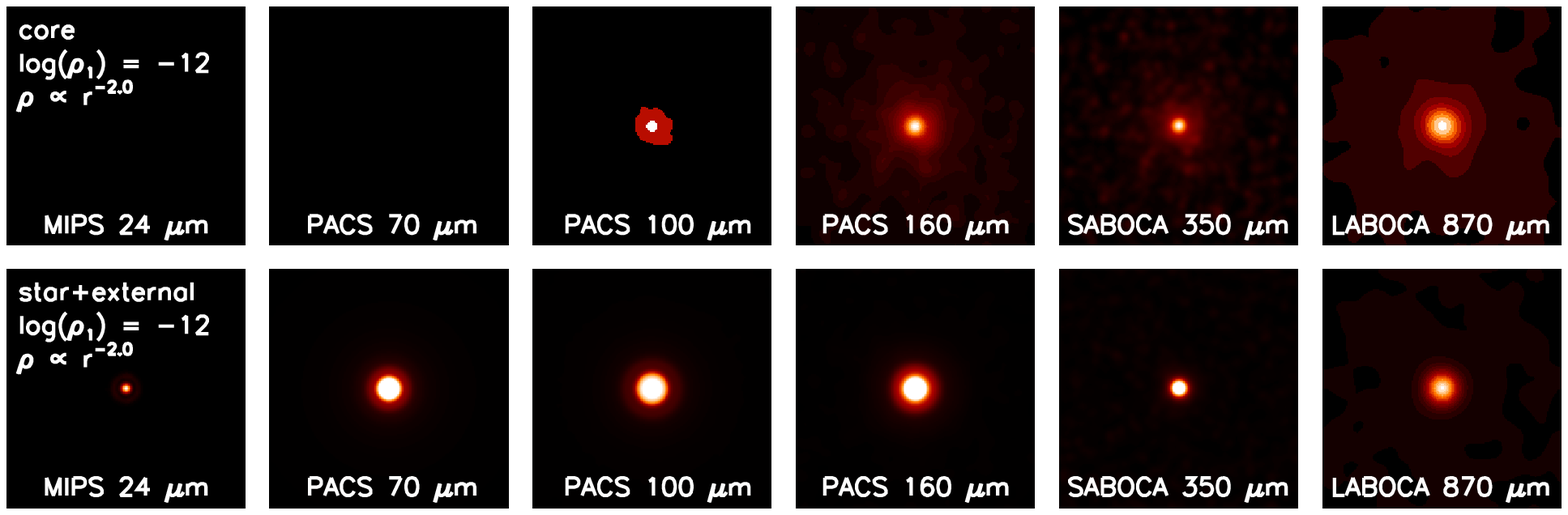}}
    \caption{Convolved model images at the indicated wavelengths,
      shown on a log scale with the same minimum and maximum flux
      levels for all panels to illustrate the shape of the model SEDs.
      Each image is 200$''$ (or 84000~AU) on a side.  The top row
      shows the ``core'' model images, with log$(\rho_1) = -12$ and a
      density profile shape of $\rho \propto r^{-2.0}$, corresponding
      to the top left panel in Figure~\ref{fig:hypsr}.  The bottom row
      shows the "star$+$external'' model, with the same density
      paramaters as the top row, corresponding to the bottom right
      panel in Figure~\ref{fig:hypsr}.}
    \label{fig:modimg}
  \end{center}
\end{figure*}    

\subsection{Modified black--body fits to the PBRs}

An individual detailed fit to each protostar is beyond the scope of
this work considering the vast amount of parameter space needed to
model protostellar SEDs (i.e., source luminosity, envelope density,
envelope rotation, outflow cavity geometry, external heating, outer
envelope structure).  Furthermore, unlike most of the HOPS protostars,
whose properties can be constrained from a combination of far--IR
photometry, 5~\micron\ to 40~\micron\ \spitzer/IRS spectra, and Hubble
near--IR imaging \citep{fischer10,fischer12}, the properties of the
SEDs must currently be derived from 5 to 6 photometry points at long
wavelength.  While the above modeling and analysis shows that the
internal source is important, the longer wavelength fluxes are probing
the bulk of the envelope mass, expected to mostly be at a single
temperature.  Furthermore, for density profiles in the range of
$\rho(r) \propto r^{-3/2}$ or $r^{-2}$, as assumed above, we expect
that most of the envelope will be located at large radii.  We
therefore perform modified black--body fits to the $\lambda \geq
70$~\micron\ SEDs listed in Tables~\ref{tab:phot} and
\ref{tab:photapex}. For this analysis we use the beam flux
measurements for the sub--millimeter 350~\micron\ and
870~\micron\ portion of the SED. The results of the analysis are
presented in Table~\ref{tab:mbb} and the model SEDs are plotted with
the data in Figure~\ref{fig:allseds}.

Before fitting the long--wavelength SEDs of the sources, we apply
color corrections to the \herschel\ 70~\micron, 100~\micron, and
160~\micron\ fluxes. Following \citep{laun12}, these photometric color
corrections have been derived iteratively from the slopes of the PACS
SEDs, using polynomial fits to the values in Table~2 of the PACS
calibration release note ``PACS Photometer Passbands and Colour
Correction Factors for Various Source SEDs'' from April 12, 2011.  The
color corrections for the APEX data are assumed to be negligible.

The form of the modified black--body function is given by
\begin{equation}
  S_{\nu} = \Omega\,B_{\nu}(\nu,T_{\rm d})\,(1-e^{-\tau(\nu)}),
\end{equation}
where $\Omega$\ is the solid angle of the emitting element,
$B_{\nu}(T_{\rm d})$ is the Planck function at a dust temperature
$T_{\rm d}$, and $\tau(\nu)$\ is the optical depth at frequency $\nu$.
Here, the optical depth is given by $\tau(\nu) = N_{\rm H}\,m_{\rm
  H}\,R_{gd}^{-1}\,\kappa(\nu)$, where $N_{\rm H} = 2\,\times\,N({\rm
  H_2}) + N({\rm H})$ is the total hydrogen column density, $m_{\rm
  H}$ in the proton mass, $\kappa_{\nu}$ is the assumed dust opacity
law from \citet{ossen94}, and $R_{gd}$ is the gas--to--dust ratio,
assumed to be 110~\citep{sodroski97}.  The best--fit total masses
M$_{\rm tot}$ reported in Table~\ref{tab:mbb} have been multiplied by
an additional factor of 1.36 to account for helium and metals.
Furthermore, in Table~\ref{tab:mbb} we also report the peak wavelength
of the best--fit modified black--body model.  Finally, we estimate
L$_{\rm smm}$ from the model SED, where L$_{\rm smm}$ is integrated
over $\lambda \geq 350~\micron$.  

If a given source SED has coverage over fewer than 4 long--wavelength
points, we do not fit a model to the SED.  While all the PBRs sources
satisfy this criterion, all the new candidate protostars and HOPS
sources do not (see \S~5.2) and are therefore not fitted.  The errors
on $T_{\rm d}$, M$_{\rm tot}$, and the thermal component of the
luminosity (L$_{\rm MBB}$) are estimated through a straight--forward
Monte Carlo method\footnote{``Offered the choice between mastery of a
  five--foot shelf of analytical statistics books and middling ability
  at performing statistical Monte Carlo simulations, we would surely
  choose to have the latter skill.''  Press, 1993, Numerical Recipes,
  page 686.}.  For each source we generate 2000 synthetic SEDs drawn
from a normal density with mean and standard deviation equal to those
of the measured SED at each wavelength.  We then fit each synthetic
SED. The reported error is equal to the standard deviation of the
resulting distribution of each parameter. These errors do not include
systematics introduced by, e.g., our dust model assumption or
variation in the gas--to--dust ratio.

We show the modified black--body fit results, along with the SEDs of
the PBRs, in Figure~\ref{fig:allseds}.  The resulting best--fit mass,
luminosity, and temperature is also indicated for each source. The
model fits the data surprising well considering that significant
temperature gradients in the envelope are expected.  Furthermore, in
all cases the 24~\micron\ point, when detected, has a much higher flux
level than the modified black--body model.  We interpret this
discrepancy as strong evidence for internal heating by a protostar.  

Excluding the 70~\micron\ point and fitting only the $\lambda \geq
100$~\micron\ SED has a minor effect on the resulting parameter
values. Without 70~\micron, the masses systematically increase by
40\%, the temperatures decrease by 5\%, and the luminosities decrease
by 7\%.  This small effect may be understood by the fact that the
100~\micron\ fluxes are well--correlated with the 70~\micron\ fluxes
for this sample, tracing similar material near the protostars.  The
temperatures at 100~\micron\ and 70~\micron\ are not dramatically
different, and most likely both points are dominated by optical--depth
effects such that the $\tau = 2/3$ surface is not significantly
different between the two wavelengths.  From \citet{hartmann09}, the
radius of the $\tau = 2/3$ surface can be roughly approximated as
$r_\lambda \propto \kappa_\lambda^2$; this relation implies that
$r_{70}/r_{100} \sim 2.5$.

Our best--fit modified black--body model always underestimates the
observed 870~\micron\ flux of all sources.  The model sub--millimeter
SEDs are always bluer than the observed SEDs.  We find that the
discrepancy is at the $0.83\pm0.26$~Jy level (or a factor of
$\sim\,3)$ excess), where the error bar represents the standard
deviation in the residual distribution.  It is likely that this
discrepancy is dominated by the larger beam size of the
870~\micron\ observations which has the effect of mixing the source
flux with that of the surrounding cold and possibly high--column
environment.  Contamination to the 870~\micron\ flux by disk emission
may also increase this discrepancy.  \citet{jorgensen09} find average
disk masses of $\sim\,0.13$~\msun\ (with a large scatter) across their
sample of Class~0 sources.  The sources in their sample that are
comparable to our PBRs, however, are those with the lowest values of
T$_{bol}$.  For reference, they found that IRAS4A1, with a T$_{bol} =
43$~K, has the largest disk mass of 0.46~\msun\ in their sample; on
the other hand, L1157, with a similar T$_{bol} = 42$~K, has a disk
mass about a factor of 4 smaller.  We estimate that a 30~K disk of
0.5~\msun\ would contribute $\sim\,0.6$~Jy to the beam flux at
870~\micron\ (assuming \citet{ossen94} dust opacities, as above).
Therefore, disk emission could indeed contribute to the observed
870~\micron\ flux but further detailed observations at high resolution
are needed to disentangle the envelope component from the possible
disk emission.  Another possible source of ambiguity in interpreting
the 870~\micron\ flux discrepancy is the model dust opacity
assumption.  Furthermore, we do not find an 870~\micron\ discrepancy
in the analysis of model images presented in the previous section.
This indicates that large disk masses may not be necessary to explain
the sub--millimeter fluxes.  We therefore emphasize that the disk
masses inferred here from the 870~\micron\ excess should be regarded
only as upper limits; further detailed investigation into the disk
properties of our sources is deferred to future work.

Independent of these issues, it is clear that a more accurate
treatment of the data would require all images to be convolved to a
matched resolution; however, this approach would have the effect of
causing non--detections for a majority of sources at the shorter
wavelengths due to the relatively large limiting beam size of our
data--set ($\sim19\arcsec$ at 870~\micron).  Homogeneously extracted
SEDs are therefore not feasible for this data--set as a
whole. Nevertheless, we test the effects of convolving the the data
before extracting the SEDs.  We choose PBRs 119019 as a test source
because it is isolated and has approximately median values for the
best--fit modified black--body temperature, luminosity, and mass.
Ignoring the 870~\micron\ data, this source is clearly detected at
70~\micron, 100~\micron, 160~\micron, and 350~\micron.  For these four
wavelengths, the largest beam size of $\sim12\arcsec$ corresponds to
the 160~\micron\ data.  We therefore convolve the 70~\micron,
100~\micron, and 350~\micron\ data to a resolution matching the
160~\micron\ observations and extract a beam--smoothed SED.  We then
fit this SED in the same way as described above.  Compared to the
non--convolved SED modified black--body fitting results, we find that
the temperature decreases by $\sim\,4$\%, the luminosity increases by
$\sim\,10$\%, and the the mass in the thermal component increases by
$\sim\,30$\%.  These systematic shifts are similar to but somewhat
larger than the errors quoted in Table~\ref{tab:mbb} ($\sim\,2$\% on
the temperature, $\sim\,10$\% on the luminosity, and $\sim\,30$\% on
the mass).  We note, however, that the errors quoted in
Table~\ref{tab:mbb} are purely random and do not include any
systematic component.  We therefore conclude that extracting SEDs from
images matched to a resolution of $\sim12\arcsec$ will not greatly
affect our results.

Modified black--body fits provide a somewhat limited means of analysis
of our sources since the model assumes a single temperature and
density along the line of sight for the emitting material.  We expect
that the assumption of a single line--of--sight temperature will cause
an underestimate of the source masses
\citep[e,g.,][]{nielbock12,laun12}.  However, radiative transfer
models have large ambiguities in the assumed source temperature and
density structure, leading to mass estimates that strongly
model--dependent.  Furthermore, the dust law that is assumed will
introduce significant uncertainties into the derived masses,
irrespective of the analysis method that is implemented.  For example,
the masses listed in Table~\ref{tab:mbb} increase by a factor of
$\sim\!$4 on average when we assume \citet{draine84} $R_V = 3.1$
ISM--like dust.  These issues indicate that the masses derived here
represent lower limits to the true envelope masses.  Nevertheless, we
consider the modified black--body fits to the measured photometry to
provide the most robust estimates of the mass that we currently have.

We note that with only 5 SED flux points at best, fitting a multiple
component (modified) black--body model cannot be justified.  Since
most of the mass is located at relatively large scales and expected to
have cold temperatures, excluding the warmer shorter wavelength data
arising from inner material will not significantly increase the
masses we derive. The modified black--body fits thus to provide an
approximate measurement of the optical--depth averaged gross
properties of the envelopes being investigated.  These issues point to
the need for a more sophisticated modeling approach that will be
carried out in future work.

\section{Discussion}

As seen in Figure~\ref{fig:modcolors}, the observed $70/24$ colors of
a protostar can be driven towards redder values through various
strongly degenerate parameters.  For example, the total column of
material along the line--of--sight (LOS) towards a given protostar can
have multiple contributions: the attenuation of the mid--IR emission
by dense foreground material, the density of the envelope, the amount
of envelope flattening, the opening angle of the outflow cavity, and
the source inclination.  Furthermore, the assumed model central
protostar mass remains largely unconstrained by observations to date
and can affect the interpretation of the $70/24$ colors.

Foreground extinction can have various contributions, such as
intervening dust between the observer and the cloud and dense material
associated with the cloud itself, such as filamentary material.  Of
these two components, the first is expected to be relatively small,
while the latter can be expected to vary from source to source by
relatively large amounts, with a corresponding effect on the observed
colors.  For example, some PBRs are located in filamentary regions
(e.g., Figures~\ref{fig:img7} and \ref{fig:img8}), while others appear
more isolated (e.g., see Appendix for Figure~\ref{fig:img6}).  We have
estimated the effects of foreground extinction levels up to a level of
$A_{\rm V} = 40$ mag, and find that the values of \lbol, \lsmm, and
\tbol\ not significantly affected.

On the other hand, the effects of source inclination are not as
straightforward to assess.  When considering the presence of flattened
rotating envelopes, disks, and outflow cavities, the source
inclination will have a large effect on the observed source SED, as
illustrated by the model tracks shown in Figure~\ref{fig:modcolors}
\citep[see also, e.g., ][]{whitney03a,debuizer05,offner12}.  

\begin{figure*}[t]
  \begin{center}
    \scalebox{0.9}{\includegraphics{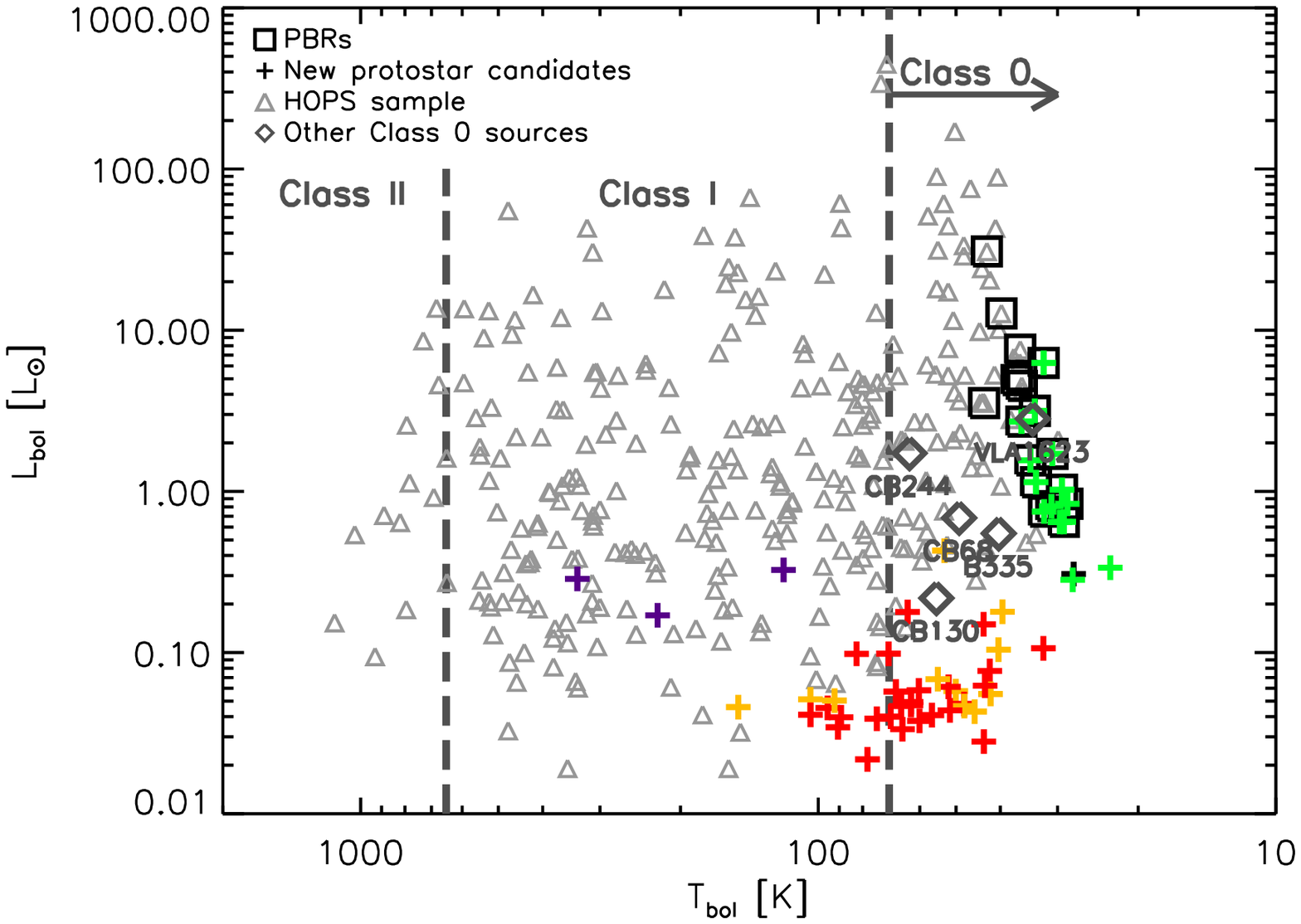}}
    \scalebox{0.9}{\includegraphics{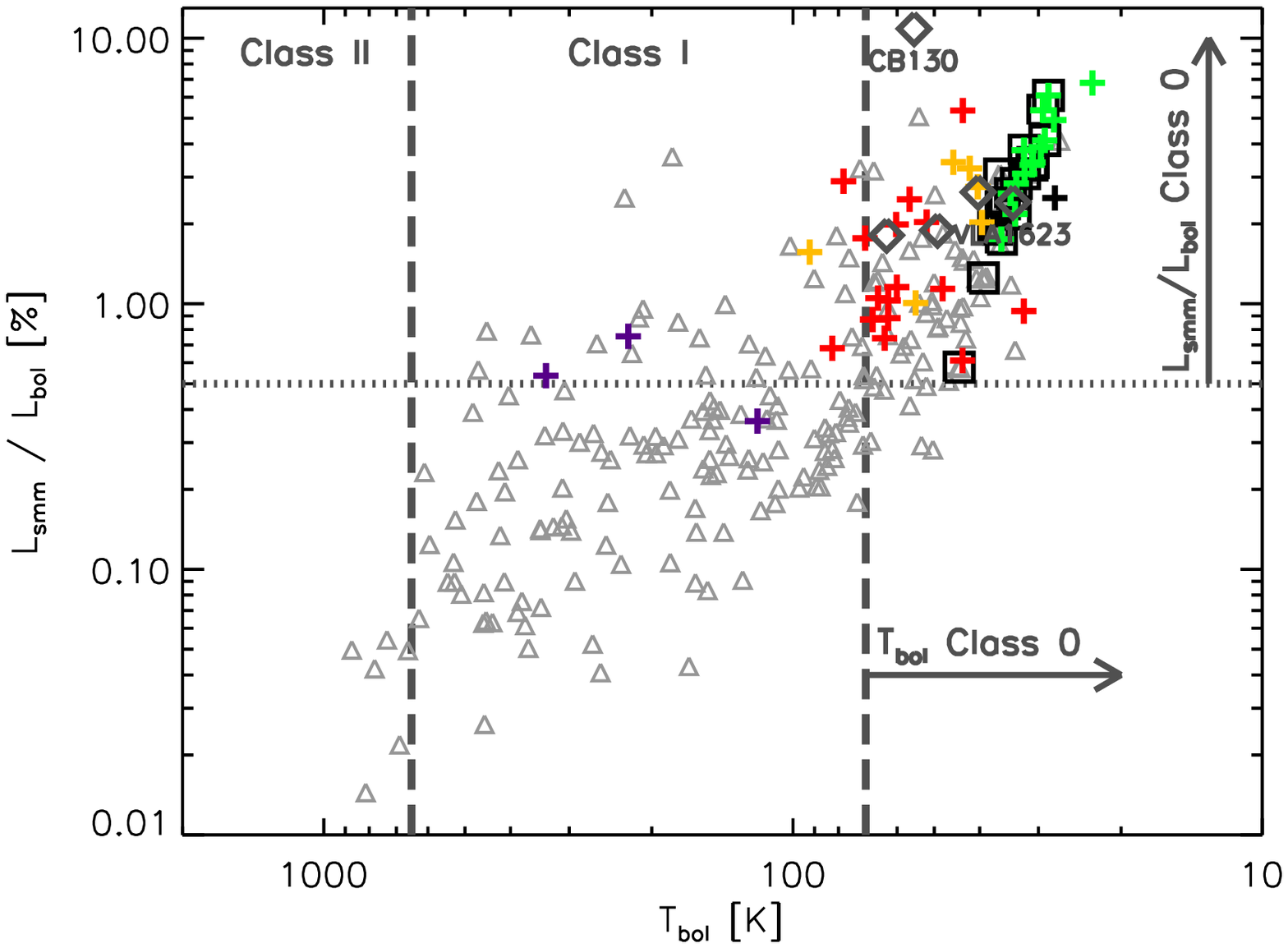}}
    \caption{Left: Bolometric luminosity vs.\ bolometric temperature
      for the new candidate protostars ($+$--symbols) and the HOPS
      protostars (light--grey triangles).  The squares indicate the
      PBRs sample, drawn from both the new candidate protostar sample
      and the HOPS protostar sample.  The dashed lines indicate the
      canonical \tbol\ divisions between protostellar Classes.  Right:
      \lsl\ vs\ bolometric temperature for the subset of sources shown
      in the left panel for which we have sufficient submillimeter
      coverage to estimate \lsmm\ (see text).  The horizontal dotted
      line indicates the \citet{andre00} proposed L$_{\rm
        smm}$/L$_{\rm bol} > 0.5$\%\ Class~0 threshold.}
    \label{fig:blt}
  \end{center}
\end{figure*}  

Therefore, it appears that the very red PBRs can be explained by
multiple effects that all result in increasingly red observed $70/24$
colors.  These very red colors may be driven by elevated envelope
densities (or equivalently, \mdotenv), high source inclinations, or
elevated levels of extinction associated with structures larger than
the envelope--protostar system.  The current data and SED coverage do
not allow us to break these degeneracies conclusively.  Furthermore,
we consider it likely that the red observed colors are not driven any
single cause, but instead are the result of several.

We have, however, designed our PBRs selection to find the densest
envelopes in Orion (c.f., Figure~\ref{fig:modcolors}).  Furthermore,
the effect of external foreground extinction is not expected to be
large at these long wavelengths, even with elevated levels of material
along the LOS (see above). Indeed, we have also shown that the PBRs
require a central heating source, indicating that the detection of a
70~\micron\ point source drives the interpretation of the sample as
Class~0 sources, irrespective of source inclination.  However, we note
that if the central masses are significantly different than the
assumed value of 0.5~\msun, then for a fixed reference envelope
density the inferred envelope infall rates need to be scaled
accordingly (see Equation~1).

To further investigate the evolutionary state of the PBRs, in
Figure~\ref{fig:blt} we show we the values of \lbol, \tbol, and
\lsmm$/$\lbol for the entire sample of new candidate protostars and
the previously identified \spitzer\ HOPS sample (Fischer et al.\ in
preparation).  In the left panel, we show \lbol\ vs.\ \tbol\ for the
entire sample of new protostar candidates, including those flagged as
possible extra--galactic contamination. We also show the four
reference Class~0 sources presented in Table~\ref{tab:mbb}.  The PBRs
sample in particular, and the entire sample of new candidate
protostars, are generally clustered around low \tbol\ values. Ignoring
inclination degeneracies and other considerations, these low \tbol
values indicate that the PBRs sample is indeed composed of young
Class~0 sources.  In the right panel, we show \tbol\ vs.\ L$_{\rm
  smm}$/L$_{\rm bol}$ for the sources with sufficient coverage to
estimate L$_{\rm smm}$ (see \S~6.2).  The PBRs, as expected if the
sample can be explained as Class~0 sources, cluster around larger
values of L$_{\rm smm}$/L$_{\rm bol}$ compared to the rest of the
sample.  \citet{andre00} proposed the L$_{\rm smm}$/L$_{\rm bol} >
0.5\%$ threshold for Class~0 sources, and all but one of the new
candidate protostars for which we can estimate L$_{\rm smm}$ fall into
this category.  Irrespective of the evolutionary indicator that is
chosen (\tbol\ or L$_{\rm smm}$/L$_{\rm bol}$ ), all of the new
candidate protostars in both the reliable and lower probability
categories (green and yellow points, respectively), would be
considered of Class~0 status.  Finally, while the PBRs $70/24 > 1.65$
color criterion causes some sources with very low values of \tbol\ and
very high values of L$_{\rm smm}$/L$_{\rm bol}$ to be missed, the
color selection is able to capture the vast majority of the most
extreme Class~0 sources at the extrema of the L$_{\rm smm}$/L$_{\rm
  bol}$ and \tbol\ distributions.

This evidence strongly supports the interpretation of the PBRs (and
indeed all the sources classified as reliable protostellar candidates)
as very dense Class~0 protostars, irrespective of the source
inclination.  On the other hand, the new candidate protostar sample,
taken as a whole, may be explained by a combination of the effects
described above: high inclination, high densities, and extreme values
of foreground extinction, along with elevated levels of extragalactic
contamination.  Of particular interest is the possibility that some of
the sources classified as low probability protostars at low
\lbol\ values may be confirmed as bona fide protostars with future
observations \citep[see ][ for a detailed discussion of the
  significance of such sources]{offner11}; this will be investigated
in future work (Stutz et al.\ in preparation).

To definitely measure the inclinations of our sample of sources and to
therefore determine envelope densities more accurately we require
millimeter line emission maps at high resolution from e.g., ALMA,
along with single dish observations of high density tracers.  A
vigorous follow--up campaign is therefore underway to more firmly
place these protostars within the context of star formation in the
Orion clouds. We are observing ammonia spectra toward the full sample
of protostar candidates to verify the presence of dense molecular gas
and determine kinetic temperatures. A \herschel\ PACS range
spectroscopy program toward 8 PBRs will characterize the energetics of
outflows and UV heating on small--scales. The outflows (CO), dust
continuum, and surrounding dense molecular gas (N$_2$H$^+$) are being
observed in the millimeter to determine the source inclinations,
outflow opening angles, inner envelope properties, and the kinematics
of the larger--scale dense gas.

\section{Conclusions}

We have discovered a sample of 55 new candidate protostars in Orion
with \herschel, as part of the HOPS Open Time Key Programme scan--map
observations at 70~\micron\ and 160~\micron.  We conclude that:

\noindent $\bullet$ The new candidate protostars are either very
faint or undetected at Spitzer wavelengths.  We find 34 sources with
24~\micron\ magnitudes that are greater than 7.0 and 21 sources that are
undetected in the MIPS 24~\micron\ band.

\noindent $\bullet$ We analyze the IRAC colors and the broad shape of
the SEDs between 3.6~\micron\ and 160~\micron. Based on this analysis,
we classify the sample as follows: 27\% (15 sources) are considered
reliable protostars, 18\% (10 sources) are considered lower
probability protostars, 47\% (26 sources) are classified as
extragalactic contamination, including AGN, 3 sources have IRAC colors
consistent with stellar photospheres but Herschel and APEX SEDs
consistent with cold dust emission, and 1 source does not have IRAC
coverage. We find that the subset of sources without
24~\micron\ detections dominates the number of sources categorized as
most reliable protostellar candidates by a factor of $\sim\,3$,
suggesting that sources with no short--wavelength detections and only
PACS 70~\micron\ and longer wavelength detections are much less likely
to be of extragalactic origin.

\noindent $\bullet$ We combine the new protostar candidate sample with
the previously identified \spitzer\ HOPS sample and find that 18
sources have $70/24$ colors greater than 1.65.  These are the reddest
protostars known in Orion, 11 of which are newly identified
\herschel\ sources in the reliable protostar category listed above.
We name these sources ``PACS Bright Red sources'', or PBRs. Compared
to the other protostars in the HOPS fields, the PBRs populate the
extrema in the distributions of standard evolutionary diagnostics,
having both the largest L$_{\rm smm}$/L$_{\rm bol}$ ratios and lowest
\tbol\ values.  The PBR source SEDs and peak SED wavelengths are
consistent with the hypothesis that the PBRs do indeed represent a
population composed of Class~0 sources with the densest envelopes in
Orion.

\noindent $\bullet$ A comparison to radiative transfer models of
rotating, collapsing protostellar envelopes with outflow cavities show
that the $70/24 > 1.65$ color limit selects sources with envelope
densities with log $\rho_1 / ({\rm g\,cm}^{-3}) \gtrsim -13.4$,
irrespective of inclination effects. While the $70/24 > 1.65$ color
selects sources with dense envelopes, this color criterion does not
find all dense sources above the threshold of of $\rho_1 / ({\rm
  g\,cm}^{-3}) = -13.4$. Therefore this selection should be used in
conjunction with other evolutionary indicators, namely millimeter and
sub--millimeter measurements, to determine the nature of the observed
sources.

\noindent $\bullet$ Our modeling of the PBRs SEDs reveals that these
sources are not consistent with being externally heated starless
cores; the presence of a 70~\micron\ point source requires that the
sample be interpreted as dense envelopes containing embedded
protostars.

\noindent $\bullet$ The fraction of known protostars that are PBRs
varies from 1\% in the Orion A cloud to 17\% in the Orion B cloud,
with an an average fraction over the Orion complex of 5\%. These
numbers suggest that if the PBRs represent a distinct phase in
protostellar evolutions, protostars spend on average 5\% of their
lifetime in the PBRs phase.  Most of the PBRs in the Orion B cloud are
concentrated in dense gas near the NGC2068, NGC2071, and NGC2023
nebulae.  These regions of dense gas are also known for a lack of more
evolved pre--main sequence stars. The high percentage of PBRs in Orion
B suggests either that the regions containing the PBRs may currently
be undergoing more vigorous star--formation than other regions of
Orion, or that the PBRs lifetime is longer in these regions.

\noindent $\bullet$ The sources with $70/24 < 1.65$ colors and
faint 70~\micron\ fluxes must be confirmed as either protostellar or
contamination sources before their significance can be assessed. If
confirmed as the former, however, this sample would constitute an
important population of very low--luminosity, cold protostars
previously unobserved, most interesting from the point of view of
constraining the faint end of the luminosity function of
protostars.

\noindent $\bullet$ We expect that a comparable number of very red
protostars will be found with \herschel\ in more near--by
star--forming regions.  We caution that a careful treatment of
possible extragalactic contamination must be implemented to understand
the broader significance of such sources.

\acknowledgments 

The authors gratefully acknowledge help from H\'{e}l\`{e}ne Roussel in
the production of {\it Scanamorphos} PACS maps.  Furthermore, we are
grateful to Oskari Miettinen for sharing the reduced SABOCA map of the
090003 region and to Joel Green for providing the SED VLA1623–-243.
AMS kindly acknowledges helpful and insightful discussions with Ralf
Launhardt.  The work of AMS was supported by the Deutsche
Forschungsgemeinschaft priority program 1573 ("Physics of the
Interstellar Medium").  JT acknowledges support provided by NASA
through Hubble Fellowship grant \#HST- HF-51300.01-A awarded by the
Space Telescope Science Institute, which is operated by the
Association of Universities for Research in Astronomy, Inc., for NASA,
under contract NAS 5-26555.  The National Radio Astronomy Observatory
is a facility of the National Science Foundation operated under
cooperative agreement by Associated Universities, Inc.  This
publication is based on data acquired with the Atacama Pathfinder
Experiment (APEX). APEX is a collaboration between the
Max-Planck-Institut f\"{u}r Radioastronomie, the European Southern
Observatory, and the Onsala Space Observatory.  The \herschel
spacecraft was designed, built, tested, and launched under a contract
to ESA managed by the Herschel/Planck Project team by an industrial
consortium under the overall responsibility of the prime contractor
Thales Alenia Space (Cannes), and including Astrium (Friedrichshafen)
responsible for the payload module and for system testing at
spacecraft level, Thales Alenia Space (Turin) responsible for the
service module, and Astrium (Toulouse) responsible for the telescope,
with in excess of a hundred subcontractors.  PACS has been developed
by a consortium of institutes led by MPE (Germany) and including UVIE
(Austria); KU Leuven, CSL, IMEC (Belgium); CEA, LAM (France); MPIA
(Germany); INAF-IFSI/OAA/OAP/OAT, LENS, SISSA (Italy); IAC
(Spain). This development has been supported by the funding agencies
BMVIT (Austria), ESA-PRODEX (Belgium), CEA/CNES (France), DLR
(Germany), ASI/INAF (Italy), and CICYT/MCYT (Spain).  HCSS / HSpot /
HIPE is a joint development (are joint developments) by the Herschel
Science Ground Segment Consortium, consisting of ESA, the NASA
Herschel Science Center, and the HIFI, PACS and SPIRE consortia.  We
also use the Spitzer Space Telescope and the Infrared Processing and
Analysis Center (IPAC) Infrared Science Archive, which are operated by
JPL/Caltech under a contract with NASA.  This research has made use of
the the SIMBAD database and VizieR catalogue access tool, operated at
CDS, Strasbourg, France.  Support for this work was provided by the
National Aeronautics and Space Administration (NASA) through awards
issued by the Jet Propulsion Laboratory, California Institute of
Technology (JPL/Caltech).

%%\clearpage

\appendix

\section{Previous PBRs detections}

We searched the SIMBAD and VizieR \citep{ochsenbein00} services for
previous identifications of the PBRs.  We restricted our search to a
radius of 20$\arcsec$ from the 70~\micron\ source coordinates.  The
results are summarized in Table~\ref{tab:obs}.  This list is likely
incomplete and is intended to provide a resource and rough
guide to some of the previous detections of these sources.

$-$ 061012, 119019, and 097002: None found. 

$-$ HOPS169: Known protostar and outflow bipolar outflow V380 Ori NE
\citep[e.g.,][]{davis00}.  \citet{stanke02} detected source 59 in
their 2.12~\micron\ catalog of Orion A, offset by 6$\arcsec$ from our
source coordinates. \citet{davis09} also detected an outflow about
2$\arcsec$ away from our coordinates. \citet{nutter07} classified this
source as hosting a young stellar object (YSO) and measured an
850~\micron--flux--derived mass of 2.8~\msun\ (assuming a temperature of
20~K and a distance of 400~pc).

$-$ 019003: Since this source is located in the very crowded and
complex ONC filament, it is often not clear which previous
identifications may be associated with it in particular.
\citet{tsujimoto03} listed a near--IR source located 3.35$\arcsec$
away from our coordinates.  \citet{nutter07} detected a source offset
by 8.4$\arcsec$; they classified it as hosting a YSO and measure a
mass of 6.2~\msun.  While it is not clear if the detection is
associated with the PBRs we consider it probable.  \citet{chini97} and
\citet{nielbock03} likely detected this source in their 1.3~mm maps
near FIR1a in OMC--2, although their beam sizes were too large
identify the source unambiguously.

$-$ 082005: This source was previously classified as a starless core
\citep[e.g.,][]{johnstone06,nutter07,mookerjea99}.  The
\citet{nutter07} location of the source is about 8$\arcsec$ away from
our source with an associated mass of 3.3~\msun. \citet{mookerjea99}
measured a mass of 4.3~\msun\ for this source.  

$-$ HOPS372 and 082012: These sources have coordinates that are
significantly offset from the \citet{nutter07} coordinates; the two
sources are unresolved and lie about 13$\arcsec$ away from their
catalog entree, with a mass of 12.7~\msun.  The \citet{mookerjea99}
analysis derived a mass of 1~\msun\ for 082012 (MM1) and
7.4~\msun\ for HOPS372 (MM2), from modified black--body fits to
long--wavelength SEDs.

$-$ 090003: \citet{miettinen09} detected this source (SMM3) in their
850~\micron\ map, and concluded that it is a promising Class~0
candidate based on the shape of the SED. They calculated a mass of
about 7.5~\msun\ for this source.  \citet{miettinen10} and
\citet{miettinen12} also observed this source in various molecular
line transitions and with SABOCA at 350~\micron; their measured flux
for this source is $S^{peak}_{350} = 3.63$~Jy~beam$^{-1}$.  We include
their flux measurement and the 350~\micron\ map in our analysis.

$-$ HOPS358: \citet{strom76}, \citet{strom86}, and
\citep{reipurth99} detected the Herbig--Haro complex HH24--26, with HH25
located about 10$\arcsec$ away from HOPS358.  This source is also
included in the \citet{wu04} high velocity outflow catalog.
\citet{nutter07} classified a nearby source (11$\arcsec$ away) as a
starless core with a mass of 6.3~\msun.

$-$ 091015 and 091016: These sources were detected by
\citet[e.g.,][]{lis99} at 1.3~mm and 350~\micron\ (as sources 5 and 6). 
The reported masses are $\sim\!2$~\msun\ for each source.  Both
sources are classified as starless by \citet{nutter07}, with catalog
masses of 0.8~\msun\ (091015) and 1.3~\msun\ (091016).

$-$ HOPS373 and 093005: Both sources were classified as starless by
\citet{nutter07}, with reported masses of 4.2~\msun\ (HOPS373) and
3.9~\msun\ (093005).  \citet{motte01} observed this region at 450 and
850~\micron\ and also classified 093005 as a starless core, based on the
lack of evidence for an embedded source.  Source HOPS373, on the other
hand, is known to be driving a CO outflow \citep{gibb00}, and therefore
has been classified as a candidate Class~0 source.  Furthermore,
\citet{haschick83} identified a water maser near the HOPS373 location.  

$-$ 302002: \citet{nutter07} reported a mass of 2.7~\msun\ and
classified this source as protostellar. \citet{phillips01} classify
this source (LBS18S) as pre--protostellar, however.

$-$ HOPS359: \citet{nutter07} classified this source as protostellar
and measure a mass of 2.7~\msun.  

$-$ HOPS341: This source is strongly blended with HOPS340.  We have listed
some detections of the combined system in Table~\ref{tab:obs}.

$-$ HOPS354: \citet{reipurth08} described this source in the context of
the L1622 cloud.  \citet{bally09} also detected this source in their
analysis of \spitzer\ IRAC images but did not analyze it in detail.  

\clearpage

\section{Gallery of images of PACS Bright Red sources}

\begin{figure*}[h!]
  \begin{center}
    \scalebox{0.72}{\includegraphics{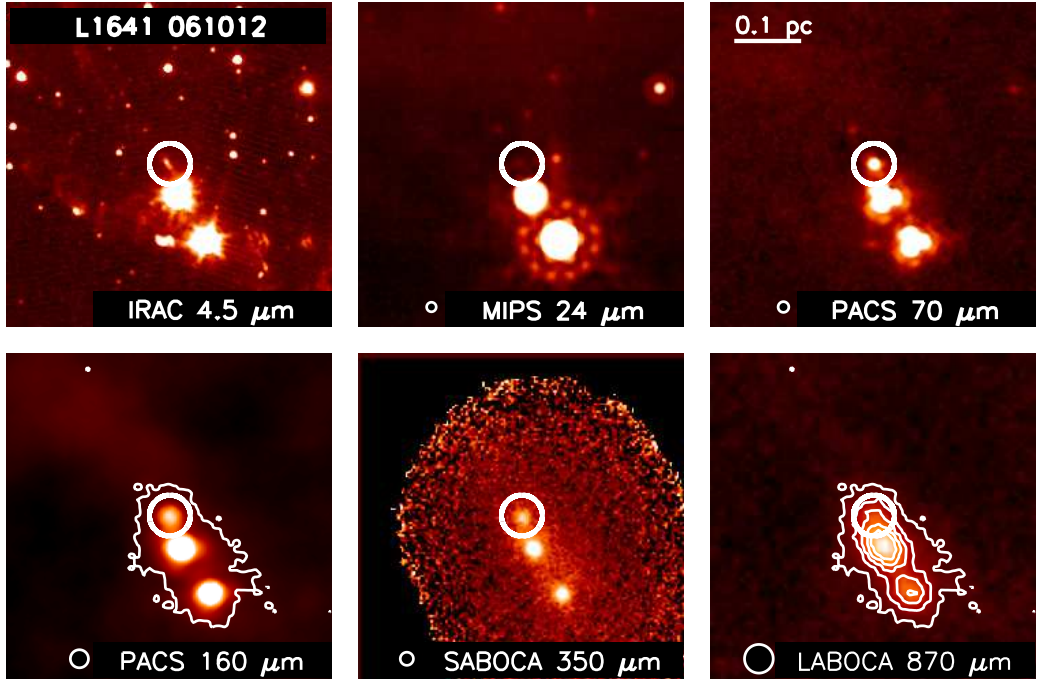}}
    \caption{Same as Figure~\ref{fig:img7}, showing $4\arcmin \times
      4\arcmin$ images of PBRs 061012. The IRAC--band emission
      associated with the source is clearly visible at 4.5~\micron.
      Contours indicate the 870~\micron\ emission levels at \{0.1,
      0.25, 0.4, 0.55, 0.7\}~Jy\,beam$^{-1}$; the long wavelength
      sum--mm data trace the cold envelope material associated with
      the source. The 160~\micron\ panel is shown with the lowest
      870~\micron\ emission contour. }
    \label{fig:img1}
  \end{center}
\end{figure*}    

\begin{figure*}
  \begin{center}
    \scalebox{0.72}{\includegraphics{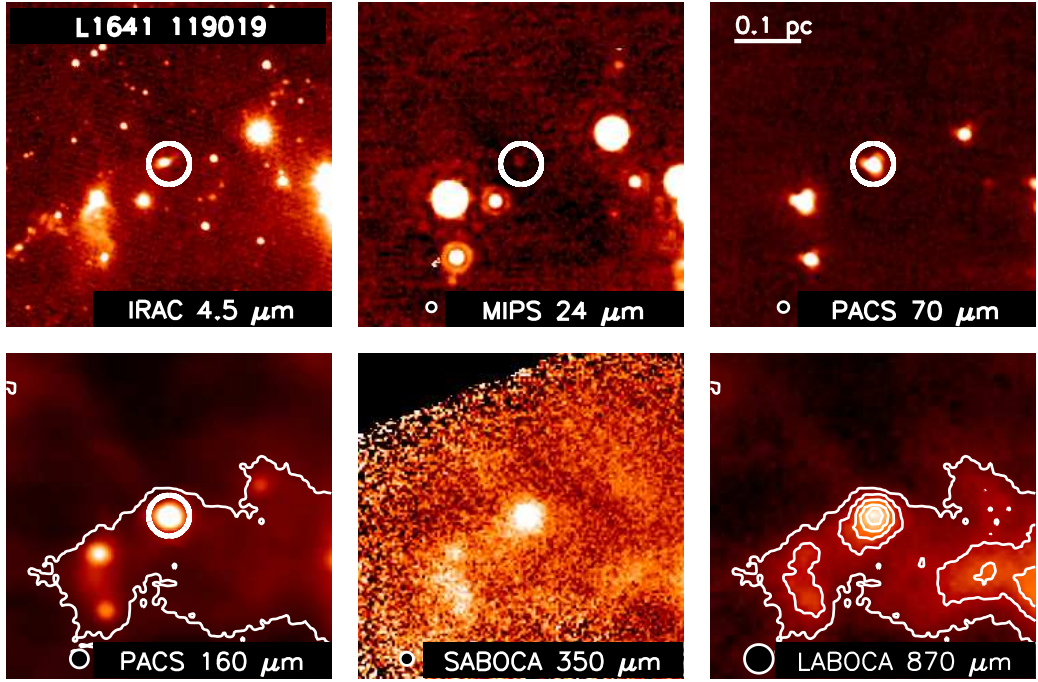}}
    \caption{Same as Figure~\ref{fig:img7}, showing $4\arcmin \times
      4\arcmin$ images and SED of PBRs 119019.  Contours indicate the
      870~\micron\ emission levels at \{0.1, 0.2, 0.3, 0.4, 0.5\}
      Jy\,beam$^{-1}$.  This source has prominent IRAC emission and
      is located in an IRAC 8~\micron\ and MIPS 24~\micron\ shadow.}
    \label{fig:img2}
  \end{center}
\end{figure*}

\begin{figure*}
  \begin{center}
    \scalebox{0.72}{\includegraphics{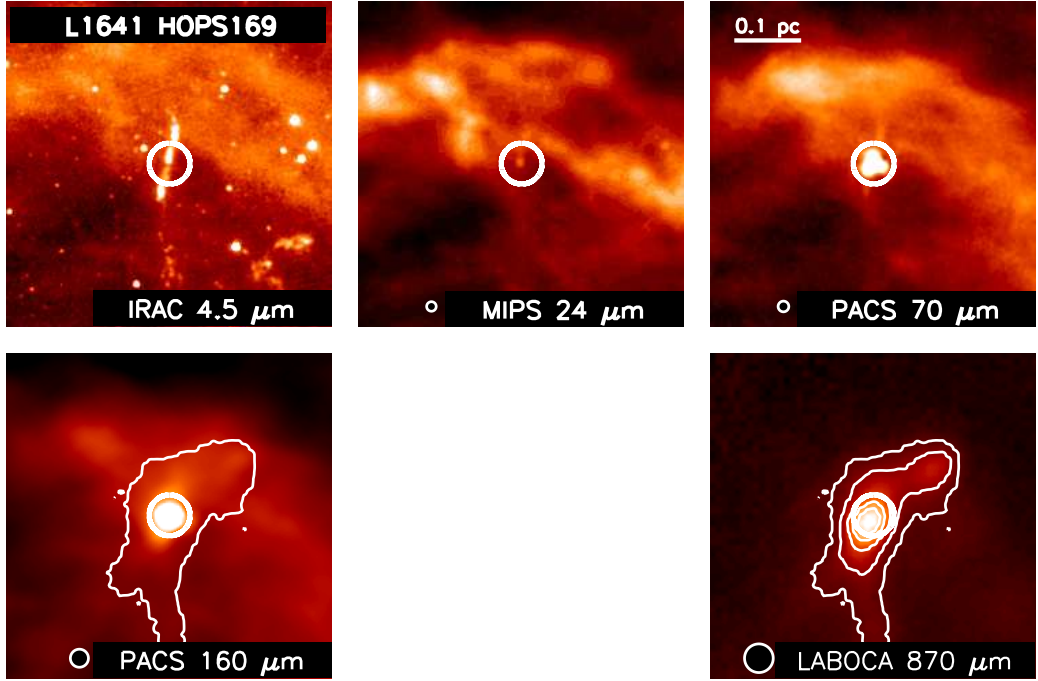}}
    \caption{Same as Figure~\ref{fig:img7}, showing $4\arcmin \times
      4\arcmin$ images and SED of HOPS169.  Contours indicate the
      870~\micron\ emission levels at \{0.1, 0.2, 0.4, 0.6, 0.8\}
      Jy\,beam$^{-1}$.  This source has clear outflow activity, traced
      by the IRAC emission, and appears to be at high inclination.}
    \label{fig:img3}
  \end{center}
\end{figure*}

\begin{figure*}
  \begin{center}
    \scalebox{0.72}{\includegraphics{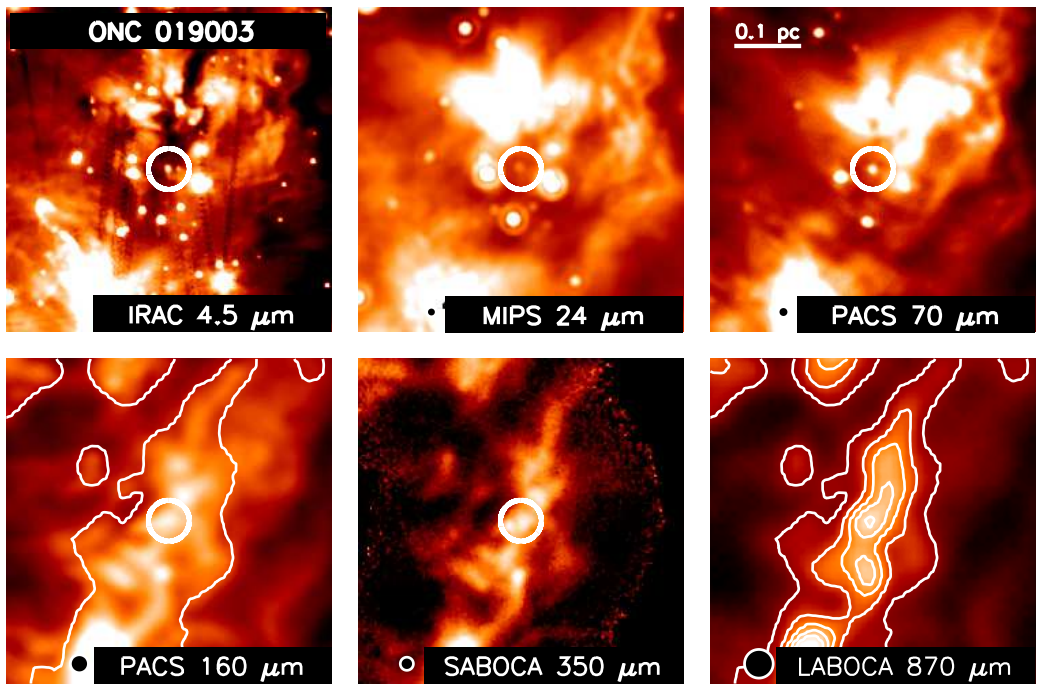}}
    \caption{Same as Figure~\ref{fig:img7}, showing $4\arcmin \times
      4\arcmin$ images and SED of PBR source 019003.  Contours
      indicate the 870~\micron\ emission levels at \{0.5, 1.0, 1.5,
      2.0, 2.5\} Jy\,beam$^{-1}$.  This source has indications outflow
      activity, as traced by the IRAC emission. The
      \herschel\ photometry may be strongly affected by blending due
      to the source location in a very dense filament.}
    \label{fig:img4}
  \end{center}
\end{figure*}

%%\clearpage

\begin{figure*}
  \begin{center}
    \scalebox{0.72}{\includegraphics{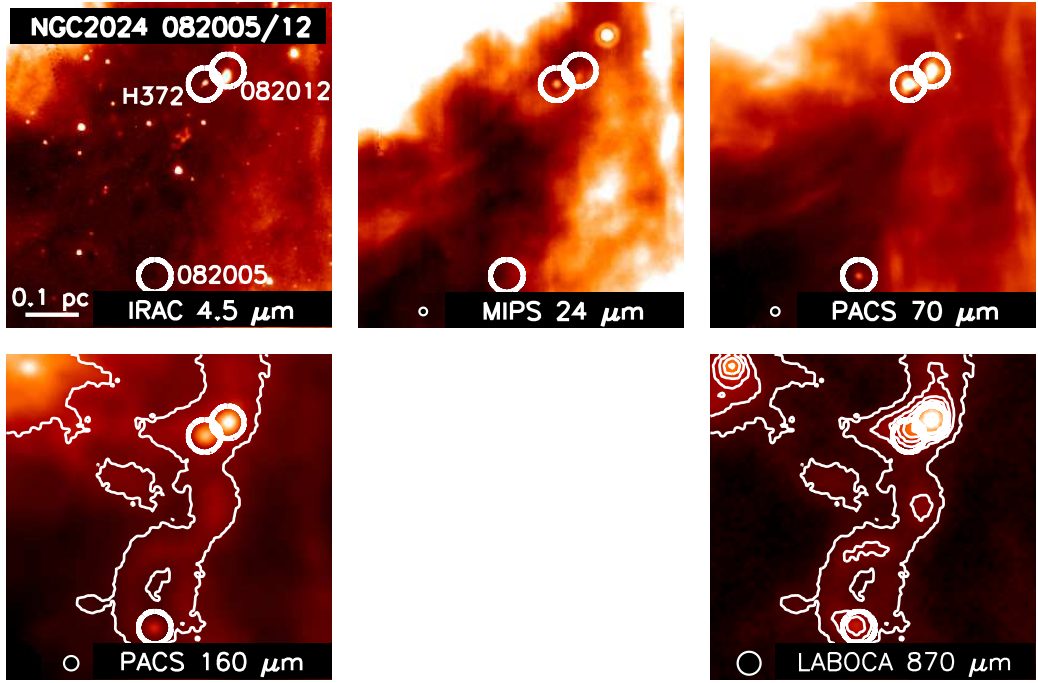}}
    \caption{Top: Same as Figure~\ref{fig:img7}, showing $5\arcmin
      \times 5\arcmin$ images of three red sources: HOPS372 and 082012
      (top) and 082005 (bottom).  Contours indicate the
      870~\micron\ emission levels at \{0.25, 0.5, 0.75, 1.0, 1.25,
      1.5\} Jy\,beam$^{-1}$.  No IRAC emission is detected for 082005;
      however, this source is located in dense filamentary material
      traced by the sub-mm emission and an 8~\micron\ absorption
      feature.}
    \label{fig:img5}
  \end{center}
\end{figure*}

\begin{figure*}
  \begin{center}
    \scalebox{0.72}{\includegraphics{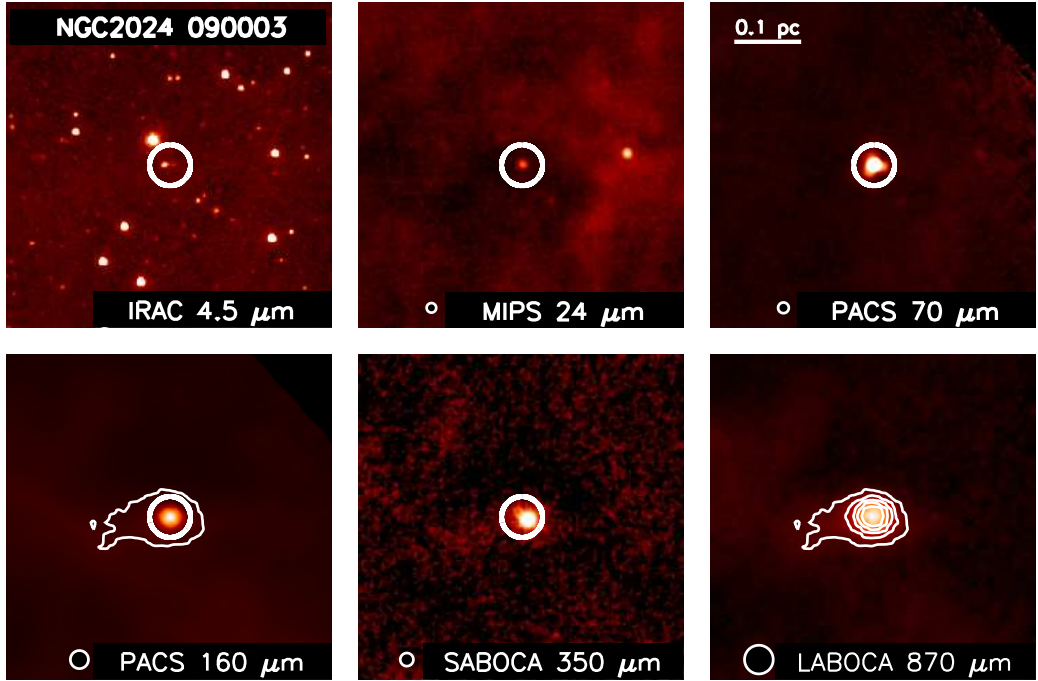}}
    \caption{Same as Figure~\ref{fig:img7}, showing $4\arcmin \times
      4\arcmin$ images and SED of PBRs 090003.  Contours indicate the
      870~\micron\ emission levels at \{0.25, 0.5, 0.75, 1.0, 1.25,
      1.5\} Jy\,beam$^{-1}$. The IRAC data show faint indications of
      extended emission, possibly associated with an outflow activity
      or a second source.  The 350~\micron\ image and SED point are from
      \citet{miettinen12}}
    \label{fig:img6}
  \end{center}
\end{figure*} 

%%\clearpage

\begin{figure*}
  \begin{center}
    \scalebox{0.72}{\includegraphics{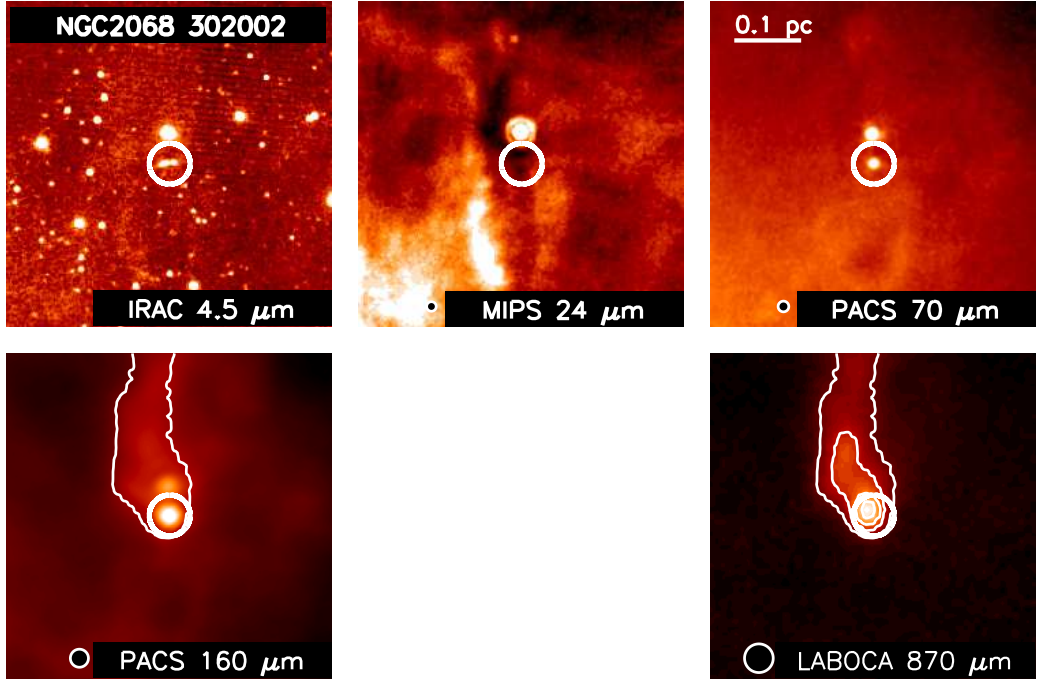}}
    \caption{Same as Figure~\ref{fig:img7}; $4\arcmin \times 4\arcmin$
      images of PBRs 302002.  Contours indicate the
      870~\micron\ emission levels at \{0.25, 0.5, 0.75, 1.0, 1.25,
      1.5\} Jy\,beam$^{-1}$, tracing the cometary globule shaped
      region. The IRAC data show indications of emission associated
      with outflow activity; furthermore, the 4.5~\micron\ data show
      evidence that this source is observed at high inclination.  This
      source is the second most massive source in our sample, with a
      best--fit M$_{env} = 1.7$~\msun.}
    \label{fig:img9}
  \end{center}
\end{figure*}

\begin{figure*}
  \begin{center}
    \scalebox{0.72}{\includegraphics{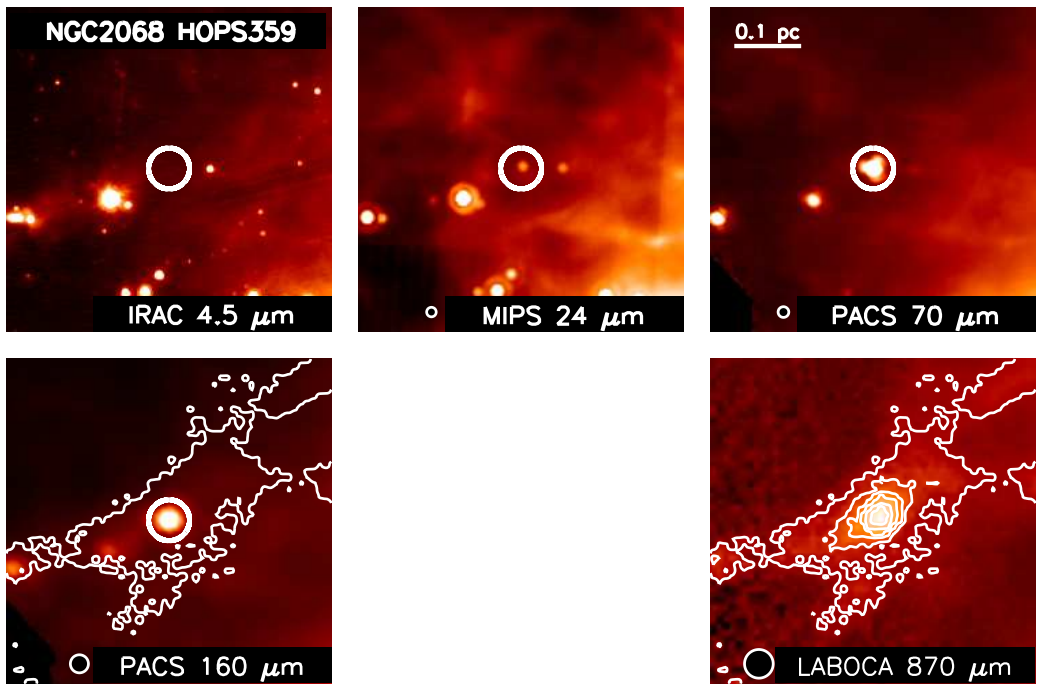}}
    \caption{Same as Figure~\ref{fig:img7}; $4\arcmin \times 4\arcmin$
      images of HOPS359.  Contours indicate the 870~\micron\ emission
      levels at \{0.25, 0.5, 0.75, 1.0, 1.25, 1.5\} Jy\,beam$^{-1}$. }
    \label{fig:img10}
  \end{center}
\end{figure*}

%%\clearpage

\begin{figure*}
  \begin{center}
    \scalebox{0.72}{\includegraphics{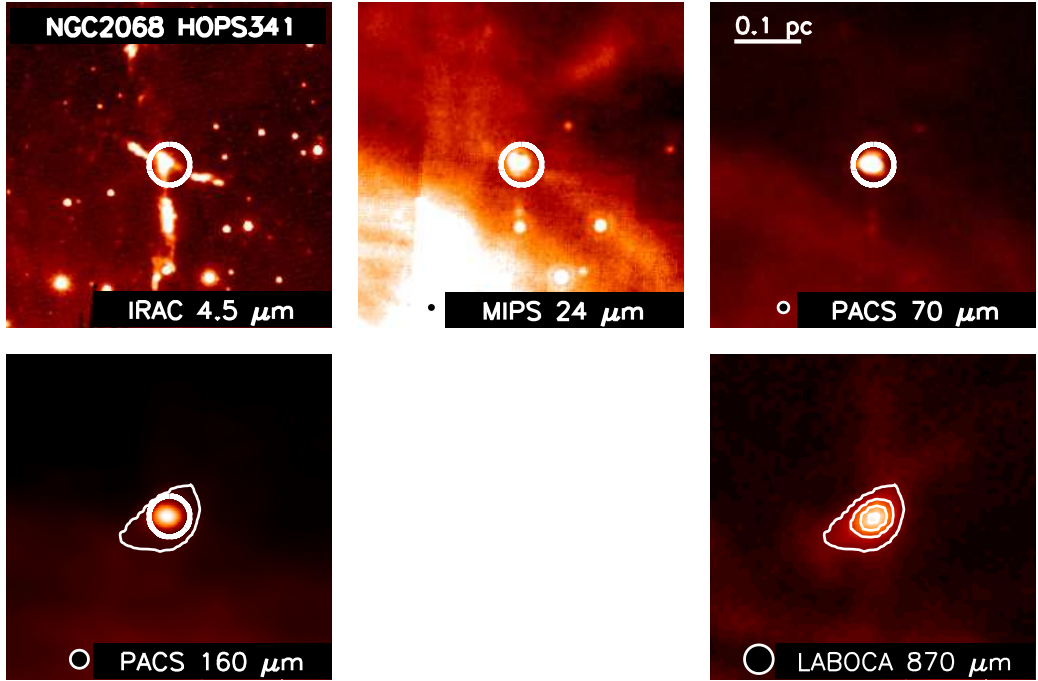}}
    \caption{Same as Figure~\ref{fig:img7}; $4\arcmin \times 4\arcmin$
      images of HOPS341.  Contours indicate the 870~\micron\ emission
      levels at \{0.25, 0.5, 0.75, 1.0\} Jy\,beam$^{-1}$. The
      $\times$--shaped morphology of this source in the IRAC bands
      indicated that tit is a binary.  Indeed, the photometry of this
      source is strongly blended with HOPS340; nevertheless we include
      this source in our sample for completeness.}
    \label{fig:img11}
  \end{center}
\end{figure*}

\begin{figure*}
  \begin{center}
    \scalebox{0.72}{\includegraphics{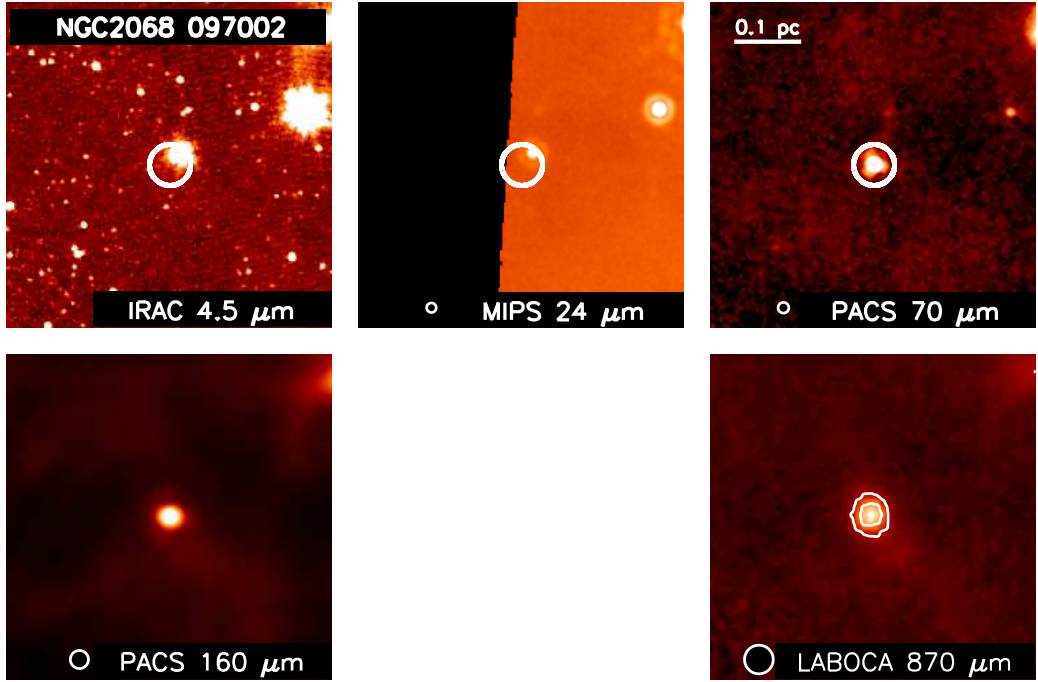}}
    \caption{Same as Figure~\ref{fig:img7}; $4\arcmin \times 4\arcmin$
      images of PBRs 097002.  Contours indicate the
      870~\micron\ emission levels at \{0.25, 0.5\} Jy\,beam$^{-1}$.}
    \label{fig:img12}
  \end{center}
\end{figure*}

%%\clearpage

\begin{figure*}
  \begin{center}
    \scalebox{0.72}{\includegraphics{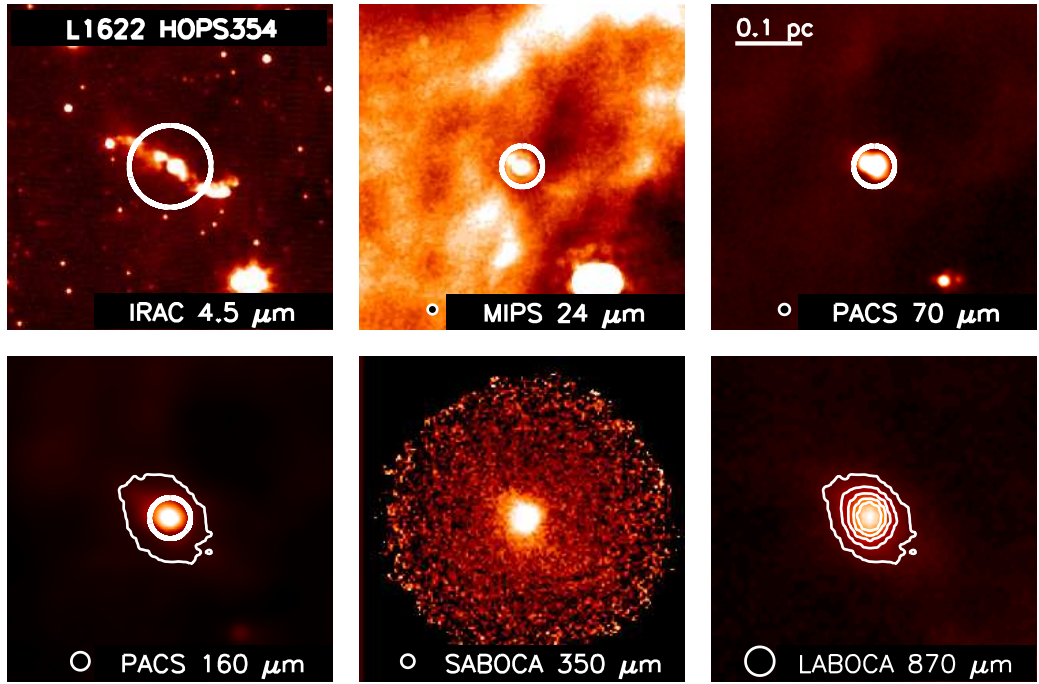}}
    \caption{Same as Figure~\ref{fig:img7}; $4\arcmin \times 4\arcmin$
      images of HOPS354.  Contours indicate the 870~\micron\ emission
      levels at \{0.25, 0.5, 0.75, 1.0, 1.25\} Jy\,beam$^{-1}$. The
      IRAC images display clear indications of outflow activity;
      furthermore the source appears highly inclined.}
    \label{fig:img13}
  \end{center}
\end{figure*}

%%\clearpage

\clearpage

\begin{deluxetable*}{llllllc}
\tablecaption{Summary of HOPS \herschel\ PACS L1641 observations}
\tablewidth{0pt}
\tablehead{
\colhead{HOPS} 
& \colhead{R.A.$^a$} 
& \colhead{Decl.$^a$} 
& \colhead{Field} 
& \colhead{AOR ID} 
& \colhead{OD} 
& \colhead{Map size} \\
\colhead{group name} 
& \colhead{h:m:s}
& \colhead{\degree:$\arcmin$;$\arcsec$} 
& \colhead{} 
& \colhead{} 
& \colhead{} 
& \colhead{$\arcmin\times\arcmin$}
}
\startdata
54 & 5:42:38.570 & $-$8:50:18.67 & L1641 & 1342218796(7) & 704 & 8$\times$8  \\
53 & 5:43:06.770 & $-$8:46:09.56 & L1641 & 1342218735(6) & 703 & 8$\times$8  \\
60 & 5:41:29.690 & $-$8:41:28.59 & L1641 & 1342215359(60) & 662 & 8$\times$8  \\
55 & 5:42:50.490 & $-$8:40:54.73 & L1641 & 1342218798(9) & 704 & 8$\times$8  \\
56 & 5:42:52.750 & $-$8:37:20.98 & L1641 & 1342205256(7) & 502 & 8$\times$8  \\
117 & 5:41:33.250 & $-$8:36:41.62 & L1641 & 1342218790(1) & 704 & 8$\times$8  \\
59 & 5:42:55.540 & $-$8:32:48.26 & L1641 & 1342218794(5) & 704 & 8$\times$8  \\
58 & 5:43:09.580 & $-$8:29:27.13 & L1641 & 1342218788(9) & 704 & 8$\times$8  \\
61 & 5:42:47.760 & $-$8:16:50.72 & L1641 & 1342205254(5) & 502 & 8$\times$8  \\
67 & 5:40:20.130 & $-$8:14:05.78 & L1641 & 1342227078(9) & 831 & 8$\times$8  \\
119 & 5:40:47.820 & $-$8:10:28.38 & L1641 & 1342206322(3) & 516 & 8$\times$8  \\
62 & 5:42:47.370 & $-$8:10:08.76 & L1641 & 1342218792(3) & 704 & 8$\times$8  \\
63 & 5:41:35.440 & $-$8:08:22.49 & L1641 & 1342218800(1) & 704 & 8$\times$8  \\
66 & 5:40:54.500 & $-$8:06:08.98 & L1641 & 1342215361(2) & 662 & 8$\times$8  \\
118 & 5:41:27.790 & $-$8:04:03.70 & L1641 & 1342205250(1) & 502 & 8$\times$8  \\
78 & 5:38:51.480 & $-$8:01:27.44 & L1641 & 1342228169(70) & 844 & 5$\times$5  \\
64 & 5:41:49.950 & $-$8:01:26.51 & L1641 & 1342205252(3) & 502 & 5$\times$5  \\
69 & 5:40:38.330 & $-$8:00:36.00 & L1641 & 1342227080(1) & 831 & 5$\times$5  \\
65 & 5:41:24.880 & $-$8:00:02.34 & L1641 & 1342215591(2) & 663 & 8$\times$8  \\
72 & 5:40:20.710 & $-$7:56:01.36 & L1641 & 1342218733(4) & 703 & 8$\times$8  \\
70 & 5:40:40.530 & $-$7:54:39.82 & L1641 & 1342228167(8) & 844 & 5$\times$5  \\
121 & 5:41:23.440 & $-$7:54:39.26 & L1641 & 1342205248(9) & 502 & 8$\times$8  \\
68 & 5:41:19.660 & $-$7:50:41.03 & L1641 & 1342227848(9) & 842 & 5$\times$5  \\
74 & 5:40:17.690 & $-$7:49:29.88 & L1641 & 1342218731(2) & 703 & 8$\times$8  \\
71 & 5:40:40.150 & $-$7:49:18.71 & L1641 & 1342228163(4) & 844 & 8$\times$8  \\
320 & 5:40:58.890 & $-$7:48:02.05 & L1641 & 1342228165(6) & 844 & 5$\times$5  \\
73 & 5:40:42.910 & $-$7:45:01.91 & L1641 & 1342228425(6) & 847 & 5$\times$5  \\
75 & 5:40:24.620 & $-$7:43:08.26 & L1641 & 1342227082(3) & 831 & 5$\times$5  \\
76 & 5:40:26.090 & $-$7:37:32.02 & L1641 & 1342205246(7) & 502 & 5$\times$5  \\
79 & 5:39:57.200 & $-$7:30:19.89 & L1641 & 1342205244(5) & 502 & 8$\times$8  \\
77 & 5:40:44.670 & $-$7:29:54.46 & L1641 & 1342228427(8) & 847 & 5$\times$5  \\
123 & 5:40:08.780 & $-$7:27:27.68 & L1641 & 1342228161(2) & 844 & 5$\times$5  \\
26 & 5:39:24.640 & $-$7:26:13.81 & L1641 & 1342218729(30) & 703 & 8$\times$8  \\
25 & 5:39:56.200 & $-$7:24:53.71 & L1641 & 1342215589(90) & 663 & 8$\times$8  \\
313 & 5:39:33.300 & $-$7:22:57.36 & L1641 & 1342227084(5) & 831 & 5$\times$5  \\
28 & 5:38:56.470 & $-$7:20:44.32 & L1641 & 1342227086(7) & 831 & 8$\times$8  \\
30 & 5:38:44.050 & $-$7:11:49.89 & L1641 & 1342204254(5) & 484 & 8$\times$8  \\
29 & 5:39:06.710 & $-$7:11:12.80 & L1641 & 1342204252(3) & 484 & 8$\times$8  \\
32 & 5:38:01.100 & $-$7:07:37.01 & L1641 & 1342227045(6) & 830 & 8$\times$8  \\
312 & 5:38:46.540 & $-$7:05:37.46 & L1641 & 1342205242(3) & 502 & 5$\times$5  \\
31 & 5:38:44.870 & $-$7:00:37.03 & L1641 & 1342204256(7) & 484 & 8$\times$8  \\
33 & 5:38:20.090 & $-$6:59:04.85 & L1641 & 1342228171(2) & 844 & 5$\times$5  \\
35 & 5:37:24.460 & $-$6:58:32.77 & L1641 & 1342227314(5) & 833 & 5$\times$5  \\
34 & 5:37:59.990 & $-$6:57:27.50 & L1641 & 1342205240(1) & 502 & 8$\times$8  \\
101 & 5:37:17.090 & $-$6:49:49.33 & L1641 & 1342227312(3) & 833 & 5$\times$5  \\
36 & 5:37:52.390 & $-$6:47:18.67 & L1641 & 1342227088(9) & 831 & 8$\times$8  \\
38 & 5:36:22.050 & $-$6:45:41.23 & L1641 & 1342205238(9) & 502 & 8$\times$8  \\
40 & 5:36:26.650 & $-$6:38:27.74 & L1641 & 1342227094(5) & 831 & 8$\times$8  \\
37 & 5:37:17.280 & $-$6:36:18.18 & L1641 & 1342227090(1) & 831 & 8$\times$8  \\
43 & 5:35:50.020 & $-$6:34:53.40 & L1641 & 1342227310(1) & 833 & 5$\times$5  \\
50 & 5:34:15.880 & $-$6:34:32.70 & L1641 & 1342217748(9) & 686 & 8$\times$8  \\
39 & 5:36:41.330 & $-$6:34:00.08 & L1641 & 1342227092(3) & 831 & 5$\times$5  \\
41 & 5:36:19.440 & $-$6:29:06.79 & L1641 & 1342227316(7) & 833 & 5$\times$5  \\
45 & 5:35:34.120 & $-$6:26:41.70 & L1641 & 1342215593(4) & 663 & 8$\times$8  \\
311 & 5:34:39.860 & $-$6:25:14.16 & L1641 & 1342203649(50) & 470 & 5$\times$5  \\
42 & 5:36:22.460 & $-$6:23:39.14 & L1641 & 1342205236(7) & 502 & 8$\times$8  \\
44 & 5:36:36.980 & $-$6:14:57.98 & L1641 & 1342204258(9) & 484 & 5$\times$5  \\
51 & 5:35:22.180 & $-$6:13:06.24 & L1641 & 1342227318(9) & 833 & 5$\times$5  \\
47 & 5:36:17.260 & $-$6:11:11.00 & L1641 & 1342227324(5) & 833 & 5$\times$5  \\
49 & 5:35:52.000 & $-$6:10:01.85 & L1641 & 1342227322(3) & 833 & 5$\times$5  \\
52 & 5:35:33.210 & $-$6:06:09.65 & L1641 & 1342227320(1) & 833 & 5$\times$5  \\
48 & 5:36:31.360 & $-$6:01:16.81 & L1641 & 1342217444(5) & 685 & 8$\times$8  \\
5 & 5:35:07.960 & $-$5:56:56.40 & L1641 & 1342204248(9) & 484 & 8$\times$8  \\
6 & 5:35:24.560 & $-$5:55:33.42 & L1641 & 1342227328(9) & 833 & 5$\times$5  \\
7 & 5:36:19.020 & $-$5:55:25.46 & L1641 & 1342227326(7) & 833 & 5$\times$5  \\
8 & 5:35:04.400 & $-$5:51:00.76 & L1641 & 1342217446(7) & 685 & 8$\times$8  \\
9 & 5:33:30.710 & $-$5:50:41.03 & L1641 & 1342217750(1) & 686 & 8$\times$8  \\
10 & 5:36:10.100 & $-$5:50:08.34 & L1641 & 1342227096(7) & 831 & 5$\times$5  \\
12 & 5:34:46.830 & $-$5:42:28.72 & L1641 & 1342204246(7) & 484 & 8$\times$8  \\
13 & 5:35:17.340 & $-$5:42:14.51 & L1641 & 1342227098(9) & 831 & 5$\times$5  \\
14 & 5:34:30.440 & $-$5:37:47.44 & L1641 & 1342204244(5) & 484 & 8$\times$8  
\enddata
\label{tab:all_aors_1}
\tablenotetext{a}{Field center coordinates.}  
\end{deluxetable*}

\begin{deluxetable*}{lcccccc}
\tablecaption{Summary of HOPS \herschel\ PACS ONC, NGC2024, NGC2068, and L1622 observations}
\tablewidth{0pt}
\tablehead{
\colhead{HOPS} 
& \colhead{R.A.$^a$} 
& \colhead{Decl.$^a$} 
& \colhead{Region} 
& \colhead{AOR ID} 
& \colhead{OD} 
& \colhead{Map size} \\
\colhead{group name} 
& \colhead{h:m:s}
& \colhead{\degree:$\arcmin$;$\arcsec$} 
& \colhead{} 
& \colhead{} 
& \colhead{} 
& \colhead{$\arcmin\times\arcmin$}
}
\startdata
15 & 5:35:06.620 & $-$5:35:05.68 & ONC & 1342205234(5) & 502 & 8$\times$8  \\
308 & 5:33:45.870 & $-$5:32:58.09 & ONC & 1342204433(4) & 487 & 5$\times$5  \\
16 & 5:34:43.990 & $-$5:32:11.21 & ONC & 1342217448(9) & 685 & 8$\times$8  \\
17 & 5:35:16.320 & $-$5:29:32.60 & ONC & 1342217450(1) & 685 & 8$\times$8  \\
18 & 5:33:55.730 & $-$5:22:39.97 & ONC & 1342217752(3) & 686 & 8$\times$8  \\
200 & 5:35:19.270 & $-$5:14:46.49 & ONC & 1342205232(3) & 502 & 8$\times$8  \\
130 & 5:35:24.710 & $-$5:09:06.02 & ONC & 1342205228(9) & 502 & 8$\times$8  \\
135 & 5:35:26.280 & $-$5:06:35.24 & ONC & 1342205226(7) & 502 & 8$\times$8  \\
19 & 5:35:23.300 & $-$5:00:35.73 & ONC & 1342204250(1) & 484 & 8$\times$8  \\
20 & 5:35:13.770 & $-$4:54:57.33 & ONC & 1342217758(9) & 686 & 8$\times$8  \\
21 & 5:34:32.340 & $-$4:53:54.26 & ONC & 1342217754(5) & 686 & 8$\times$8  \\
306 & 5:35:32.280 & $-$4:46:48.47 & ONC & 1342191970(1) & 300 & 5$\times$5  \\
24 & 5:35:23.340 & $-$4:40:10.45 & ONC & 1342217756(7) & 686 & 8$\times$8  \\
80 & 5:40:51.710 & $-$2:26:48.62 & NGC2024 & 1342226729(30) & 826 & 5$\times$5  \\
81 & 5:41:28.940 & $-$2:23:19.36 & NGC2024 & 1342226733(4) & 826 & 5$\times$5  \\
82 & 5:41:23.740 & $-$2:16:51.10 & NGC2024 & 1342228913(4) & 858 & 8$\times$8  \\
83 & 5:41:42.180 & $-$2:16:26.20 & NGC2024 & 1342227049(50) & 830 & 8$\times$8  \\
85 & 5:42:02.620 & $-$2:07:45.70 & NGC2024 & 1342226735(6) & 826 & 5$\times$5  \\
86 & 5:41:43.560 & $-$1:53:28.42 & NGC2024 & 1342227047(8) & 830 & 8$\times$8  \\
89 & 5:42:27.680 & $-$1:20:01.00 & NGC2024 & 1342205220(1) & 502 & 5$\times$5  \\
90 & 5:43:04.370 & $-$1:16:11.60 & NGC2024 & 1342228376(7) & 849 & 8$\times$8  \\
91 & 5:46:06.690 & $-$0:13:05.15 & NGC2068 & 1342205218(9) & 502 & 8$\times$8  \\
92 & 5:46:14.210 & $-$0:05:26.84 & NGC2068 & 1342205216(7) & 502 & 5$\times$5  \\
93 & 5:46:40.830 & $+$0:00:30.52 & NGC2068 & 1342215363(4) & 662 & 8$\times$8  \\
94 & 5:46:39.580 & $+$0:04:16.61 & NGC2068 & 1342228365(6) & 848 & 5$\times$5  \\
302 & 5:46:28.320 & $+$0:19:49.40 & NGC2068 & 1342228374(5) & 849 & 5$\times$5  \\
303 & 5:47:24.810 & $+$0:20:59.68 & NGC2068 & 1342227966(7) & 843 & 8$\times$8  \\
96 & 5:47:08.970 & $+$0:21:52.86 & NGC2068 & 1342215587(8) & 663 & 8$\times$8  \\
128 & 5:46:56.220 & $+$0:23:42.41 & NGC2068 & 1342218727(8) & 703 & 8$\times$8  \\
301 & 5:45:53.590 & $+$0:25:27.30 & NGC2068 & 1342216450(1) & 675 & 5$\times$5  \\
97 & 5:47:58.060 & $+$0:35:30.12 & NGC2068 & 1342227969(70) & 843 & 8$\times$8  \\
98 & 5:47:31.850 & $+$0:38:05.77 & NGC2068 & 1342227971(2) & 843 & 8$\times$8  \\
300 & 5:47:42.990 & $+$0:40:57.50 & NGC2068 & 1342205214(5) & 502 & 5$\times$5  \\
0 & 5:54:15.240 & $+$1:43:15.59 & L1622 & 1342215365(6) & 662 & 8$\times$8  \\
1 & 5:54:55.370 & $+$1:45:03.08 & L1622 & 1342218780(1) & 704 & 8$\times$8  \\
3 & 5:54:23.540 & $+$1:49:17.78 & L1622 & 1342218703(4) & 702 & 8$\times$8  \\
4 & 5:54:36.260 & $+$1:53:54.00 & L1622 & 1342218778(9) & 704 & 8$\times$8  \\
\enddata
\label{tab:all_aors_2}
\tablenotetext{a}{Field center coordinates.}  
\tablecomments{Note: the ONC field contains the extended Orion Nebula
  region, the NGC1977 region, and OMC2/3; the NGC2024 field contains
  the NGC2024 HII region and the NGC2023 reflection nebula.  The
  NGC2068 field includes the NGC2068 and NGC2071 reflection nebulae as
  well as LBS23 region.}
\end{deluxetable*}

\begin{deluxetable*}{lccccccccccc}
%%\rotate
\tablecaption{\herschel\ protostar candidate coordinates and photometry}
\tablewidth{0pt}
\tablehead{
\colhead{Source} 
& \colhead{group} 
& \colhead{R.A.$^a$} 
& \colhead{Decl.$^a$} 
& \colhead{Field} 
& \colhead{$24~\micron$} 
& \colhead{$70~\micron$} 
& \colhead{$160~\micron$} 
& \colhead{flag$^b$} 
& \colhead{870~$\micron$} 
& \colhead{T$_{\rm bol}$} 
& \colhead{L$_{\rm bol}$}\\
\colhead{} 
& \colhead{name} 
& \colhead{h:m:s} 
& \colhead{\degree:$\arcmin$;$\arcsec$} 
& \colhead{} 
& \colhead{[mJy]} 
& \colhead{[mJy]} 
& \colhead{[mJy]} 
& \colhead{} 
& \colhead{detection$^c$}
& \colhead{[K]} 
& \colhead{[L$_{\odot}$]} 
}
\startdata
{\bf 061012} & 061 & 05:42:48.87 & $-$08:16:10.70 & L1641 & $\leq$1.14 & 703$\pm$35 & 5634$\pm$845 & 1 & \nodata & 32.1$\pm$0.9 & 0.75$\pm$0.06  \\
{\bf 119019} & 119 & 05:40:58.47 & $-$08:05:36.10 & L1641 & 1.46$\pm$0.2 & 1604$\pm$80. & 10745$\pm$1612 & 1 & Yes & 34.4$\pm$0.8 & 1.56$\pm$0.14  \\
026011 & 026 & 05:39:17.00 & $-$07:24:26.64 & L1641 & $\leq$1.30 & 78.4$\pm$3.9 & 2516$\pm$377 & 1 & Yes & 23.0$\pm$1.0 & 0.34$\pm$0.03  \\
313006 & 313 & 05:39:30.75 & $-$07:23:59.40 & L1641 & $\leq$1.25 & 138$\pm$7. & 2198$\pm$330 & 1 & Yes & 27.8$\pm$1.0 & 0.28$\pm$0.03  \\
029003 & 029 & 05:39:13.15 & $-$07:13:11.69 & L1641 & 1.19$\pm$0.2 & 29.5$\pm$1.5 & 365$\pm$55 & 1 & \nodata & 42.0$\pm$1.5 & 0.06$\pm$0.01  \\
{\bf 019003} & 019 & 05:35:23.92 & $-$05:07:53.46 & ONC & $\leq$16.2 & 2506$\pm$127 & 31577$\pm$4737 & 1 & Yes & 33.6$\pm$1.1 & 3.16$\pm$0.31  \\
{\bf 082005} & 082 & 05:41:29.40 & $-$02:21:17.06 & NGC2024 & $\leq$3.13 & 506$\pm$25 & 9308$\pm$140 & 1 & Yes & 29.3$\pm$0.8 & 1.02$\pm$0.11  \\
{\bf 082012} & 082 & 05:41:24.94 & $-$02:18:08.54 & NGC2024 & 6.51$\pm$0.4 & 4571$\pm$229 & 51254$\pm$7688 & 1 & Yes & 32.2$\pm$0.9 & 6.27$\pm$0.65  \\
{\bf 090003} & 090 & 05:42:45.23 & $-$01:16:14.18 & NGC2024 & 4.74$\pm$0.3 & 3286$\pm$164 & 16937$\pm$2541 & 1 & Yes & 36.0$\pm$0.8 & 2.71$\pm$0.24  \\
{\bf 091015} & 091 & 05:46:07.65 & $-$00:12:20.73 & NGC2068 & $\leq$1.29 & 648$\pm$32 & 6615$\pm$992 & 1 & Yes & 30.9$\pm$0.8 & 0.81$\pm$0.07  \\
{\bf 091016} & 091 & 05:46:09.97 & $-$00:12:16.85 & NGC2068 & $\leq$1.14 & 431$\pm$22 & 5108$\pm$766 & 1 & Yes & 29.1$\pm$0.9 & 0.65$\pm$0.06  \\
{\bf 093005} & 093 & 05:46:27.75 & $-$00:00:53.81 & NGC2068 & $\leq$1.14 & 1427$\pm$71. & 12131$\pm$1820 & 1 & Yes & 30.8$\pm$0.9 & 1.71$\pm$0.15  \\
{\bf 302002} & 302 & 05:46:28.24 & 00:19:27.00 & NGC2068 & $\leq$1.14 & 302$\pm$15 & 7187$\pm$108 & 1 & Yes & 28.6$\pm$0.9 & 0.84$\pm$0.09  \\
{\bf 097002} & 097 & 05:48:07.76 & 00:33:50.79 & NGC2068 & $\leq$1.14 & 1049$\pm$52. & 7993$\pm$120 & 1 & Yes & 33.4$\pm$0.9 & 1.14$\pm$0.11  \\
300001 & 300 & 05:47:43.36 & 00:38:22.43 & NGC2068 & $\leq$7.19 & 478$\pm$24 & 5042$\pm$756 & 1 & Yes & 29.6$\pm$0.9 & 0.65$\pm$0.06  \\
\hline
068006 & 068 & 05:41:11.79 & $-$07:53:35.09 & L1641 & 9.19$\pm$0.2 & 44.8$\pm$2.2 & 146$\pm$22 & 2 & \nodata & 149.7$\pm$5.06 & 0.05$\pm$0.01  \\
038002 & 038 & 05:36:11.11 & $-$06:49:11.29 & L1641 & $\leq$1.25 & 36.4$\pm$1.8 & 537$\pm$81 & 2 & \nodata & 40.4$\pm$1.0 & 0.10$\pm$0.01  \\
037003 & 037 & 05:37:00.35 & $-$06:37:10.95 & L1641 & 8.85$\pm$0.2 & 725$\pm$36 & 3798$\pm$570 & 2 & \nodata & 53.1$\pm$1.7 & 0.43$\pm$0.03  \\
037008 & 037 & 05:37:34.31 & $-$06:35:20.33 & L1641 & $\leq$1.39 & 27.7$\pm$1.4 & 321$\pm$48 & 2 & \nodata & 45.5$\pm$1.9 & 0.04$\pm$0.01  \\
092011 & 092 & 05:46:26.17 & $-$00:04:45.31 & NGC2068 & 2.31$\pm$0.2 & 77.8$\pm$3.9 & 363$\pm$54 & 2 & \nodata & 104.1$\pm$5.63 & 0.05$\pm$0.01  \\
093001 & 093 & 05:46:56.32 & $-$00:03:14.73 & NGC2068 & $\leq$1.14 & 21.8$\pm$1.1 & 217$\pm$33 & 2 & Yes & 47.9$\pm$0.7 & 0.05$\pm$0.01  \\
302004 & 302 & 05:46:16.55 & 00:21:35.09 & NGC2068 & 2.87$\pm$0.2 & 108$\pm$5. & 283$\pm$42 & 2 & \nodata & 54.8$\pm$1.8 & 0.07$\pm$0.01  \\
096023 & 096 & 05:46:53.23 & 00:22:10.05 & NGC2068 & $\leq$2.13 & 147$\pm$7. & 1691$\pm$254 & 2 & \nodata & 39.6$\pm$1.4 & 0.18$\pm$0.02  \\
301003 & 301 & 05:46:02.15 & 00:23:29.86 & NGC2068 & 8.08$\pm$0.2 & 42.6$\pm$2.1 & 231$\pm$35 & 2 & \nodata & 92.1$\pm$3.8 & 0.05$\pm$0.01  \\
000011 & 000 & 05:54:32.10 & 01:42:54.92 & L1622 & $\leq$1.25 & 28.5$\pm$1.4 & 147$\pm$22 & 2 & \nodata & 50.1$\pm$0.5 & 0.06$\pm$0.01  \\
\hline
053002 & 053 & 05:43:24.07 & $-$08:49:03.75 & kOri & $\leq$1.38 & 33.5$\pm$1.7 & 190$\pm$28 & 3 & \nodata & 43.2$\pm$0.3 & 0.06$\pm$0.01  \\
117004 & 117 & 05:41:40.40 & $-$08:41:40.60 & L1641 & 1.17$\pm$0.2 & 25.3$\pm$1.3 & 234$\pm$35 & 3 & \nodata & 56.4$\pm$2.2 & 0.04$\pm$0.01  \\
117014 & 117 & 05:41:29.28 & $-$08:36:14.60 & L1641 & 1.20$\pm$0.2 & 31.0$\pm$1.6 & 237$\pm$36 & 3 & \nodata & 32.2$\pm$2.4 & 0.11$\pm$0.01  \\
119016 & 119 & 05:40:40.54 & $-$08:05:55.00 & L1641 & 1.97$\pm$0.2 & 42.7$\pm$2.1 & 148$\pm$22 & 3 & \nodata & 62.7$\pm$1.7 & 0.05$\pm$0.01  \\
121011 & 121 & 05:41:37.90 & $-$07:55:44.35 & L1641 & 1.46$\pm$0.2 & 74.1$\pm$3.7 & 193$\pm$29 & 3 & \nodata & 51.6$\pm$1.5 & 0.04$\pm$0.01  \\
025044 & 025 & 05:39:56.80 & $-$07:19:21.40 & L1641 & $\leq$1.18 & 23.1$\pm$1.2 & 150$\pm$22 & 3 & \nodata & 42.2$\pm$0.2 & 0.08$\pm$0.01  \\
030013 & 030 & 05:38:59.58 & $-$07:10:31.92 & L1641 & 1.43$\pm$0.2 & 25.9$\pm$1.3 & 153$\pm$23 & 3 & \nodata & 48.0$\pm$1.0 & 0.05$\pm$0.01  \\
031003 & 031 & 05:38:55.35 & $-$07:05:29.17 & L1641 & 2.96$\pm$0.2 & 49.9$\pm$2.5 & 175$\pm$26 & 3 & \nodata & 67.7$\pm$1.8 & 0.06$\pm$0.01  \\
034010 & 034 & 05:37:54.76 & $-$06:56:59.65 & L1641 & $\leq$1.14 & 61.3$\pm$3.1 & 249$\pm$37 & 3 & \nodata & 43.5$\pm$0.7 & 0.15$\pm$0.01  \\
031037 & 031 & 05:38:28.20 & $-$06:56:40.16 & L1641 & $\leq$1.14 & 64.3$\pm$3.2 & 396$\pm$59 & 3 & \nodata & 51.9$\pm$2.4 & 0.06$\pm$0.01  \\
034014 & 034 & 05:38:14.92 & $-$06:53:03.56 & L1641 & 1.89$\pm$0.2 & 47.4$\pm$2.4 & 235$\pm$35 & 3 & \nodata & 65.5$\pm$3.5 & 0.03$\pm$0.01  \\
036003 & 036 & 05:38:05.97 & $-$06:50:58.91 & L1641 & 1.74$\pm$0.2 & 67.4$\pm$3.4 & 291$\pm$44 & 3 & \nodata & 74.6$\pm$3.1 & 0.04$\pm$0.01  \\
036006 & 036 & 05:37:45.07 & $-$06:50:02.39 & L1641 & 2.83$\pm$0.2 & 34.8$\pm$1.7 & 143$\pm$22 & 3 & \nodata & 60.1$\pm$1.8 & 0.04$\pm$0.01  \\
036011 & 036 & 05:37:42.72 & $-$06:47:08.31 & L1641 & 2.68$\pm$0.2 & 41.8$\pm$2.1 & 159$\pm$24 & 3 & Yes & 65.9$\pm$1.9 & 0.05$\pm$0.01  \\
037011 & 037 & 05:37:37.80 & $-$06:34:43.35 & L1641 & 2.42$\pm$0.2 & 58.0$\pm$2.9 & 409$\pm$61 & 3 & \nodata & 69.6$\pm$3.0 & 0.04$\pm$0.01  \\
050006 & 050 & 05:34:23.01 & $-$06:32:58.00 & L1641 & 2.82$\pm$0.2 & 45.5$\pm$2.3 & 313$\pm$47 & 3 & \nodata & 60.3$\pm$2.3 & 0.06$\pm$0.01  \\
037013 & 037 & 05:37:22.00 & $-$06:32:56.48 & L1641 & 2.93$\pm$0.2 & 52.0$\pm$2.6 & 180$\pm$27 & 3 & \nodata & 62.7$\pm$1.8 & 0.05$\pm$0.01  \\
041001 & 041 & 05:36:28.68 & $-$06:30:42.13 & L1641 & 2.72$\pm$0.2 & 47.5$\pm$2.4 & 262$\pm$39 & 3 & \nodata & 90.7$\pm$4.8 & 0.03$\pm$0.01  \\
009001 & 009 & 05:33:32.55 & $-$05:53:34.25 & ONC & 4.83$\pm$0.3 & 74.9$\pm$3.8 & 245$\pm$37 & 3 & \nodata & 95.4$\pm$2.6 & 0.05$\pm$0.01  \\
021010 & 021 & 05:34:19.63 & $-$04:53:23.54 & ONC & 11.3$\pm$0.5 & 316$\pm$16 & 678$\pm$10 & 3 & \nodata & 63.8$\pm$2.2 & 0.18$\pm$0.01  \\
303017 & 303 & 05:47:49.01 & 00:20:26.47 & NGC2068 & 4.64$\pm$0.2 & 107$\pm$5. & 634$\pm$95 & 3 & \nodata & 70.2$\pm$3.8 & 0.10$\pm$0.01  \\
303023 & 303 & 05:47:45.58 & 00:21:14.68 & NGC2068 & 4.36$\pm$0.2 & 65.9$\pm$3.3 & 234$\pm$35 & 3 & \nodata & 104.1$\pm$4.81 & 0.04$\pm$0.01  \\
098001 & 098 & 05:47:25.77 & 00:33:37.43 & NGC2068 & 1.73$\pm$0.2 & 9.79$\pm$0.5 & 231$\pm$35 & 3 & \nodata & 43.5$\pm$2.1 & 0.03$\pm$0.01  \\
097003 & 097 & 05:47:45.88 & 00:34:12.71 & NGC2068 & 1.37$\pm$0.2 & 13.0$\pm$0.7 & 137$\pm$21 & 3 & \nodata & 78.1$\pm$4.4 & 0.02$\pm$0.01  \\
000003 & 000 & 05:54:17.28 & 01:40:18.68 & L1622 & 8.63$\pm$0.4 & 165$\pm$8. & 386$\pm$58 & 3 & \nodata & 82.5$\pm$3.3 & 0.10$\pm$0.01  \\
000010 & 000 & 05:54:37.27 & 01:42:52.39 & L1622 & 5.10$\pm$0.3 & 64.7$\pm$3.2 & 215$\pm$32 & 3 & \nodata & 89.2$\pm$3.9 & 0.04$\pm$0.01  \\
\hline
069001 & 069 & 05:40:46.20 & $-$08:04:35.12 & L1641 & 5.13$\pm$0.2 & 325$\pm$16 & 722$\pm$11 & 4 & \nodata & 336.1$\pm$14.7 & 0.29$\pm$0.01  \\
026001 & 026 & 05:39:18.49 & $-$07:27:52.37 & L1641 & 4.71$\pm$0.2 & 223$\pm$11 & 598$\pm$90 & 4 & \nodata & 224.4$\pm$11.2 & 0.17$\pm$0.01  \\
306004 & 306 & 05:35:24.66 & $-$04:49:43.53 & ONC & $\leq$124 & 435$\pm$22 & 637$\pm$96 & 4 & \nodata & 119.1$\pm$3.56 & 0.33$\pm$0.01  \\
\hline
006006 & 006 & 05:35:11.47 & $-$05:57:05.09 & L1641 & 5.06$\pm$0.2 & 126$\pm$6. & 4027$\pm$604 & 5 & Yes & 27.7$\pm$0.6 & 0.31$\pm$0.04
\enddata
\label{tab:pbrscat}
\tablenotetext{a}{PACS 70~\micron\ source coordinates.}
\tablenotetext{b}{This column indicates if a source is flagged as
  a reliable protostar (value $=1$), if the source is
  considered less likely be a protostar (value $=2$), if the source is
  flagged as extragalactic contamination (value $=3$), if the SED shape
  is with IRAC photospheric emission (value $=4$), and
  finally, one source has no IRAC coverage (value $= 5$).}
\tablenotetext{c}{This column indicates if the source has a strong
  870~\micron\ detection; sources with no data are either are assigned
  upper limits or have no coverage}
\tablecomments{Sources indicated in bold are those with
    $log (\lambda F_\lambda70 / \lambda F_\lambda24) > 1.65$.}
\tablecomments{Note: the ONC field contains the extended Orion Nebula
  region, the NGC1977 region, and OMC2/3; the NGC2024 field contains
  the NGC2024 HII region and the NGC2023 reflection nebula.  The
  NGC2068 field includes the NGC2068 and NGC2071 reflection nebulae as
  well as LBS23 region.}
\end{deluxetable*}

\begin{deluxetable*}{lccccccccc}
%%\rotate
\tablecaption{\spitzer\ and \herschel\ photometric properties of PACS bright red sources}
\tablewidth{0pt}
\tablehead{
\colhead{Source} 
& \colhead{HOPS} 
& \colhead{R.A.$^{a}$} 
& \colhead{Decl.$^{a}$} 
& \colhead{Field} 
& \colhead{$24~\micron$} 
& \colhead{$70~\micron$} 
& \colhead{$100~\micron$} 
& \colhead{$160~\micron$} 
& \colhead{log $70/24$$^b$}\\
\colhead{} 
& \colhead{group name} 
& \colhead{h:m:s} 
& \colhead{\degree:$\arcmin$:$\arcsec$} 
& \colhead{} 
& \colhead{[mJy]} 
& \colhead{[mJy]} 
& \colhead{[mJy]} 
& \colhead{[mJy]} 
& \colhead{}
}
\startdata
061012  &  061  &  05:42:48.8  &  $-$08:16:10.70  &  L1641  & $\leq$1.14 & 703$\pm$35 & 2018$\pm$517 & 5634$\pm$845 & $\geq$2.31  \\
119019  &  119  &  05:40:58.4  &  $-$08:05:36.10  &  L1641  & 1.46$\pm$0.2 & 1604$\pm$80. & 5789$\pm$148 & 10745$\pm$1612 & 2.56  \\
HOPS169  &  040  &  05:36:36.0  &  $-$06:38:54.02  &  L1641  & 4.80$\pm$0.5 & 5001$\pm$250 & 15753$\pm$4033 & 29975$\pm$4496 & 2.54  \\
019003  &  019  &  05:35:23.9  &  $-$05:07:53.46  &  ONC  & $\leq$16.2 & 2506$\pm$127 & 4711$\pm$121 & 31577$\pm$4737 & $\geq$1.71  \\
082005  &  082  &  05:41:29.4  &  $-$02:21:17.06  &  NGC2024  & $\leq$3.13 & 506$\pm$25 & 3003$\pm$769 & 9308$\pm$140 & $\geq$1.73  \\
HOPS372  &  082  &  05:41:26.3  &  $-$02:18:21.08  &  NGC2024  & 12.0$\pm$1.2 & 6178$\pm$309 & 16217$\pm$4151 & 31090$\pm$4664 & 2.24  \\
082012  &  082  &  05:41:24.9  &  $-$02:18:08.54  &  NGC2024  & 6.51$\pm$0.4 & 4571$\pm$229 & 20357$\pm$5211 & 51254$\pm$7688 & 2.37  \\
090003  &  090  &  05:42:45.2  &  $-$01:16:14.18  &  NGC2024  & 4.74$\pm$0.3 & 3286$\pm$164 & 10914$\pm$2794 & 16937$\pm$2541 & 2.36  \\
HOPS358  &  091  &  05:46:07.2  &  $-$00:13:30.86  &  NGC2068  & 422$\pm$42 & 60681$\pm$3041 & 104322$\pm$26706 & 123207$\pm$18481 & 1.68\\
091015  &  091  &  05:46:07.6  &  $-$00:12:20.73  &  NGC2068  & $\leq$1.29 & 648$\pm$32 & 2543$\pm$651 & 6615$\pm$992 & $\geq$2.22  \\
091016  &  091  &  05:46:09.9  &  $-$00:12:16.85  &  NGC2068  & $\leq$1.14 & 431$\pm$22 & 1977$\pm$506 & 5108$\pm$766 & $\geq$2.10  \\
HOPS373  &  093  &  05:46:30.7  &  $-$00:02:36.80  &  NGC2068  & 14.1$\pm$1.4 & 5258$\pm$263 & 20188$\pm$5168 & 36724$\pm$5509 & 2.10  \\
093005  &  093  &  05:46:27.7  &  $-$00:00:53.81  &  NGC2068  & $\leq$1.14 & 1427$\pm$71. & 5373$\pm$138 & 12131$\pm$1820 & $\geq$2.62  \\
302002  &  302  &  05:46:28.2  &  00:19:27.00  &  NGC2068  & $\leq$1.14 & 302$\pm$15 & 3101$\pm$794 & 7187$\pm$108 & $\geq$1.95  \\
HOPS359  &  303  &  05:47:24.8  &  00:20:58.24  &  NGC2068  & 22.8$\pm$2.3 & 20758$\pm$1039 & 43592$\pm$1116 & 60409$\pm$9061 & 2.48  \\
HOPS341  &  128  &  05:47:00.9  &  00:26:20.76  &  NGC2068  & 14.0$\pm$1.4 & 3001$\pm$301 & 15138$\pm$3875 & 25213$\pm$3782 & 1.86  \\
097002  &  097  &  05:48:07.7  &  00:33:50.79  &  NGC2068  & $\leq$1.14 & 1049$\pm$52. & 4163$\pm$107 & 7993$\pm$120 & $\geq$2.49  \\
HOPS354  &  000  &  05:54:24.1  &  01:44:20.15  &  L1622  & 28.9$\pm$2.9 & 8492$\pm$426 & 37423$\pm$9580 & 39258$\pm$5889 & 1.99  
\enddata
\label{tab:phot}
\tablenotetext{a}{Object coordinates are derived from the PACS
  70~\micron\ images.} 
\tablenotetext{b}{log $\lambda F_\lambda70 / \lambda F_\lambda24$}  
\tablecomments{Note: the ONC field contains the extended Orion Nebula
  region, the NGC1977 region, and OMC2/3; the NGC2024 field contains
  the NGC2024 HII region and the NGC2023 reflection nebula.  The
  NGC2068 field includes the NGC2068 and NGC2071 reflection nebulae as
  well as LBS23 region.}
\end{deluxetable*}

\begin{deluxetable*}{lcccccc}
\tablecaption{APEX 350 and 870~\micron\ photometry of PACS bright red sources}
\tablewidth{0pt}
\tablehead{
\colhead{Source} 
& \colhead{$350~\micron$} 
& \colhead{$350~\micron$$^{a}$} 
& \colhead{$350~\micron$$^{b}$} 
& \colhead{$870~\micron$} 
& \colhead{$870~\micron$$^{a}$} 
& \colhead{$870~\micron$$^{b}$} \\
\colhead{}
& \colhead{[Jy beam$^{-1}$]} 
& \colhead{[Jy]} 
& \colhead{[Jy]} 
& \colhead{[Jy beam$^{-1}$]} 
& \colhead{[Jy]}
& \colhead{[Jy]} 
}
\startdata
061012 & 2.60 & 3.93 & 2.53 & $\leq$0.7 & $\leq$1.3 & $\leq$1.3  \\
119019 & 3.38 & 5.27 & 3.63 & 0.5 & 0.9 & 0.6  \\
HOPS169 & \nodata & \nodata & \nodata & 1.0 & 1.6 & 1.4  \\
019003 & 8.29 & 19.6 & 7.18 & 2.5 & 5.0 & 3.0  \\
082005 & \nodata & \nodata & \nodata & 0.8 & 1.6 & 1.0  \\
HOPS372 & \nodata & \nodata & \nodata & $\leq$1.4 & $\leq$4.0 & $\leq$4.0  \\
082012 & \nodata & \nodata & \nodata & 2.7 & 4.3 & 3.6  \\
090003 & 3.63$^{c}$ & \nodata & \nodata & 1.7 & 2.2 & 1.9  \\
HOPS358 & 13.0 & 22.4 & 17.3 & 1.8 & 3.3 & 2.6  \\
091015 & 2.14 & 3.15 & 2.26 & 0.6 & 1.3 & 0.5  \\
091016 & 2.50 & 3.76 & 2.81 & 0.7 & 1.2 & 0.8  \\
HOPS373 & 9.02 & 12.9 & 10.5 & 1.5 & 2.5 & 2.0  \\
093005 & 6.58 & 10.1 & 6.74 & 1.4 & 2.6 & 1.6  \\
302002 & \nodata & \nodata & \nodata & 1.1 & 1.7 & 1.5  \\
HOPS359 & \nodata & \nodata & \nodata & 1.8 & 2.9 & 2.2  \\
HOPS341 & \nodata & \nodata & \nodata & 1.0 & 1.5 & 1.3  \\
097002 & \nodata & \nodata & \nodata & 0.6 & 0.9 & 0.8  \\
HOPS354 & 13.8 & 25.9 & 18.7 & 1.5 & 2.7 & 2.2
\enddata
\label{tab:photapex}
\tablenotetext{a}{Source flux measured in an aperture with
  radius equal to 1$\,\times\,$FWHM, where the FWHM$\,=
  7.34\arcsec$ and $19.0\arcsec$ at 350 and 870~\micron,
  respectively.}
\tablenotetext{b}{Source flux measured in the same aperture as $(a)$
  but with with local sky subtraction over radii equal to \{1.5,
  2.0\}$\times$FWHM.}
\tablenotetext{c}{350~\micron\ point from \citet{miettinen12}.}
\end{deluxetable*}

\begin{deluxetable*}{lccccccl}
%%\rotate
\tablecaption{Orion PBRs observed properties: detections, environment, and previous detections}
\tablewidth{0pt}
\tablehead{
\colhead{Source} 
& \colhead{R.A.} 
& \colhead{Decl.} 
& \colhead{4.5~\micron} 
& \colhead{High} 
& \colhead{24~\micron } 
& \colhead{Note$^{b}$}  
& \colhead{Selected }  \\ 
\colhead{} 
& \colhead{h:m:s} 
& \colhead{\degree:$\arcmin$:$\arcsec$} 
& \colhead{detection} 
& \colhead{incl.$^{a}$} 
& \colhead{detection} 
& \colhead{}
& \colhead{references}
}
\startdata
061012 & 05:42:48.8 & $-$08:16:10.7  & yes & \nodata & no  & ns    & \nodata \\
119019 & 05:40:58.4 & $-$08:05:36.1  & yes & \nodata & yes & f,i   & \nodata \\
HOPS169 & 05:36:36.0 & $-$06:38:54.0 & yes & yes     & yes & f,i   & Le88, Mo91, Za97, Da00, St02, Jo06, Nu07, Ba09, Me12\\
019003 & 05:35:24.0 & $-$05:07:50.1  & yes & \nodata & no  & f,c   & Me90, Ch97, Ni03, Ts03, Nu07, Me12\\
082005 & 05:41:29.4 & $-$02:21:17.1  & no  & \nodata & no  & f,i   & La91, La96, Mo99, Jo06, Nu07\\
HOPS372 & 05:41:26.3 & $-$02:18:21.1 & yes & \nodata & yes & f,ns & La91, La96, Mo99, Jo06, Nu07, Me12\\ 
082012 & 05:41:24.9 & $-$02:18:08.5  & no  & \nodata & yes & f,ns & La91, La96, Mo99, Wu04, Jo06, Nu07\\
090003 & 05:42:45.2 & $-$01:16:14.2  & yes & \nodata & yes & i     & Mi09 \\
HOPS358 & 05:46:07.2 & $-$00:13:30.9 & yes & \nodata & yes & f,c   & St86, Li99, Re99, Mi01, Wu04, Nu07, Me12\\
091015 & 05:46:07.7 & $-$00:12:19.1  & no  & \nodata & no  & f,ns & Li99, Mi01, Nu07 \\
091016 & 05:46:10.0 & $-$00:12:15.4  & no  & \nodata & no  & f,ns & Li99, Mi01, Nu07 \\
HOPS373 & 05:46:30.7 & $-$00:02:36.8 & yes & \nodata & yes & f,i   & Ha83, La91, Gi00, Mo01, Mi01, Nu07, Me12\\
093005 & 05:46:27.7 & $-$00:00:51.5  & yes & \nodata & no  & f,i   & La91, Mo01, Mi01, Nu07\\
302002 & 05:46:28.2 & $+$00:19:28.4  & yes & yes     & no  & f,ns   & La91, Jo01, Ph01, Mo01, Sa03, Nu07\\
HOPS359 & 05:47:24.8 & $+$00:20:58.2 & no  & \nodata & yes & f,i   & La91, Mo01, Nu07, Me12\\
HOPS341 & 05:47:00.9 & $+$00:26:20.8 & yes & yes     & yes & b     & Jo01, Mo01, Nu07, Sa10, Me12 \\
097002 & 05:48:07.7 & $+$00:33:52.7  & no  & \nodata & no  & b/ns     & \nodata\\
HOPS354 & 05:54:24.2 & $+$01:44:20.1 & yes & yes     & yes & b     & Re08, Ba09, Me12
\enddata
\label{tab:obs}
\tablenotetext{a}{Sources with indications of an observed high
  inclination orientation from the 4.5~\micron\ images.}
\tablenotetext{b}{ns = nearby source and indicates that we find a
  source close ($\lesssim 30\arcsec$) to the target; f = filament and
  indicates that a significant level of elongation or extended
  emission is seen in the 870~\micron\ data; i = isolated and
  indicates that no near--by source is observed; c = crowed,
  indicating that the source resides in region possibly with multiple
  sources, extended emission, or both and whose photometry may be
  affected by source blending; finally, b = binary, are sources that
  may have an unresolved secondary source and therefore blended
  photometry.}
\tablecomments{Le88: \citet{levreault88}; Mo91: \citet{morgan91};
  ZA97: \citet{zavagno97}; Da00: \citet{davis00}; St02:
  \citet{stanke02}; Da02: \citet{davis09}; Nu07: \citet{nutter07} and
  \citep{difra08}; Jo06: \citet{johnstone06}; Ts03:
  \citet{tsujimoto03}; Ch97: \citet{chini97}; Ni03:
  \citet{nielbock03}; Me90: \citet{mezger90}; Mo99:
  \citet{mookerjea99}; La96: \citet{laun96b}; La91: \citet{lada91};
  Wu04: \citet{wu04}; Mi09: \citet{miettinen09}; Re99
  \citet{reipurth99}; St86: \citet{strom86}; Li99: \citet{lis99};
  Mi01: \citet{mitchell01}; Mo01: \citet{motte01}; Gi00:
  \citet{gibb00}; Ha83: \citet{haschick83}; Ph01: \citet{phillips01};
  Sa03: \citet{savva03}; Jo01: \citet{johnstone01}; Re08:
  \citet{reipurth08}; Ba09: \citet{bally09}; Sa10: \citet{sadavoy10}; 
  Me12: \citet{megeath12}}
\end{deluxetable*}

\begin{deluxetable}{llccccc}
\tablecaption{Numbers and fractions of new sources and red PBRs found by region}
\tablewidth{0pt}
\tablehead{
\colhead{Region} 
& \colhead{Field} 
& \colhead{Total} 
& \colhead{New} 
& \colhead{New} 
& \colhead{PBRs} 
& \colhead{PBRs} \\
\colhead{} 
& \colhead{} 
& \colhead{number} 
& \colhead{number} 
& \colhead{$\%$} 
& \colhead{number} 
& \colhead{$\%$} 
}
\startdata
All     &     &    355     &   56   &    16$\%$    &    18   &     5$\%$ \\
 & & & & & \\
Orion A  &     &    274     &   35   &    13$\%$    &     4   &     1$\%$ \\
 &   L1641$^a$&    200     &   31   &    16$\%$    &     3   &     2$\%$ \\
 &    ONC     &     74     &    4   &     5$\%$    &     1   &     1$\%$ \\
& & & & & \\
Orion B  &     &     81     &   21   &    26$\%$    &    14   &    17$\%$ \\
 &    NGC2024 &     17     &    3   &    18$\%$    &     4   &    24$\%$ \\
 &    NGC2068 &     52     &   15   &    29$\%$    &     9   &    17$\%$ \\
 &    L1622   &     12     &    3   &    25$\%$    &     1   &     8$\%$ 
\enddata
\label{tab:frac}
\tablenotetext{a}{$\kappa$Ori sources have been included here.} 
\end{deluxetable}

\begin{deluxetable*}{lccccccc}
\tablecaption{Orion PBRs and comparison Class~0 sources: observed properties and modified black--body fit parameters}
\tablewidth{0pt}
\tablehead{
\colhead{Source} 
& \colhead{T$_{\rm bol}$} 
& \colhead{L$_{\rm bol}$} 
& \colhead{L$_{\rm smm}$/L$_{\rm bol}^{a}$} 
& \colhead{T$_{\rm MBB}^{b}$} 
& \colhead{L$_{\rm MBB}^{b}$} 
& \colhead{M$_{\rm MBB}^{b,c}$} 
& \colhead{$\lambda_{\rm peak,MBB}^{b}$}\\
\colhead{}
& \colhead{[K]} 
& \colhead{[L$_{\odot}$]} 
& \colhead{$\%$}
& \colhead{[K]} 
& \colhead{[L$_{\odot}$]} 
& \colhead{[$\msun$]}
& \colhead{[$\micron$]}
}
\startdata
061012$^d$ & 32.1$\pm$0.9 & 0.75$\pm$0.06 & 3.1 & 19.2$\pm$0.4 & 0.68$\pm$0.07 & 0.20$\pm$0.04 & 133  \\
119019 & 34.4$\pm$0.9 & 1.56$\pm$0.14 & 2.6 & 19.9$\pm$0.4 & 1.35$\pm$0.11 & 0.32$\pm$0.06 & 127  \\
HOPS169 & 35.4$\pm$0.9 & 4.50$\pm$0.42 & 2.4 & 20.4$\pm$0.5 & 3.85$\pm$0.41 & 0.81$\pm$0.20 & 127  \\
019003$^d$ & 33.6$\pm$1.1 & 3.16$\pm$0.29 & 1.1 & 21.4$\pm$0.8 & 1.52$\pm$0.30 & 0.24$\pm$0.18 & 121  \\
082005$^d$ & 29.3$\pm$0.8 & 1.02$\pm$0.11 & 5.4 & 17.0$\pm$0.4 & 1.02$\pm$0.13 & 0.61$\pm$0.15 & 151  \\
HOPS372 & 36.9$\pm$1.0 & 4.90$\pm$0.45 & 1.9 & 21.3$\pm$0.6 & 3.96$\pm$0.43 & 0.65$\pm$0.17 & 121  \\
082012 & 32.2$\pm$0.9 & 6.27$\pm$0.65 & 3.8 & 18.6$\pm$0.4 & 5.89$\pm$0.71 & 2.20$\pm$0.52 & 139  \\
090003 & 36.0$\pm$0.8 & 2.71$\pm$0.24 & 1.8 & 21.4$\pm$0.4 & 2.06$\pm$0.15 & 0.33$\pm$0.06 & 121  \\
HOPS358 & 44.3$\pm$0.9 & 30.6$\pm$2.21 & 0.6 & 27.2$\pm$1.0 & 19.5$\pm$1.23 & 0.78$\pm$0.14 & 88.  \\
091015$^d$ & 30.9$\pm$0.9 & 0.81$\pm$0.07 & 3.3 & 18.7$\pm$0.4 & 0.72$\pm$0.06 & 0.24$\pm$0.04 & 136  \\
091016$^d$ & 29.1$\pm$0.9 & 0.65$\pm$0.06 & 4.1 & 17.9$\pm$0.3 & 0.61$\pm$0.06 & 0.26$\pm$0.05 & 142  \\
HOPS373 & 36.0$\pm$0.9 & 5.20$\pm$0.48 & 2.3 & 20.2$\pm$0.4 & 4.21$\pm$0.32 & 0.93$\pm$0.16 & 127  \\
093005$^d$ & 30.8$\pm$0.9 & 1.71$\pm$0.15 & 3.4 & 18.8$\pm$0.4 & 1.57$\pm$0.15 & 0.52$\pm$0.10 & 136  \\
302002$^d$ & 28.6$\pm$0.9 & 0.84$\pm$0.10 & 6.1 & 16.2$\pm$0.3 & 0.82$\pm$0.11 & 0.63$\pm$0.15 & 157  \\
HOPS359 & 39.3$\pm$0.9 & 12.6$\pm$1.00 & 1.3 & 23.4$\pm$0.7 & 9.56$\pm$0.84 & 0.91$\pm$0.23 & 115  \\
HOPS341 & 36.3$\pm$1.1 & 3.62$\pm$0.37 & 3.1 & 19.1$\pm$0.5 & 3.16$\pm$0.35 & 0.97$\pm$0.24 & 133  \\
097002$^d$ & 33.4$\pm$0.9 & 1.14$\pm$0.11 & 2.8 & 19.4$\pm$0.5 & 0.99$\pm$0.11 & 0.27$\pm$0.07 & 130  \\
HOPS354 & 37.4$\pm$0.8 & 7.53$\pm$0.74 & 1.9 & 21.0$\pm$0.4 & 5.84$\pm$0.45 & 1.05$\pm$0.19 & 124  \\
\hline
B\,335$^e$    & 40.3$\pm$1.3 & 0.55$\pm$0.04 & 2.6 & 20.0$\pm$0.7 & 0.47$\pm$0.04 & 0.11$\pm$0.02 & 130  \\
CB\,130$^e$   & 55.2$\pm$3.3 & 0.22$\pm$0.01 & 10.9 & 12.7$\pm$0.5 & 0.17$\pm$0.01 & 0.55$\pm$0.11 & 199  \\
CB\,244--SMM1$^{e}$ & 63.2$\pm$2.6 & 1.72$\pm$0.14 & 1.8 & 20.3$\pm$0.8 & 1.10$\pm$0.11 & 0.24$\pm$0.04 & 127 \\
CB\,68$^e$     & 49.2$\pm$1.4 & 0.68$\pm$0.06 & 1.9 & 19.8$\pm$1.1 & 0.42$\pm$0.05 & 0.10$\pm$0.02 & 130  \\
VLA1623--243$^f$   & 34.0$\pm$1.2 & 2.82$\pm$0.32 & 2.4 & 21.3$\pm$0.5 & 2.55$\pm$0.24 & 0.48$\pm$0.09 & 124 
\enddata
\label{tab:mbb}
\tablenotetext{a}{Our L$_{\rm smm}$/L$_{\rm bol}$ may be
  underestimated as the L$_{\rm smm}$ values are measured from the
  modified black--body fits to the SEDs (see text).}
\tablenotetext{b}{Best--fit modified black--body parameters; the
  parameters shown here are derived from fitting the $\lambda \geq
  70$~\micron\ SED points.  Excluding the 70~\micron\ point and
  fitting only the $\lambda \geq 100$~\micron\ points systematically
  increases the derived masses by 20\% on average, decreases the
  temperature by 5\% on average, and decreases the luminosity by 7\%
  on average.}
\tablenotetext{c}{The masses derived here will increase by a factor of
  $\sim\!$4 if we assume an ISM--type dust model \citep{draine84}.
  Furthermore, we assume a gas--to--dust ratio of 110 (see text).}
\tablenotetext{d}{The values of T$_{\rm bol}$ and L$_{\rm bol}$ for
  these sources should be considered upper limits because we include
  the 24~\micron\ upper limits in our analysis for these sources.}
\tablenotetext{e}{Additional well--studied isolated Class~0 sources
  from \citet{laun12} and \citet{stutz10} are shown for comparison;
  the distances to these sources lie between 100~pc and 250~pc.}
\tablenotetext{f}{SED from J.\ Green and DIGIT team,
  private communication, 2012, and Green et al., in prep.}
\end{deluxetable*}

\end{document}